\documentclass[onecolumn,preprintnumbers,amsmath,amssymb]{revtex4}
\usepackage{graphicx}
\usepackage{amsfonts}
\usepackage{amssymb}
\usepackage{rotating}
\usepackage{booktabs}
\usepackage{xcolor}
\usepackage{soul}
\usepackage{color}
\usepackage{slashed}
\usepackage{multirow}
\usepackage{makecell}
\usepackage{epsf}
\usepackage{ulem}
\usepackage{cancel}
\usepackage{color,bm}
\usepackage[colorinlistoftodos]{todonotes}
\usepackage[colorlinks=true,citecolor=cyan,urlcolor=blue,bookmarks=true,bookmarks=true,bookmarksopen=true,bookmarksnumbered=true,bookmarksopenlevel=3]{hyperref}

\usepackage{diagbox}

\definecolor{airforceblue}{rgb}{0.36, 0.54, 0.66}
\definecolor{steelblue}{rgb}{0.27, 0.51, 0.71}
\definecolor{amber}{rgb}{1.0, 0.49, 0.0}

\def\comment#1{}

\begin{document}

\title{Inclusive diffractive $\rm \eta_{c}$ production in pp, pA and AA modes at the LHC}
\date{\today}
\author{\textsc{Tichouk}}
\author{\textsc{Hao Sun}\footnote{Corresponding author: haosun@mail.ustc.edu.cn \hspace{0.2cm} haosun@dlut.edu.cn}}
\author{\textsc{Xuan Luo}}
\affiliation{Institute of Theoretical Physics, School of Physics, Dalian University of Technology, \\ No.2 Linggong Road, Dalian, Liaoning, 116024, P.R.China }

\begin{abstract}

In this paper, the inclusive Pomeron-Pomeron, Reggeon-Reggeon, Pomeron-Reggeon as well as gluon-Pomeron(-Reggeon) and photon-Pomeron(-Reggeon) interactions for the $\rm \eta_{c}$ at the LHC energies  have been examined in proton-proton, proton-nucleus and nucleus-nucleus collision modes. The cross section has been computed based on the NRQCD factorization and Regge theory formalism. 
The cross exchange of Pomeron-Reggeon contribution is important in $\rm pp$ and $\rm pA$ modes. 
The Pomeron-Pomeron contribution is significant in $\rm AA$ mode. The Pomeron contribution is considerable for $\rm AA$ and $\rm pA$ modes in single diffractive process where $\rm A$ undergoes the diffractive process. Reggeon contribution is sizable in $\rm pp$ and $\rm pA$ modes where $\rm p$ only undergoes the diffractive process. 
The Pomeron and Reggeon contributions in photon-Pomeron and-Reggeon process remain smaller than that of gluon-Pomeron and-Reggeon processes. 
Our results show that the experimental study of Reggeon, Pomeron and their cross exchange can be carried out in certain kinematic windows with the specific choice of the mode at the LHC. The investigation can be useful to better constrain the Reggeon and Pomeron parton content. The inclusive process serves as the background to related exclusive processes which should be predicted.

\vspace{0.5cm}
\end{abstract}
\maketitle
\setcounter{footnote}{0}

\section{INTRODUCTION}
\label{Intro}

The recent experimental cross-section results of the prompt $\rm \eta_{c} (1S)$ hadroproduction from LHCb  \cite{Aaij:2014bga,Aaij:2019gsn} in $\rm pp$ collisions  have opened a window for the pseudoscalar quarkonia production investigation. The data of a considerable interest can be used to probe the interplay between the long and short distance QCD regimes of the strong interactions within a controlled parameter environment. This interplay is a valuable tool to test the ideas and methods of the QCD physics of bound states for example effective field theories, lattice QCD and NRQCD \cite {HarlandLang:2009qe}. To understand this released experimental data, the examination of direct $\rm \eta_{c}$ hadroproduction at leading order (LO) in $\alpha_{s}$ within nonrelativistic QCD (NRQCD) framework  has been carried out in Refs \cite{Biswal:2010xk,Likhoded:2014fta,Mathews:1998nk,Hao:1999kq,Lansberg:2019adr} and at  next-to-leading order (NLO) in Refs \cite{Butenschoen:2014dra,Han:2014jya,Lansberg:2017ozx,Feng:2019zmn}. They have achieved good agreement with almost all the experimental measurements on quarkonia hadroproduction and almost clarified the ambiguity on the long distance matrix elements determination \cite{Zhang:2014ybe,Feng:2019zmn,Lansberg:2017ozx}. However, the polarization prediction in hadroproduction within conventional NRQCD calculations compared to the world's data  \cite{Butenschoen:2014dra} has become rather puzzling, for example  the study of small or no polarization in $\rm J/\psi$ meson prompt production \cite{Aaij:2013nlm} remains inexplicable within the available theoretical framework\cite{Brambilla:2010cs}. The overall scenario was even known as challenging \cite{Butenschoen:2014dra} because the accessible theory lost its power of prediction  by a huge factor off the measured cross section \cite{Han:2014jya}. Therefore, the further information on the long-distance matrix elements and the heavy-quark spin-symmetry \cite{Butenschoen:2012px,Chao:2012iv} are required through the investigation of  $\rm \eta_c$ hadroproduction and photon-induced production in collinear momentum space with off-shell matrix elements or transverse momentum dependent\cite{Echevarria:2019ynx,Fleming:2019pzj,Ma:2012hh}, transverse momentum space with off-shell matrix elements \cite{Baranov:2019joi} and the potential model in the transverse momentum space with off-shell gluons \cite{Babiarz:2019mag}.

It is expected that the $\rm \eta_c$ production can be also important to investigate the soft interactions at the LHC through a variety of diffractive processes in one common framework, i.e. single diffraction, double Pomeron or Reggeon exchange, the double Pomeron-Reggeon cross exchange and central exclusive production in which no quantum numbers are swapped between interacting particles at high energies as well as two-photon exchange. The two distinct characterizations of soft interactions are the exclusive and inclusive events \cite{PhysRevD.101.054035}. The central exclusive process can occur in quantum electrodynamics (QED) via two photon exchange from the two incoming hadrons (proton or nucleus) in ultraperipheral heavy ion collisions where nothing else is produced except the leading hadrons and the central produced object. In QCD, the exclusive diffraction arises via two gluon exchange (soft pomeron) between the two incoming hadrons or quarks \cite{Dechambre:2011py}. From the pair of gluons, one of them perturbatively couples to the hard process and the second gluon plays the role of a soft screening of color, permitting the diffraction to take place\cite{Royon:2006kn}. These exclusive collisions are topologically characterized by two empty regions in pseudo-rapidity called large rapidity gaps, separating the intact very forward hadron from the central massive or light produced object in the final state. The Pomeron and photon are considered as a color singlet object. The intact hadrons are hadrons which have lost a small fraction of their energy and are thus scattered at very small angle with respect to the beam direction. The forward hadron tagging detectors are  inserted close to the beam pipe at a large distance from the interaction point  and can move close to the beam, when the beam is stable, to observe intact outgoing hadrons \cite{Albrow:2008pn,Trzebinski:2015bra} after the interaction. The total energy of Pomeron and photon are consumed to form the leading hadrons and the central object. There is no energy loss and Pomeron remnants\cite{Royon:2018hvi}. So, the exclusive process has the best experimental signature. The leading hadrons take most of the beam hadron momentum and the total colliding energy available is expended in the collision. The exclusive production of the light and heavy ($\rm \chi_{cJ}$, $\rm \eta_c$, $\rm J/\psi$, Higgs boson, W pair and so on) central produced object via two gluon exchange have been studied in Durham model with tagged hadron or antihadron \cite{Harland-Lang:2014lxa,Pasechnik:2009bq,Pasechnik:2009qc,HarlandLang:2009qe,N.Cartiglia:2015gve} implemented in the Monte Carlo event generators SuperChic \cite{Harland-Lang:2015cta} and FPMC (Forward Physics Monte Carlo)\cite{Boonekamp:2011ky}. The Durham model presents typical features based on Regge theory and it relies on the particle spin and parity. These quantum numbers are modified by the loop integration around the internal gluon transverse momentum, the non-zero transverse momentum outgoing hadrons effects and the screening corrections coming from multi-Pomeron exchanges \cite{HarlandLang:2010ep}.The Durham model would offer a useful source of spin-parity information about the centrally produced system and an important test of the overall theoretical formalism. It has been noted the purely exclusive production predictions can be valuable to investigate the features of the produced quarkonium states or other objects. Therefore, the clear understanding of exclusive production comes from the consideration of the inclusive production\cite{Boonekamp:2002vg} which acts as their useful background. 

Unlike the exclusive process, the inclusive diffractive process is identified  by  at least one non-exponentially suppressed large rapidity gap due to the presence of Pomeron or Reggeon remnants  escorted with soft gluon radiations and hadron remnants. The rapidity gap  in inclusive  diffractive process  is smaller than that of the exclusive one due to a large loss of hadron energy. The colliding color singlet Pomerons or Reggeons are hadron-like systems composed of quarks and gluons. The inclusive process can be the result of the double Pomeron or Reggeon exchange or Pomeron-Reggeon cross exchange interaction (double diffraction shortened as DD) where Pomeron or Reggeon can emit quark or gluon to produce the central object accompanied with two intact forward hadrons, remnants and two rapidity gaps in the final state. The both hadrons are not destroyed. Moreover, it can be the result of gluon-Pomeron and gluon-Reggeon  interactions called the single diffraction (SD). One of two hadrons is completely destroyed and there are presence of one intact forward hadron, remnants, one large rapidity gap and the produced object in the final state. There are also strongly inclusive competing production channels that can  happen through photon-Pomeron and photon-Reggeon induced interactions \cite{Goncalves:2018yxc,Basso:2017mue,Goncalves:2015cik} where one of two hadrons is entirely damaged. In the final state, there exist intact hadrons, two rapidity gaps, remnants and produced object.  The rapidity gap measurement and hadron tagging are the two common techniques applied in collider experiments to select diffractive events. The diffractive process with the forward hadron tagging detector is a better understanding of the structure of the Pomeron and Reggeon. The comparison and the ratio  of the inclusive SD, DD and ND  cross sections are also useful to investigate the structure of the Pomeron and Reggeon, in terms of its quark and gluon content, in hard scattering processes\cite{Acosta:2003xi,Gallinaro:2004it}. The inclusive double, single and induced diffractive processes are studied in Regge theory. The photon spectrum is described by the equivalent hadron approximation proportional to $Z^{2}$. In this approximation, the electromagnetic field generated by the fast moving hadrons can be considered as an intense photon beam. The photons exchanged by the colliding hadrons are almost on their mass shell (low virtualities $\rm Q^{2}$). The photon can either interact elastically with the entire hadron (coherent reaction) \cite{Adam:2018tdm} or interact inelastically with individual nucleon (incoherent reaction) \cite{Klein:1999qj}. The incoherent interaction has a larger average $\rm t$, and occurs at a somewhat lower rate. Thus, the released experimental and diffractive data samples  by the LHC \cite{White:2008zzg,Aaltonen:2010qe,Aaltonen:2012tha}  have brought the diffractive production theory to be thoroughly studied in different models. For example, the inclusive diffractive processes have been examined in Regge theory or resolved-Pomeron model in Refs\cite{Ingelman:1984ns,Goncalves:2017bmo,Royon:2006by,Luszczak:2016csq} under the perturbative QCD and the soft diffractive physics. Certain  papers have pointed out and addressed the non negligible Reggeon contribution \cite{Luszczak:2014mta,Luszczak:2014cxa}. The other existing theory as such Donnachie-Landshoff \cite{Kochelev:1999wv,Donnachie:1987gu,Royon:2006kn} model and Bialas-Landshoff model\cite{Rangel:2006mm} have also investigated diffractive production of particles. Studies of diffractive events should bring more insight into description of Pomeron and Reggeon  to verify predictions of the Regge model. At high energies, the dominant hadronic signal comes from non diffractive and inclusive process via gluon-gluon fusion where the color charge partice is exchanged between the interacting hadrons \cite{Staszewski:2014hua}. In non-diffractive (ND) events for which final state particle production occurs in the entire pseudorapidity space available, rapidity gaps are exponentially suppressed as a function of gap size  \cite{Gallinaro:2004it}. The large rapidity gap in non-diffractive is due to statistical fluctuations of the distance between neighboring particles. In this normal production, the hadrons are completely destroyed and escorted by the central product object.
 
\begin{figure}[htp]
	\centering
	\includegraphics[height=3.8cm,width=4.4cm]{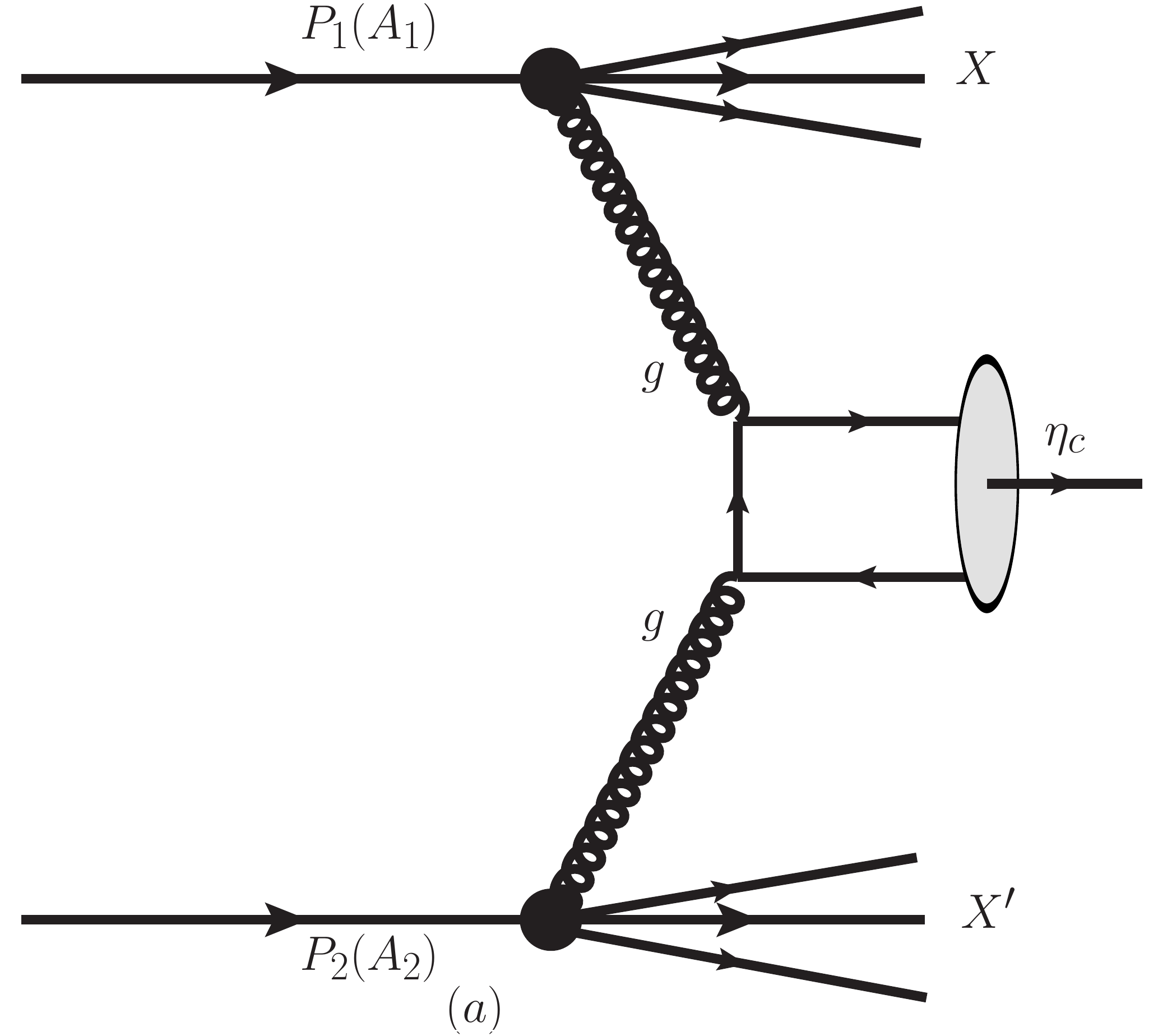}
	\includegraphics[height=3.8cm,width=4.4cm]{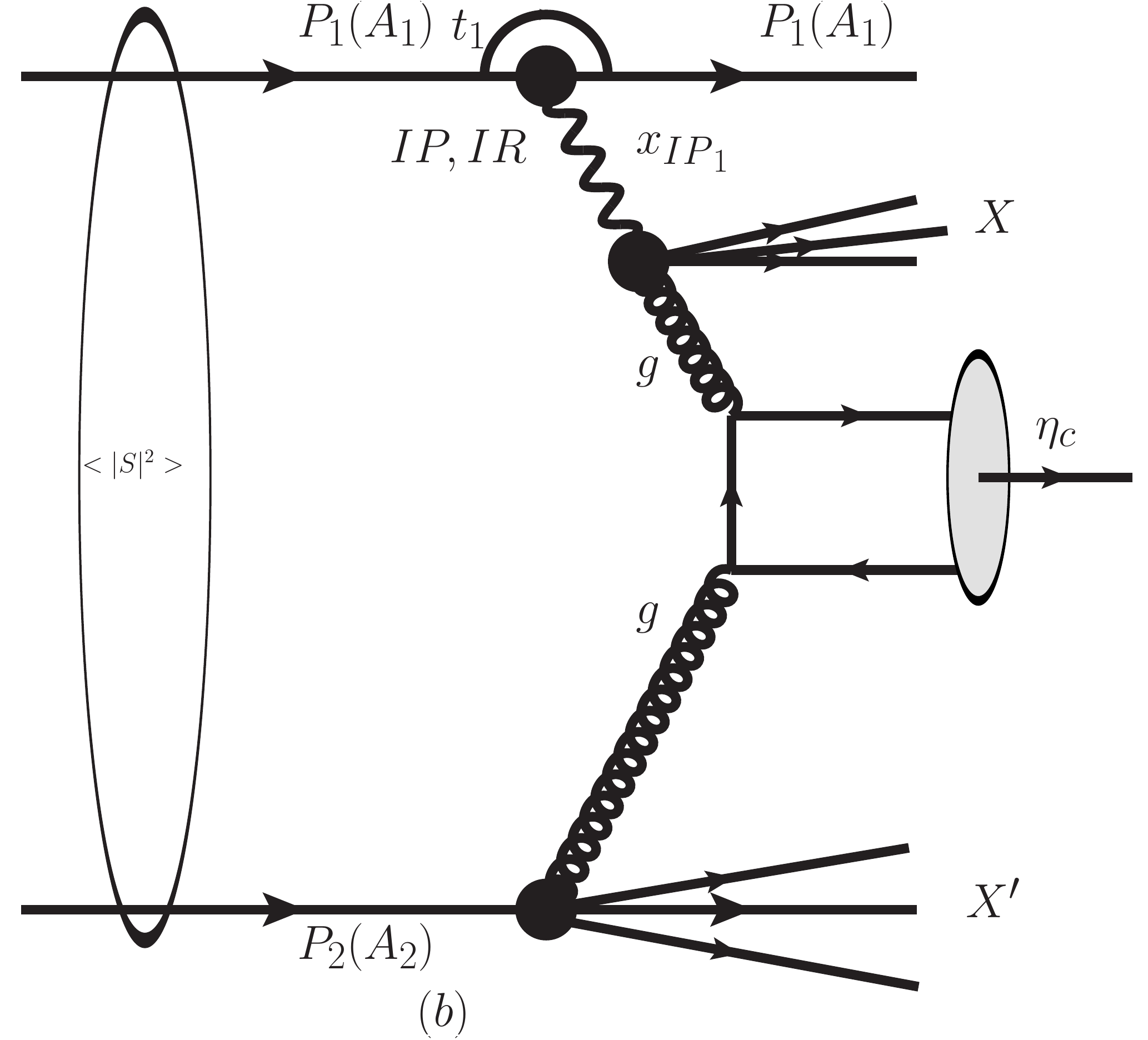}
	\includegraphics[height=3.8cm,width=4.4cm]{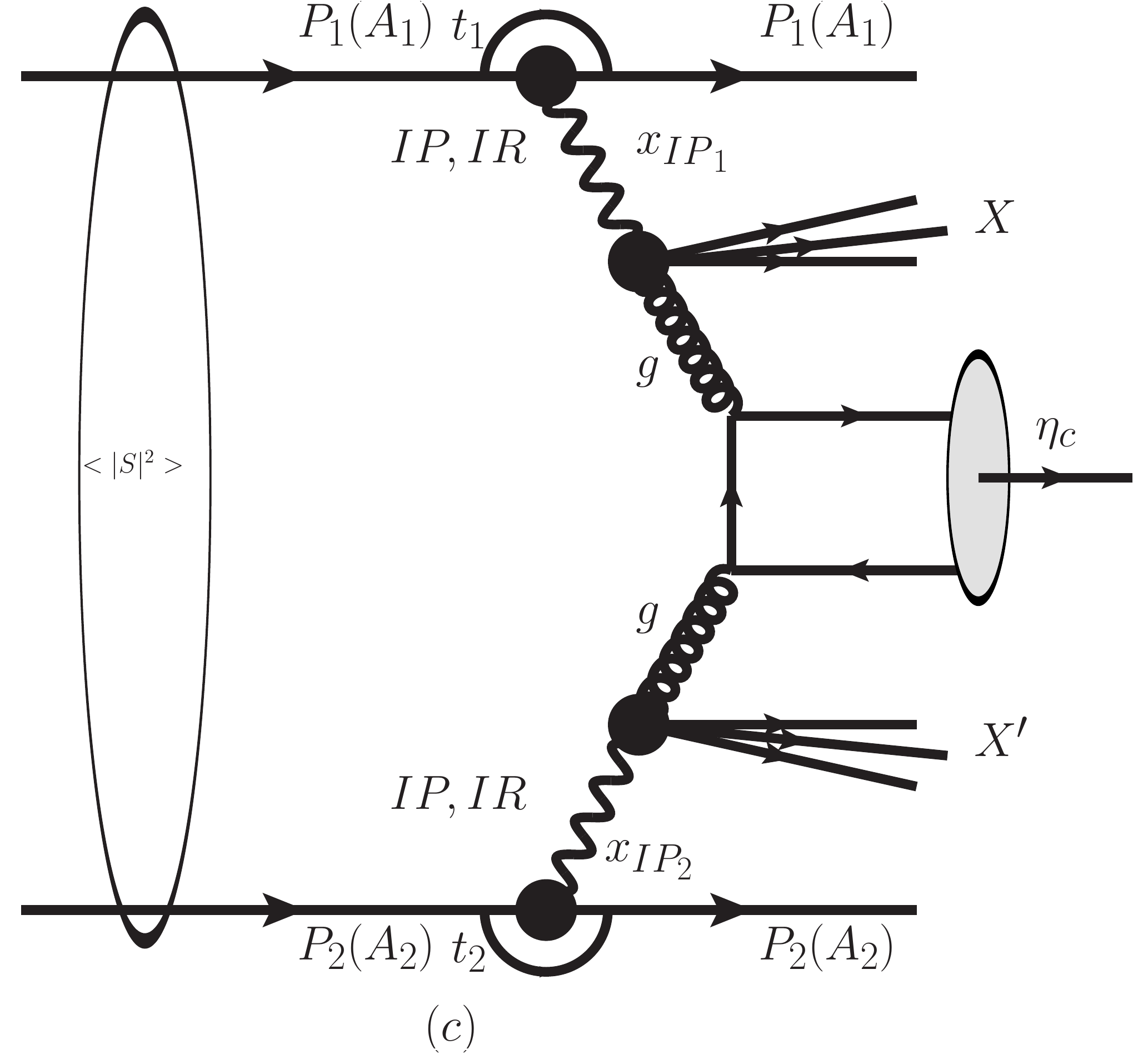}
	\includegraphics[height=3.8cm,width=4.4cm]{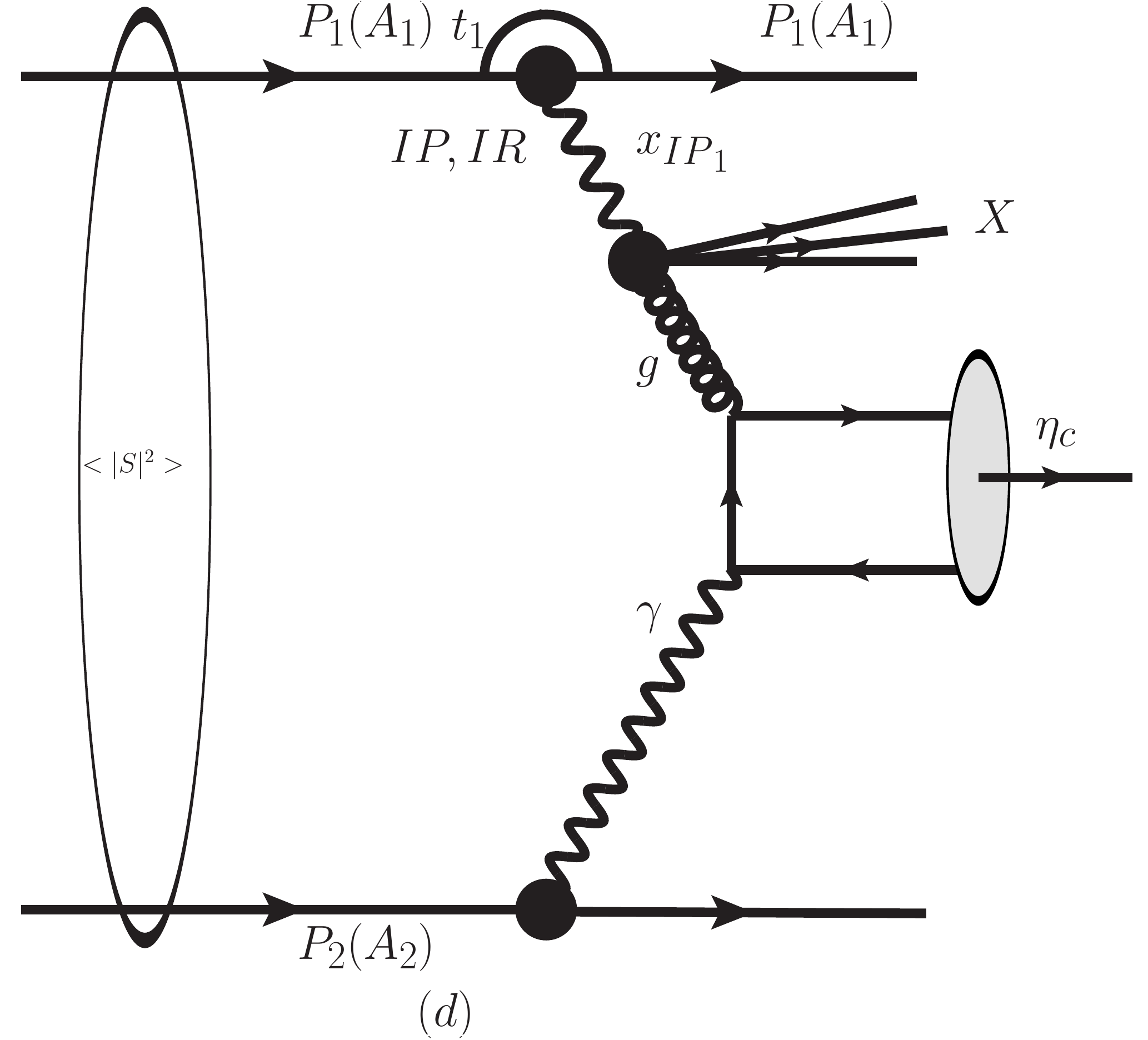}
	\caption{ \normalsize 
		Diagram representing single $\rm \eta_{c}$ quarkonium hadron-hadron, hadron-nucleus and nucleus-nucleus production in non diffractive (ND) (a), single diffractive SD (b), double diffractive DD (c) and diffractive photon - hadron (IN) (d) interactions.}
	\label{fig1:limits}
\end{figure}

In this paper, we have estimated the total inclusive cross section for $\rm \eta_{c}$ meson in
the single and double diffractive processes along with non diffractive process by updating the Ref.\cite{PhysRevD.101.054035} in $\rm pp$ mode to $\rm pA$ and $\rm AA$ modes in the first place. We have also added the induced inclusive diffractive process, photon-Pomeron and photon-Reggeon, in $\rm pp$, $\rm pA$ and $\rm AA$ modes in the second place. The  detailed comparison of the SD and DD to ND cross section are carried out and important to probe the structure of the Pomeron and Reggeon, in terms of its  gluon content, in hard scattering processes. These inclusive processes are the background contributions of the exclusive processes and their detailed determinations are required. They are sensitive to gluon content of Pomeron (Reggeon) and the Pomeron (Reggeon) themselves are sensitive to the gluon distribution in the hadron. The production model is based on collinear momentum space in NRQCD formalism with Regge theory. The related diagrams are illustrated in Fig.\ref{fig1:limits}. 

The paper is outlined into three sections including the introduction in Section \ref{Intro}. The  formalism framework for the leading order cross section of $\rm \eta_{c}$ hadro and induced productions at the LHC are clearly described in Section \ref{FRAMEWORK}. The input parameters and discussed numerical results are shown in Section \ref{NUMERICAL}. A summary is briefly given in Section \ref{Conclusion}.

\section{Calculation FRAMEWORK }
\label{FRAMEWORK}
\subsection{Cross section formalism}

The non diffractive (ND), single diffractive (SD), induced process (IN) and double diffractive (DD) hadron-hadron, hadron-nucleus and nucleus-nucleus reactions read as
\begin{eqnarray}\nonumber
&&\rm ND:\ \ \ h_{1}h_{2}\to \eta_{c}\ X \\ \nonumber
&&\rm SD:\ \ \ \ h_{1}h_{2}\to h_{1}\otimes X+\eta_{c}+X{'} \\
&&\rm DD (IN):\ \ \ h_{1}h_{2}\to h_{1}\otimes X+\eta_{c}+X{'}\otimes h_{2} ,
\end{eqnarray}
where $\rm h_{1}h_{2}$ is symbolized by $\rm pp$, $\rm pA$ and $\rm AA$ in our computations. The total inclusive  double  diffractive cross section  is formulated as  the convolution of partonic cross section, diffractive  gluon distribution functions of  incident particles for the correspondent process in gluon-gluon fusion and can be written as
\begin{eqnarray}\label{a}\nonumber
\rm \sigma^{DD} (\rm h_{1}h_{2}\to \eta_{c}X)&=& \rm \langle \left\vert S\right\vert ^{2}\rangle^{DD} \sum_{i,j,n}\int_{0}^{1}\frac{dx_{i }}{
	x_{i}}\int_{0}^{1}\frac{dx_{j}}{x_{j}}\hat{\sigma}(ij\rightarrow Q\overline{Q}
[n]+X) \\
&&\rm \langle 0|\mathcal{O}^{\eta_{c}}[n]|0\rangle\rm \lbrack \mathcal{F}^{D}_{i}(x_{i},\mu ^{2})\mathcal{F}^{D}_{j}(x_{j},\mu ^{2})+(h_{1}\leftrightarrow h_{2})].
\end{eqnarray}
For the  total inclusive single  diffractive cross section case, one of the diffractive gluon distributions  is replaced by a regular conventional integrated  gluon distribution, and  superscript DD is replaced by SD. For total inclusive induced cross section case, one of the diffractive gluon distributions  is replaced by an equivalent photon  flux, and  superscript DD is replaced by IN. For total inclusive non diffractive cross section case, both of the diffractive gluon distributions  are replaced by a regular conventional integrated  gluon distribution, the survival probability factor is removed and, superscript DD is replaced by ND. The long-distance matrix element (LDME) $\rm \langle 0|\mathcal{O}_{(1,8)}^{\eta_{c} }[n]|0\rangle$  describes the hadronization of the $\rm Q\overline{Q}$ heavy pair into the colorless physical observable quarkonium state $\rm \eta_{c}$. The $\rm \hat{\sigma}(gg\rightarrow Q\overline{Q}[n])$ and $\rm \hat{\sigma}(\gamma g\rightarrow Q\overline{Q}[n])$ denote the short-distance cross sections for the partonic process $\rm gg\rightarrow Q\overline{Q}[n]$ and $\rm \gamma g\rightarrow Q\overline{Q}[n]$, respectively. They  are found by operating the covariant projection method \cite{Tichouk:2019lxk,Petrelli:1997ge}. The Fock state $\rm n$ are given as follows:  $\rm ^{1}S_{0}^{[1]}$, $\rm ^{1}S_{0}^{[8]}$ for $\rm gg\to Q\overline{Q}[n]$ and $\rm \gamma g\to Q\overline{Q}[n]$ partonic process \cite{Basu:2002rk}. The contribution of color singlet states for $\rm \eta_{c}$ quarkonium production is at leading power in velocity (v) while the color octet contribution to S-wave quarkonium production are power suppressed \cite{Kang:2013hta}. The nuclear gluon distribution is given by $\rm x\rm \mathcal{F}_{g/A}(x,\mu_{f})$=$\rm A\cdot R_{g}\cdot x\rm \mathcal{F}_{g/p}(x,\mu_{f})$, where $\rm R_{g}$ is the nuclear modification factor for gluon taking into account the nuclear shadowing effects \cite{Armesto:2006ph} and $\rm R_{g}=1$ disregards the shadowing corrections. The gluon shadowing effect can be understood as an interaction of the gluons of different nucleons in nucleus. At small $\rm x_{g}$, there can be a spatial overlap of the gluons of different nucleons which may lead to gluon recombination, producing a decrease in their density. The gluon probability density in the nucleon is smaller within the nuclear matter than for a free nucleon \cite{MartinBlanco:2016ach}. $\rm x_{i,j}$ are  Bjorken variable defined as the momentum fractions of the proton, nucleus, Pomeron or Reggeon carried by the gluon or photon. $\rm \langle \left\vert S\right\vert ^{2}\rangle$ is the gap survival probability  factor. The partonic cross section, the matrix element squared \cite{PhysRevD.88.014008,Maltoni:2004hv,Cho:1995ce,Hao:1999kq,Hao:2000ci} and the colliding energy are formulated as 
\begin{equation}
\begin{split}
\rm  \hat{\sigma}(gg\rightarrow Q\overline{Q}[n])=&\rm\frac{\pi}{M_{\eta_{c}}^{2}}\delta(\hat{s}-M_{\eta_{c}}^{2})\overline{\sum }|\mathcal{A}_{S,L}|^{2},\\
\rm  \hat{\sigma}(\gamma g\rightarrow Q\overline{Q}[n])=&\rm\frac{\pi}{M_{\eta_{c}}^{2}}\delta(\hat{s}-M_{\eta_{c}}^{2})\overline{\sum }|\mathcal{A^{\prime}}_{S,L}|^{2},
\end{split}
\end{equation}
where $\rm \hat{s}=\rm x_{1}x_{2}s$,
\begin{equation}
\begin{split}
\rm \overline{\sum } |\mathcal{A}_{S,L}|^{2}=&\rm \frac{2}{9}\frac{\pi^2\alpha^{2}_{s}}{M_{\eta_{c}}}\langle 0|\mathcal{O}_{1}^{\eta_{c}}(^{1}S_{0})|0\rangle + \frac{5}{12}\frac{\pi^2\alpha^{2}_{s}}{M_{\eta_{c}}}\langle 0|\mathcal{O}_{8}^{\eta_{c}}(^{1}S_{0})|0\rangle\\
\rm \overline{\sum } |\mathcal{A^{\prime}}_{S,L}|^{2}=& \rm \frac{8\pi^2e^{2}_{c}\alpha_{s}\alpha}{M_{\eta_{c}}}\langle 0|\mathcal{O}_{8}^{\eta_{c}}(^{1}S_{0})|0\rangle .
\end{split}
\end{equation}

\subsection{Pomeron and Reggeon structure functions, Equivalent photon flux}

As expressed by the so-called proton-vertex factorization or the Resolved Pomeron Model \cite{Ingelman:1984ns}, the collinear diffractive gluon, $\rm \mathcal{F}^{D}(x_{g}, \mu^{2}_{f}, x_{\mathbb{P}})$ is defined as a convolution of the Pomeron (Reggeon) flux emitted by the diffracted proton or nucleus, $\rm f^{h}_{\mathbb{P,R} } (x_{\mathbb{P}} )$, and the gluon distribution in the Pomeron (Reggeon), $\rm f^{\mathbb{P,R}}_{g}(\beta,Q^{2})$  where $\rm\beta(=\frac{x_{g}}{x_{\mathbb{P}}})$ is the longitudinal momentum fraction carried by the partons inside the Pomeron. The Reggeon contribution is ignored in the hard diffraction calculations of different final states in most cases. The Reggeon contribution is treated as an exchange of quark and antiquark pair and the parton content of the Reggeon is obtained from the pion structure function \cite{Aktas:2006hy}. The difference between the two contributions exists in the $\rm x_{\mathbb{P}}$ and $\rm t$ dependence of their fluxes, where the Reggeon exchange is mostly significant at high $\rm x_{\mathbb{P}}$ at the edge of the foward hadron detector acceptance, remarkably for $\rm x_{\mathbb{P}}>0.1$. $\rm x_{\mathbb{P}}$ stands also for $\xi$. The Reggeon shape of the $\rm t$ distribution is also different showing a less steep decrease than in the Pomeron case. Nevertheless, as shown in Ref.\cite{Marquet:2016ulz}, this contribution is significant in some regions of the phase space and needed to obtain a good description of the data, the collinear diffractive gluon distribution of the proton at low $\rm \beta$ and large $\rm x_{\mathbb{P}}$ \cite{Rasmussen:2015qgr,Aktas:2006hx} is parameterized in Ref.\cite{PhysRevD.101.054035} and formulated by
\begin{eqnarray}\label{654}
\rm {\mathcal{F}^{D}(x_{g},Q^{2},x_{\mathbb{P}})=\int_{x_{g}}^{1}\frac{dx_{\mathbb{P}}}{x_{IP}}
f_{\mathbb{P}}^{h}(x_{\mathbb{P}})f^{\mathbb{P}}_{g}(\frac{x_{g}}{x_{\mathbb{P}}},Q^{2})+n_{\mathrm{\mathbb{R}}
}\int_{x_{g}}^{1}\frac{dx_{\mathbb{P}}}{x_{\mathbb{P}}}f_{\mathbb{R}}^{h}(x_{\mathbb{P}})f^{\mathbb{R}}_{g}(\frac{x_{g}}{
x_{\mathbb{P}}},Q^{2})}
\end{eqnarray}
and the Pomeron and Reggeon fluxes are literally expressed by
\begin{equation}\label{m}
\begin{split}
\rm  f_{\mathbb{P,R}}^{p}(x_{\mathbb{P}})= & \rm
\int_{t_{\min }}^{t_{\max
}}dtf_{\mathbb{P},\mathbb{R}/p}(x_{\mathbb{P}},t)=  \rm
\int_{t_{\min }}^{t_{\max }}dt\frac{A_{\mathbb{P,R}}e^{B_{\mathbb{P,R}}t}
}{x_{\mathbb{P}}^{2\alpha _{\mathbb{P,R}}(t)-1}},\\
\rm  f_{\mathbb{P,R}}^{A}(x_{\mathbb{P}})= & \rm
\int_{t_{\min }}^{t_{\max
}}dtf_{\mathbb{P},\mathbb{R}/A}(x_{\mathbb{P}},t)=  \rm
R_{g}A^{2}\int_{t_{\min }}^{t_{\max }}dt\frac{A_{\mathbb{P,R}}e^{B_{\mathbb{P,R}}t}
}{x_{\mathbb{P}}^{2\alpha _{\mathbb{P,R}}(t)-1}}.F^{2}_{A}(t),
\end{split}
\end{equation}
where $\rm R_{g}$ is the suppression factor associated to the nuclear shadowing and $\rm F_{A}(t)$ is the nuclear form factor. In what follows we will assume that $\rm R_{g} = 0.15$ as in Ref.\cite{Guzey:2016tek} and that $\rm F_{A} (t) = e^{{R^{2}_{A}t}/6}$\cite{Boreskov:1992ur}, with $\rm R_{A}=1.22A^{1/3}$ fm and $\rm R_{p} = 0.7$ fm\cite{GayDucati:2010vu} being the nuclear and proton radii, respectively. The number of nucleons or  mass number  of atom is defined by $\rm A$ and the electric charge number by $\rm Z$. For the sake of classification, the nuclides ($\rm^{A}_{Z}X$) are usually divided into three broad categories: light nuclei like Beryllium ($\rm^{9}_{4}Be$), Carbon ($\rm^{12}_{6}C$) and Oxygen ($\rm^{16}_{8}O$) with mass number $\rm A<25$; intermediate nuclei like Calcium ($\rm ^{40}_{20}Ca$), Copper($\rm^{64}_{29}Cu$) and Silver ($\rm^{108}_{47}Ag$) ; and heavy nuclei like Gold ($\rm^{197}_{79}Au$), Lead ($\rm^{208}_{82}Pb$) and Uranium ($\rm^{238}_{92}U$) with mass number $\rm A>150$. The intermediate or medium nuclei are between the light and heavy nuclei \cite{shultis2016fundamentals}. Certain fraction of the pomeron energy is only available for the hard collision and the rest being carried away by a remnant or spectator jet. On every occasion a coloured parton (gluon) is pulled out of a colour-singlet object, Pomeron or Reggeon. The Pomeron structure is well restricted by the fits, fit A and fit B, which evidently reveal that its parton content is gluon dominated. Contrariwise, the HERA data do not restrain the parton distribution function of Reggeon which is therefore needed in order to get a quantitative description of the high-$x_{\mathbb{P}}$ measurements. Consequently, measurements at the LHC will permit to examine the validity of this supposition.

The analytic estimate for the equivalent photon flux \cite{Basso:2017mue} of a hadron ($\rm \mathcal{F}(\omega,b)$) with energy $\omega$ at a transverse distance $\rm b$ from the center of nucleus, defined in the plane transverse to the trajectory, can be modelled and evaluated using the Weizsacker-William method considering the requirement that photoproduction is not followed by hadronic interaction (ultra-peripheral collisions).  The spectrum \cite{Azevedo:2019hqp} can be expressed in terms of the charge form factor $\mathrm{F}(\rm q)$ as follows
\begin{eqnarray}
\rm \mathcal{F}(\omega,b)&=& \rm\frac{ Z^{2}\alpha_{em}}{\pi^{2}b^{2}v^{2}\omega}\left(\int_{0}^{\infty} u^{2}J_{1}(u)\mathrm{F}\left( \sqrt{\frac{\left( \frac{b\omega}{\gamma_{L}}\right)^{2}+u^{2} }{b^{2}}}\right)
\frac{1}{ \left( \frac{b\omega}{\gamma_{L}}\right)^{2}+u^{2}
}du\right)^{2}
\end{eqnarray}
where where $\rm J_{1}$ is the Bessel function of the first kind and $\rm q$ is the four-momentum of the quasireal photon. The Fourrier transform of the  charge form factor gives the charge distribution \cite{Povh:1995mua}. The spectrum of the  equivalent photon flux \cite{Basso:2017mue} from nucleus with the  charge monopole form  factor
$\rm \mathrm{F}(q)=\rm \Lambda^{2}/(\Lambda^{2}+q^{2})$ is parametrized as
\begin{eqnarray}\label{f1}
\rm \mathcal{F}_{\gamma/A}(\xi)&=& \rm\frac{2 Z^{2}\alpha_{em}}{\pi\xi}[\bar{\tau}
K_{0}(\bar{\tau})K_{1}(\bar{\tau})+\frac{\bar{\tau}^{2}}{2}\mathcal{W}(\bar{\tau})] 
\end{eqnarray}
where $\rm \bar{\tau}=\rm \xi \frac{\sqrt{s}}{2} b_{min}/\gamma_{L}$, $\rm b_{min}=\rm R_{h_{1}}+R_{h_{2}}$, $\rm \gamma_{L}=\rm \frac{\sqrt{s}}{2m_{p}}$ or $\rm \frac{\sqrt{s}}{2Am_{p}}$  \cite{Baltz:2007kq} and $\rm \mathcal{W}(\bar{\tau})=\rm K^{2}_{0}(\bar{\tau})-K^{2}_{1}(\bar{\tau})$. The Lorentz boost of a single beam is $\gamma_{L}$. The photon spectrum of a relativistic proton \cite{Basso:2017mue}   with a charge dipole form factor $\rm \mathrm{F}(q)=\rm 1/[1-q^{2}/0.71GeV^{2}]^{2}$
is given by
\begin{eqnarray}\label{f2}
\rm \mathcal{F}_{\gamma/p}(\xi)&=&\rm \frac{\alpha_{em}}{\pi}[\frac{1-\xi+0.5\xi^{2}}{\xi}] \times \left(\ln\Omega-\frac{11}{6}+\frac{3}{\Omega}-\frac{3}{2\Omega^{2}}+\frac{1}{3\Omega^{3}} \right),
\end{eqnarray}
where $\rm \Omega=\rm 1+[(0.71 $GeV$^{2}$)/$ \rm Q^{2}_{min}]$ and $\rm Q^{2}_{min}=\rm (\xi m_{p})^{2}/(1-\xi)$. The computation has been carried out with the use of Eqs.\ref{f1} and \ref{f2}.
The further analytic estimate for the equivalent photon flux of proton and nucleus with a charge monopole and dipole form factors \cite{Vysotsky:2018slo} is parametrized as
\begin{equation}
\begin{split}
\rm \mathcal{F}_{\gamma/A,p}(\xi)=&\rm \frac{Z^{2}\alpha _{em}}{\pi \xi}[(2a+1)\ln(1+\frac{1}{a})-2]\\
\rm \mathcal{F}_{\gamma/A,p}(\xi)=&\rm \frac{Z^{2}\alpha _{em}}{\pi \xi}[(4a+1)\ln(1+\frac{1}{a})-\frac{24a^{2}+42a+17}{6(a+1)^{2}}],
\end{split}
\end{equation}
where $\rm h$ can be either $\rm p$ ($\rm Z=1$) or $\rm Pb$ ($\rm Z=82$), $\rm a=(\omega /\Lambda _{h}\gamma_{L} )^{2}$,
$\rm \omega =\frac{\sqrt{s}}{2}\xi$, $\rm \Lambda_{p} \approx 0.20\times e^{\frac{17}{12}}$ GeV and $\rm \Lambda_{Pb}=80$ MeV \cite{Jentschura:2009mb}.

\subsection{ Gap survival probability in diffractive processes}
The existence of extra soft partonic collisions and new particles in gap rapidity gives rise to the gap survival probability \cite{Khoze:2017sdd}. The description of the gap survival probability can be accouunt for by the eikonal factor where the  additional soft incoming or outgoing hadron-hadron rescatter with multi-Pomerons exchanged, the
enhanced factor where the incoming or outgoing proton interact with the intermediate partons, the Sudakov factor or hard QCD bremsstrahlung which takes into consideration the gluon radiation from annihilation of two energetic coloured particles, the migration which is due to the change of the forward intact proton momentum. The consideration of the collinear momentum space leads to the fact that the enhanced, Sudakov and migration factors are ignored where the intermediate parton transverse momentum space and screening gluon are not under the consideration. The incoming hadrons and outgoing intact forward hadrons have almost the same direction along the $\rm z$-beam axis. Therefore, the often-used paremetrizations of the eikonal gap survival probability \cite{Luszczak:2014mta,Basso:2017mue,Muller:1990sg} in literature denoted by  $\rm \langle \left\vert S\right\vert ^{2}\rangle $ reads
 \begin{eqnarray}\nonumber
&&\rm \langle \left\vert S\right\vert ^{2}\rangle_{pp} =   \frac{B_{1}}{B_{\mathbb{P}}}(\frac{\sigma^{tot}_{pp}(s)}{2\pi B_{1}})^{-B_{1}/B_{\mathbb{P}}}\gamma(B_{1}/B_{\mathbb{P}},\frac{\sigma^{tot}_{pp}(s)}{2\pi B_{1}}),\\  
&&\rm \langle \left\vert S\right\vert ^{2}\rangle_{AA} = \frac{2Q_{1}^{2}}{Q_{0}^{2}}(A^2\frac{\sigma^{tot}_{pp}(s)}{4\pi}Q_{0}^{2})^{-2Q_{1}^{2}/Q_{0}^{2}}\gamma(\frac{2Q_{1}^{2}}{Q_{0}^{2}},A^2\frac{\sigma^{tot}_{pp}(s)}{4\pi}Q_{0}^{2}), \\ 
&&\rm \rm \langle \left\vert S\right\vert ^{2}\rangle_{pA} =2Q_{3}^{2}B_{2}(A\frac{\sigma^{tot}_{pp}(s)}{4\pi B_{2}})^{-2Q_{1}^{2}B_{2}}\gamma(2Q_{1}^{2}B_{2},A\frac{\sigma^{tot}_{pp}(s)}{4\pi B_{2}}) \nonumber. 
\end{eqnarray}
The approximative forms \cite{Luszczak:2014mta,Sun:2017jcg,Muller:1990sg,Khoze:2000wk} for the above formulas can be expressed as 
 \begin{eqnarray}\nonumber
&&\rm \rm \langle \left\vert S\right\vert ^{2}\rangle_{pp} = \frac{a}{b+ln(\sqrt{s/s_{0}})}, \\ 
 &&\rm \langle \left\vert S\right\vert ^{2}\rangle_{AA} =\langle \left\vert S\right\vert ^{2}\rangle_{pp}/AA\\
&&\rm \rm \langle \left\vert S\right\vert ^{2}\rangle_{pA} =\langle \left\vert S\right\vert ^{2}\rangle_{pp}/A \nonumber
\end{eqnarray}
where $\gamma$ is the incomplete gamma function, $\rm a = 0.126$, $\rm b=-4.688$, $\rm Q_{3}^{2} = \frac{6}{R^{2}_{A}+3B_{\mathbb{P}}}$, $\sqrt{s}$=5.02 TeV, Q$_{0,1}$= 60 MeV is an effective parameter fitted to each nucleus \cite{Drees:1988pp} 
$\rm B_{1}=\frac{B_{0}}{2}+\frac{\alpha'}{2}ln(\frac{s}{s_{0}})$ \cite{Guzey:2016tek,Basso:2017mue}, $\rm s_{0}=1$ GeV$^{2}$ for two channel model \cite{Gotsman:2005wa}, B$_{0}$=10 GeV$^{-2}$, $\alpha$'=0.25 GeV$^{-2}$, $\rm B_{2}= \frac{1}{2Q^{2}_{0}}+ \frac{B_{1}}{4}$. The total pp cross section given by the optical theorem is $\rm \sigma^{tot}_{pp}(s)= 69.3286+12.6800log(\sqrt{s})+1.2273log^{2}(\sqrt{s})$\cite{Kohara:2014waa} or 
$\rm \sigma^{tot}_{pp}(s)= 33.73+0.2838log^{2}(s)+13.67s^{-0.412}-7.77s^{-0.5626}$ mb \cite{Agashe:2014kda}. Those extra soft interactions from eikonal factor can spoil the diffractive signature \cite{Bjorken:1992er} and the Regge factorization is known to be infringed in the treatment of diffractive interactions in hadronic collisions. The 
factorization breaking can be compensated by  the gap survival probability, independent of the details of the hard process. The gap survival probability
depends on the specific interaction, the cuts approved in the experiment and represents the last element of the resolved-Pomeron model. A range of attempts have been achieved to predict those probabilities \cite{Kopeliovich:2016rts,Gotsman:2011xc,Khoze:2008cx}, however the actual values are rather uncertain. The selected value can be regarded  a lower limit given the recent available experimental results \cite{Chatrchyan:2012vc,Aad:2015xis}.

For the photon-Pomeron and -Reggeon induced collisions, $\rm \langle \left\vert S\right\vert ^{2}\rangle=1$ is assumed due to the fact that the recent LHC data for the exclusive vector meson production in photon induced interactions can be explained without the presence of a normalization factor associated to absorption effects \cite{Basso:2017mue}. However, it is important to point out that the magnitude of the rapidity gap survival probability in  photon-Pomeron and -Reggeon induced collisions  is still a debatable theme. For instance, the predicted value of $\rm \langle \left\vert S\right\vert ^{2}\rangle $ ranges from $0.8$ to $0.9$
 \cite{Schafer:2007mm} and depends on the rapidity of   meson.

\section{NUMERICAL RESULTS AND DISCUSSION}
\label{NUMERICAL}
In the following section, we discuss the numerical results of the production of $\rm \eta_{c}$ by using some physical parameters such as: the mass of $\rm \eta_{c}$ is of $\rm M_{\eta_{c}}=2.983$ GeV and the colliding energy used in this paper is $\sqrt{s}$=5.02 TeV for $\rm pp$, $\rm AA$ and $\rm pA$. We adopt the leading-order set of the MSTW2008 parametrization \cite{Martin:2009iq} for unpolarized distribution function of gluon from the proton while the gluon from nucleus is given by the EPS09 parametrization and based on a global fit of the current nuclear data \cite{Eskola:2009uj}. The 2006 H1 proton diffractive PDFs (fit A) is used for the pomeron densities inside the proton \cite{Aktas:2006hy,Aktas:2006hx} which are probed at the factorization hard scale ($\mu=Q$) chosen as $\rm \mu=m^{\eta_{c}}_{T}$, where $\rm m^{\eta_{c}}_{T}=M_{\eta_{c}}$ is the $\rm \eta_{c}$ transverse mass. Numerical calculations are carried out by an in-house monte carlo generator. The choice of the LDMEs for $\rm \eta_{c}$ is taken from \cite{Butenschoen:2011yh,Maltoni:2004hv,Yu:2017pot} and valued to $\rm \langle 0|\mathcal{O}_{1}^{\eta_{c}}(^{1}S_{0})|0\rangle$ =0.44 GeV$^3$ and $\rm \langle 0|\mathcal{O}_{8}^{\eta_{c}}(^{1}S_{0})|0\rangle$ =0.00056 GeV$^3$. The cross section and distribution in the following subsection are not multiplied by the  gap survival probability in all our results. The  parameters of the nuclei are given in Table.\ref{tab:1}

\begin{table*}[htbp]
	\begin{center}
		\begin{tabular}{|c |c | c | c |c |}\hline
			Nucleus     & $\rm Z $             &$\rm A$ &$\rm R_{A}$ (GeV$^{-1}$) &$\rm \gamma_{L}$             \\ \hline
			p    & 1 &   1& 3.54 &2675.13              \\ \hline
			Be    & 4 &   9& 12.85 &297.24              \\ \hline
			C    & 6 &   12& 14.15 &222.93              \\ \hline
			O   & 8 &   16& 15.57 &167.20             \\ \hline
			Ca   & 20 &   40& 21.13 &66.88             \\ \hline
			Cu   & 29 &   64& 24.72 &41.80             \\ \hline
			Ag   & 47 &   108& 29.43 &24.77             \\ \hline
			Au   & 79 &   197& 35.96 &13.58             \\ \hline
			Pb   & 82 &   208& 36.62 &12.86             \\ \hline
			U   & 92 &   238& 38.30 &11.24             \\ \hline				
		\end{tabular}
	\end{center}\caption{\label{tab:1}
		The nucleus parameters.}
\end{table*}

\subsection{Double diffractive production}

\subsubsection{pp collision}

The SD, DD and ND cross section predictions of $\eta_{c}$ hadroproduction of $\rm pp$ collision are evaluated in  Ref.\cite{PhysRevD.101.054035} for $0.0015<\xi <0.5$ CMS-TOTEM forward detector acceptance \cite{Albrow:2008pn}. The results are shown in Table.\ref{tab:2} to make a clear comparison with $\rm Ap$ and $\rm AA$ modes in photon-Pomeron and-Reggeon induced processes,  gluon-Pomeron and-Reggeon processes as well as Pomeron-Pomeron and Reggeon-Reggeon and its cross exchange processes. The subsequent description of $\rm pp$ is given without being multiplied by survival gap probability. The contributions of different mechanisms show that the $\mathbb{PR+RP}$ channel is dominant in DD processes. The contribution of $\mathbb{PP}$ channel remains minor over the $\mathbb{RR}$ channel.  The increase of total cross section is due to the Reggeon distribution as well as its cross exchange with the Pomeron. Thus, the Reggeon contribution and its cross exchange with Pomeron are the same order and not negligible. It has been noticed that the ND cross section is 10 times larger that the total DD cross section.   
\begin{table*}[htbp]
	\begin{center}
		\begin{tabular}{|c |c | c | c |c |c |c |c |c |c |c |}\hline
			\multicolumn{2}{|c|}{ }&\multicolumn{4}{c|}{DD}& \multicolumn{3}{|c|}{SD}&ND  \\ \hline	
			Hadron       & $\rm \langle \left\vert S\right\vert ^{2}\rangle $             & $\mathbb{PP}$              & $\mathbb{PR+RP}$& $\mathbb{RR}$ & Total&$\mathbb{P}$p &$\mathbb{R}$p &Total &gg             \\ \hline
			 pp    & 3.29$\times$10$^{-2}$ &  9.34$\times$10$^{0}$    & 3.27$\times$10$^{1}$  & 2.29$\times$10$^{1}$ & 6.50$\times$10$^{1}$&1.68$\times$10$^{2}$  & 2.59$\times$10$^{2}$ &4.27$\times$10$^{2}$ & 7.39$\times$10$^{2}$            \\ \hline
		\end{tabular}
	\end{center}\caption{\label{tab:2}
		The DD, SD and ND cross sections in $\mu$b for $\eta_{c}$ hadroproduction in $\rm pp$ mode at LHC with forward detector acceptance, $0.0015<\xi <0.5$. }
\end{table*}

\begin{figure}[htp]
\centering
\includegraphics[height=4.8cm,width=5.9cm]{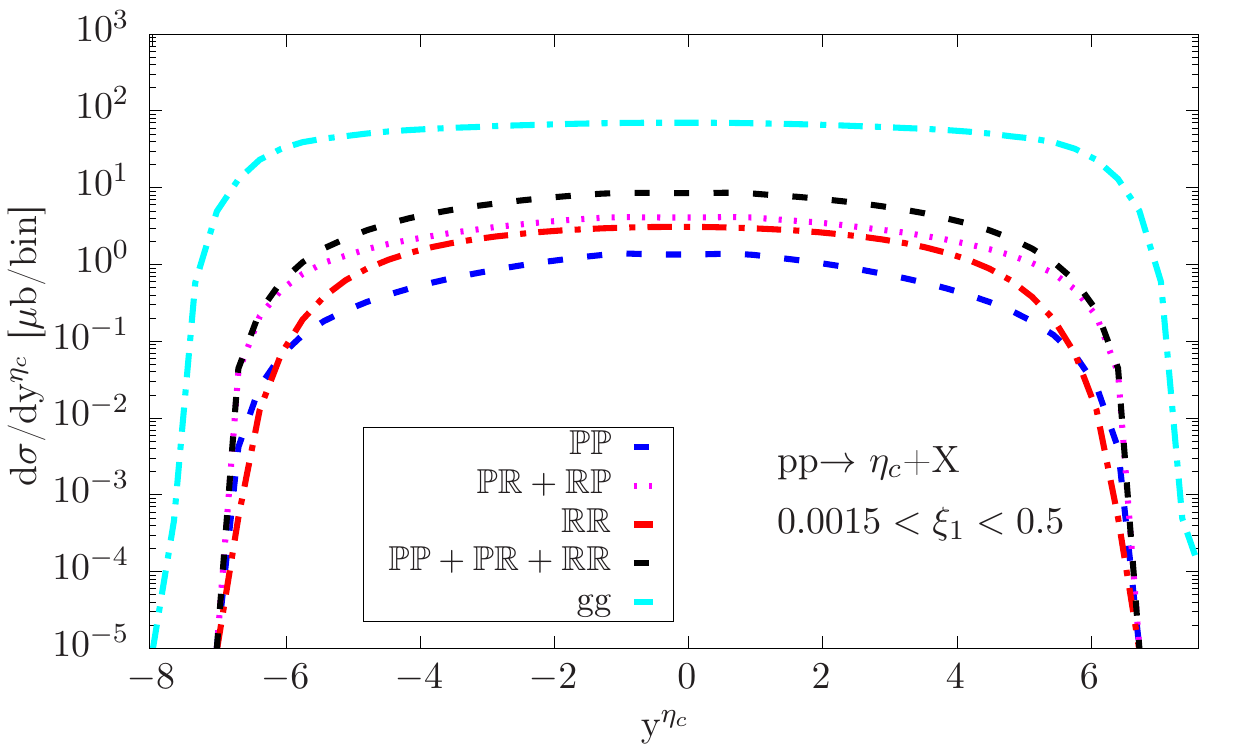}
\includegraphics[height=4.8cm,width=5.9cm]{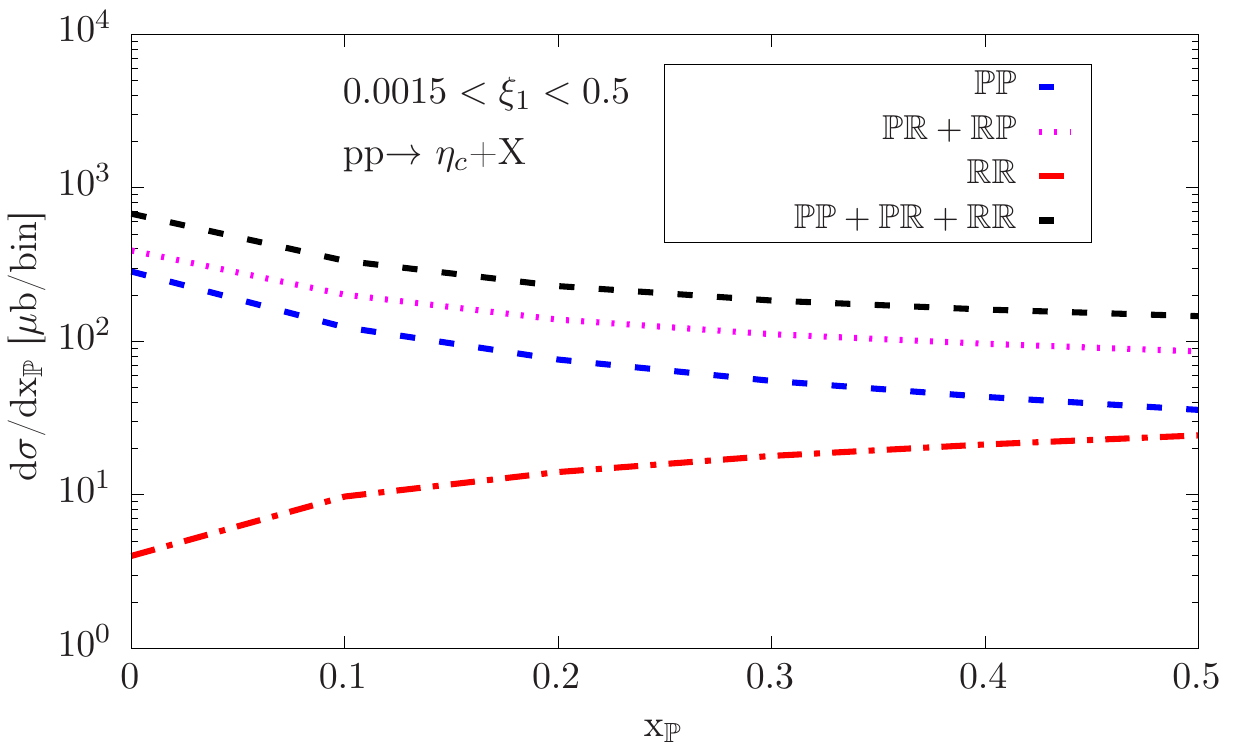}
\includegraphics[height=4.8cm,width=5.8cm]{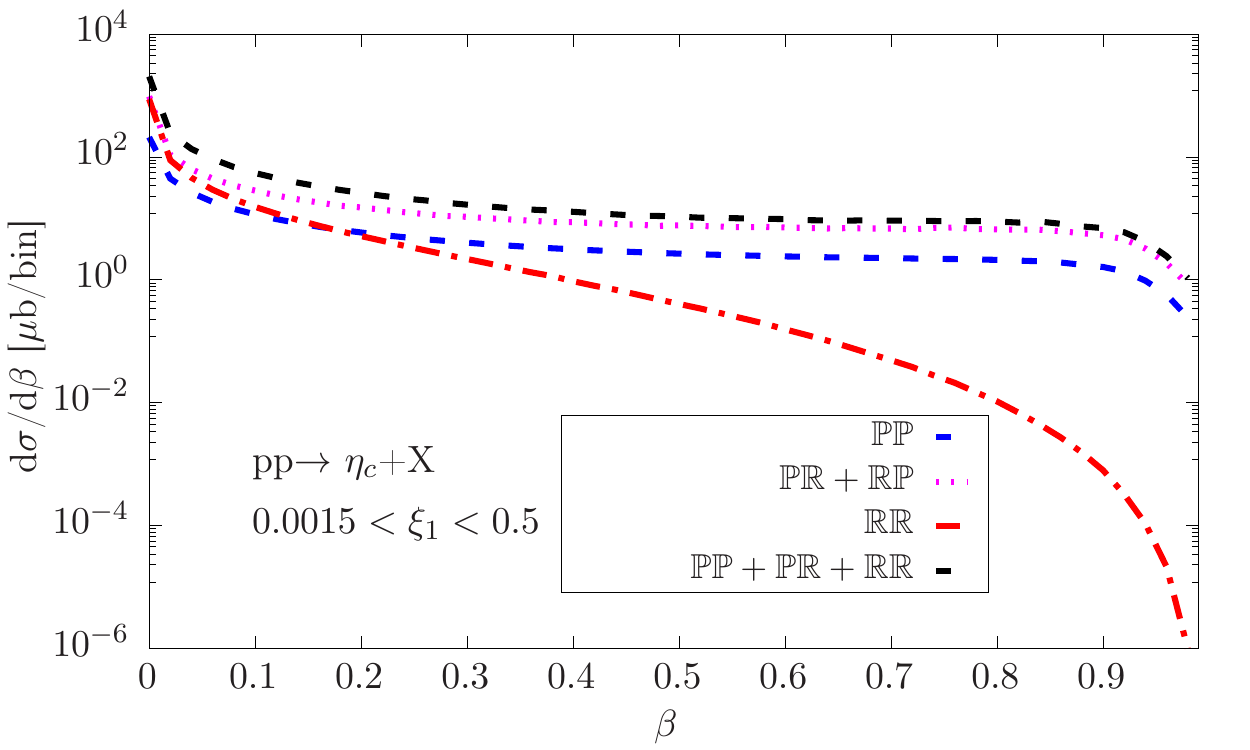}\\
\caption{\normalsize (color online) The $\rm y^{\eta_{c}}$, $x_{\mathbb{P}}$ and $\beta$ distributions for the $\mathbb{PP}$ (blue dashed line), $\mathbb{PR+RP}$ (magenta dotted line), $\mathbb{RR}$ (red dash dotted line), $\mathbb{PP+PR+RP+RR}$ (black dashed line) and $\rm gg$ (cyan dash dotted line) in DD processes for pp collision.}
\label{fig2:limits}
\end{figure}
The $\rm y^{\eta_{c}}$, $x_{\mathbb{P}}$ and $\beta$ distributions of hadroproduction in pp collision are presented in Fig.\ref{fig2:limits}. In DD processes, the initial protons ($\rm p_{i}$) with energies $\rm E_{i}$ are source of the Pomerons ($\mathbb{P}$) and Reggeons($\mathbb{R}$) with only a small squared four momentum transfer $\rm \mid t \mid$. The Pomeron and Reggeon are hadron-like objects which can emit gluon or quark along with remnants. This concept is well described by the Ingelman-Schlein (IS) picture. The incoming proton  with energy $\rm E_{i}$ turns into  intact or early intact outgoing proton  with energies $\rm E^{\prime}_{i}$. The  energy loss of incoming particle is defined as $\rm \xi_{i}=\frac{E_{i}-E^{\prime}_{i}}{E_{i}}$ and corresponds to the detector acceptance or a cut on longitudinal momentum fractions of outgoing protons. The intact outgoing proton with a reduced energy loss is observed by the forward tagging hadron detector. The forward detectors are installed symmetrically  at a distance of about 420 metres or asymmetrically \cite{Adriani:2008zz}, one at 220 metres and other at 420 metres, with respect to the interaction points.
The hard collision is triggered by the hard diffractive gluon from one proton with the other hard diffractive gluon from the other proton to form the $\eta_{c}$ meson after the hadronization process. The $\eta_{c}$ meson and  remnants are detected by the central detector which can provide the information to help the forward detector to distinguish diffractive from non-diffractive events \cite{Zhou:2018elb}. The non diffractive process is produced by  hard non diffractive gluons from the two incoming protons. The left panel shows the $\rm y^{\eta_{c}}$-rapidity distribution of $\eta_{c}$ meson from the central detector of different mechanisms in DD and ND processes.  The ND and DD distributions of are symmetric due to the emission of the same partons from similar incoming protons and modeled by the same parton distribution function (PDF). The largest distribution comes from Pomeron-Reggeon cross exchange. Moreover, the Reggeon distribution is still larger than the Pomeron one. The distributions for the DD and ND processes have maximums located at midrapidities. The middle panel presents the proton longitudinal momentum fraction distributions in which the Pomeron and Pomeron-Reggeon distributions decrease for small $x_{\mathbb{P}}$ and become almost flat for large $x_{\mathbb{P}}$. However, the Reggeon distributions increase for small $x_{\mathbb{P}}$ and also become nearly flat for large $x_{\mathbb{P}}$. The $\mathbb{PP+PR}$ distribution is more dominant over $\mathbb{PP}$. In addition, $\mathbb{RR}$ contribution is the lowest. The right panel exhibits gluon longitudinal momentum fraction with respect to the exchanged Pomeron or Reggeon distribution. The distribution of $\mathbb{PP+PR}$ and $\mathbb{PP}$ decrease for small $\beta$, become flat for intermediate values of $\beta$ and decrease for large  $\beta$. Nevertheless, $\mathbb{RR}$ contribution decreases along with $\beta$. The $\mathbb{RR}$, $\mathbb{PP+PR}$ and $\mathbb{PP}$ distributions present the concavity and convexity at the end-points. The Reggon  and Pomeron-Reggeon cross exchange distributions should be taken into consideration during the LHC measurement and should not be disregarded.

\subsubsection{AA collision}

The DD and ND cross section estimates of $\eta_{c}$ hadroproduction in  nucleus-nucleus mode for light, medium and heavy nuclei are provided in Table.\ref{tab:3}. The AA mode of different collisions shows that the cross sections increase with the increase of the atomic mass number, A, from the light to heavy nuclei. The largest contribution to the total DD cross section for the different types of nucleus comes from $\mathbb{PP}$ channels. The cross exchange $\mathbb{PR+RP}$ contribution is also important because  it keeps the same order of magnitude than the Pomeron one, for example the light nuclei and the Calcium.
The $\mathbb{PP}$ and $\mathbb{PR+RP}$ contribution to $\eta_{c}$  production are larger for heavy nuclei. The $\mathbb{RR}$ contributions to the total cross section remain lower. The ND cross section prediction is a factor $\gtrsim$ 10 larger than of the total DD prediction for $0.0015<\xi<0.5$ in $\rm AA$ mode. The ND cross section of $\eta_{c}$ in $\rm AA$ mode  is a factor $\gtrsim$ 10$^{2}$ larger than that of $\rm pp$ mode in ND process which can be explained by the enhancement of the non diffractive gluon distribution from A, i.e., proportional to $\rm A^{2}R^{2}_{g}$. The ND cross section prediction in $\rm AA$ mode is a factor $\gtrsim$ 10$^{3}$ larger than of DD cross section in $\rm pp$ mode. The DD  cross section of $\eta_{c}$ in $\rm AA$ mode  is a factor $\gtrsim$ 10 larger  than that of $\rm pp$ mode in DD  process which can be accounted for by the enhancement of the diffractive gluon distribution from A through Pomeron or Reggeon and the nuclear form factor, i.e., proportional to $\rm A^{4}R^{2}_{g}F^{4}_{A}(t)$, apart from the first light nucleus where the $\rm pp$ mode  is the same order than $\rm AA$ mode . 

\begin{table*}[htbp]
	\begin{center}
		\begin{tabular}{|c |c | c | c |c |c |c |c |}\hline
			\multicolumn{3}{|c|}{ }&\multicolumn{4}{c|}{DD}& ND  \\ \hline	
			Nuclei   & AA      & $\rm \langle \left\vert S\right\vert ^{2}\rangle $             & $\mathbb{PP}$              & $\mathbb{PR+RP}$& $\mathbb{RR}$ & Total&gg             \\ \hline
		\multirow{3}{*}{Light nucleus}& BeBe & 4.06$\times$10$^{-4}$  & 3.44$\times$10$^{1}$   &1.95$\times$10$^{1}$ &2.19$\times$10$^{0}$ &5.61$\times$10$^{1}$ &5.45$\times$10$^{4}$\\
		\multirow{3}{*}{}&CC  &  2.05$\times$10$^{-2}$  & 9.11$\times$10$^{1}$    & 4.86$\times$10$^{1}$&5.16$\times$10$^{0}$ & 1.45$\times$10$^{2}$&9.45$\times$10$^{4}$ \\
		& OO  & 6.51$\times$10$^{-3}$  &2.40$\times$10$^{2}$ &1.21$\times$10$^{2}$ & 1.21$\times$10$^{1}$&3.73$\times$10$^{2}$ &1.64$\times$10$^{5}$\\
		\cline{1-8}
		\multirow{3}{*}{Medium nucleus}& CaCa& 1.67$\times$10$^{-4}$ &5.20$\times$10$^{3}$    &2.17$\times$10$^{3}$ &1.79$\times$10$^{2}$ &7.55$\times$10$^{3}$ &9.37$\times$10$^{5}$\\
		\multirow{3}{*}{}&CuCu  &2.54$\times$10$^{-5}$   & 2.49$\times$10$^{4}$   & 9.46$\times$10$^{3}$& 7.09$\times$10$^{2}$ &3.51$\times$10$^{4}$ &2.28$\times$10$^{6}$\\
		&AgAg  &2.82$\times$10$^{-6}$   &1.41$\times$10$^{5}$ &4.83$\times$10$^{4}$ &3.25$\times$10$^{3}$ &1.92$\times$10$^{5}$ & 6.10$\times$10$^{6}$\\
		\cline{1-8}
		\multirow{3}{*}{Heavy nucleus}& AuAu   & 2.83$\times$10$^{-7}$  &1.03$\times$10$^{6}$ &3.12$\times$10$^{5}$ &1.85$\times$10$^{4}$ &1.36$\times$10$^{6}$ &1.87$\times$10$^{7}$\\
		\multirow{3}{*}{}&PbPb  &2.29$\times$10$^{-7}$   & 1.23$\times$10$^{6}$   &3.69$\times$10$^{5}$ &2.17$\times$10$^{4}$ &  1.62$\times$10$^{6}$ &2.07$\times$10$^{7}$\\
		& UU & 1.28$\times$10$^{-7}$  &1.92$\times$10$^{6}$ &5.59$\times$10$^{5}$ &3.20$\times$10$^{4}$ & 2.51$\times$10$^{6}$&2.66$\times$10$^{7}$\\
			\hline
		\end{tabular}
	\end{center}\caption{\label{tab:3}
		The DD and ND cross sections in $\mu$b for  $\eta_{c}$  hadroproduction in $\rm AA$ mode at LHC with forward detector acceptance, $0.0015<\xi<0.5$.}
\end{table*}

\begin{figure}[htp]
\centering
\includegraphics[height=4.8cm,width=5.9cm]{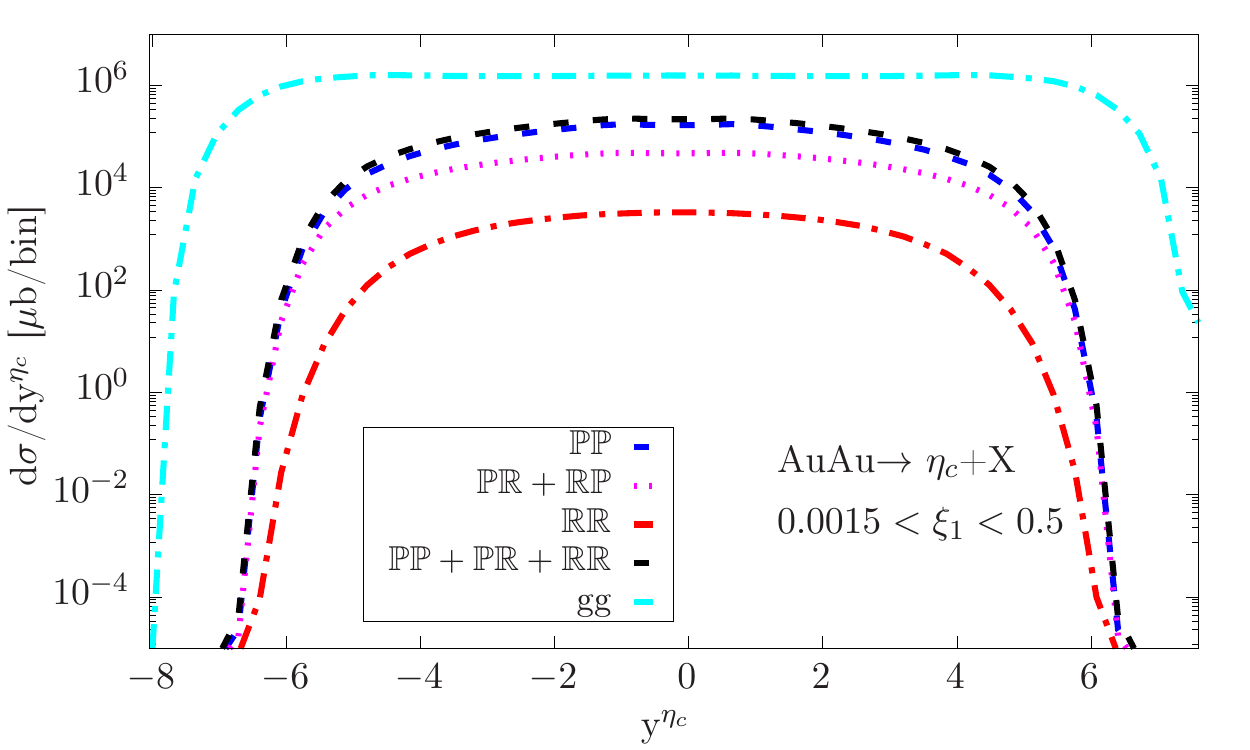}
\includegraphics[height=4.8cm,width=5.9cm]{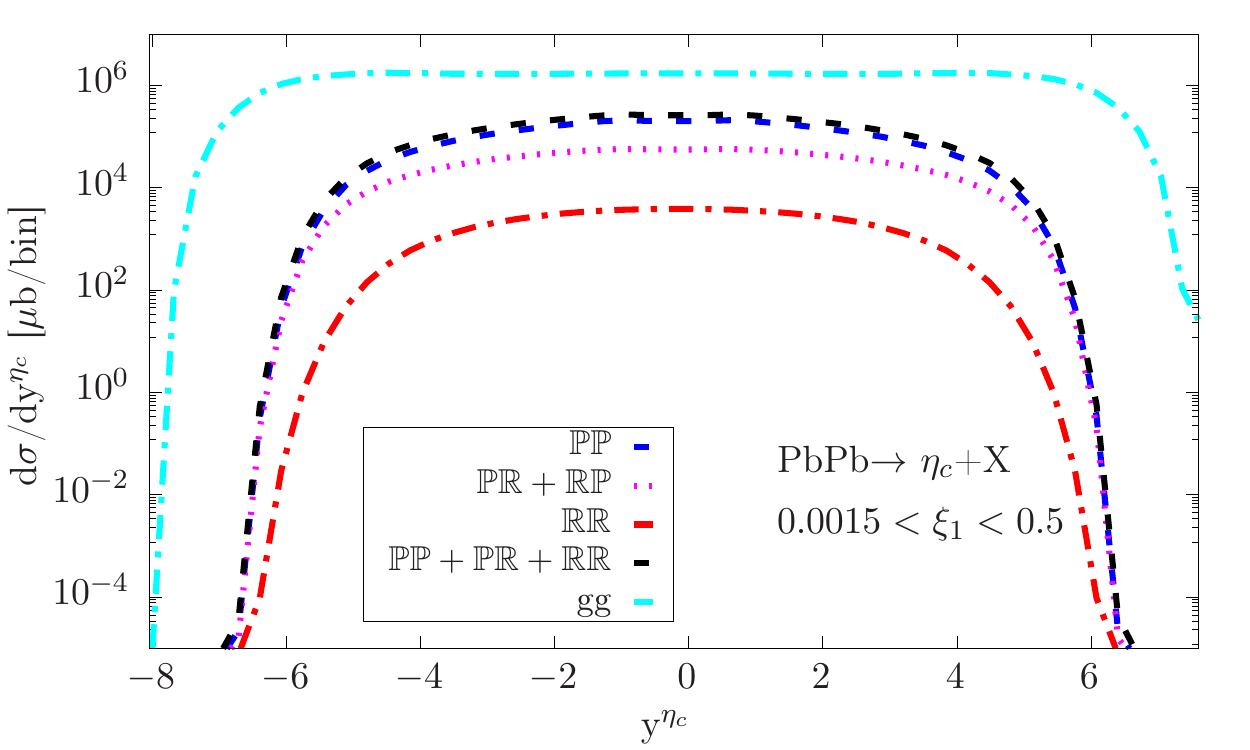}
\includegraphics[height=4.8cm,width=5.9cm]{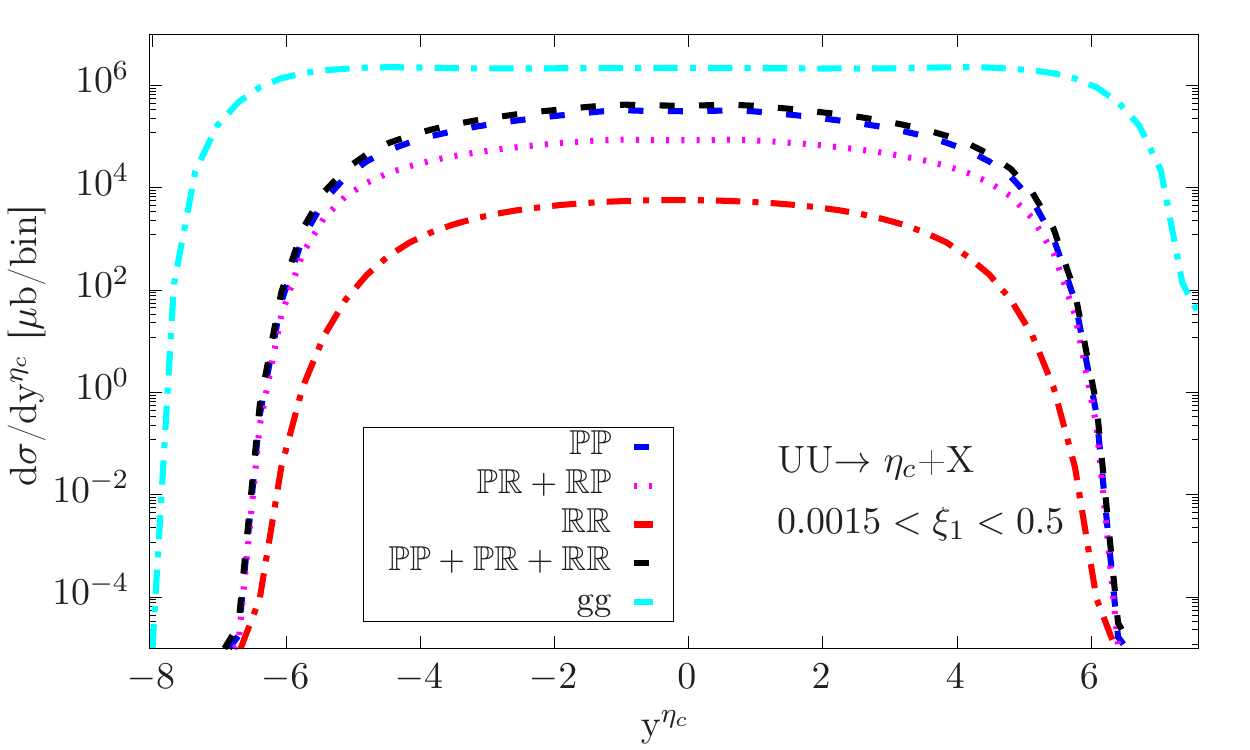}
\caption{\normalsize (color online) The $\rm y^{\eta_{c}}$ distributions for the $\mathbb{PP}$ (blue dashed line), $\mathbb{ PR+RP}$ (purple dash dotted line), $\mathbb{ RR}$ (red dash dotted line), $\mathbb{PP+PR+RP+RR}$ (black dotted line) and $\rm gg$ (cyan dotted line) in DD processes for AA collisions.}
\label{fig3:limits}
\end{figure}
The rapidity distribution of $\eta_{c}$ production in $\rm AA$ mode for heavy nuclei are only displayed in Fig.\ref{fig3:limits} because of its large cross section. We have noticed that the rapidity distributions for light and medium nuclei keep the similar shape and behavior like the heavy one. Therefore their distributions are not shown here. In the DD process, the two incoming nuclei $\rm A$  radiate  gluons via Pomeron or Reggeon. The remnant is also emitted from Pomeron or Reggeon despite the emission of hard diffractive gluon. The incoming nuclei  loss their energies and turn into intact or early intact nuclei. The emitted  hard diffractive gluons from both nuclei interact each other to produce $\eta_{c}$ meson after hadronization, accompanied with remnants.  For the non diffractive processes, the nuclei radiate gluons  without emission of Pomeron and Reggeon and turn into remnant system $\rm X$  which is an ensemble of unknown particles. The collision of the both non diffractive gluons generates the $\eta_{c}$ meson. Therefore, there are presence of the remnants and $\eta_{c}$ meson in the final state, but there are no intact particles. The forward tagging nucleus detector aims at detecting the intact nuclei in DD process while the central detector targets the $\eta_{c}$ meson and remnants in DD and ND processes. The DD and ND distributions are symmetric and keeps the maximums at mid-rapidity. The $\mathbb{PP}$ distributions are dominant over $\mathbb{PR+RP}$ and $\mathbb{RR}$ ones. The $\mathbb{PP}$ and $\mathbb{PR+RP}$ distributions are comparable for light nucleus and Calcium.
 The $\mathbb{PR+RP}$ contribution can be ignored for some medium (Copper and Silver) and heavy nucleus. But its contributions can not be put aside for the light nucleus and Calcium. By measuring these distributions, we should be able to investigate the Reggeon-Pomeron contribution at the LHC data for a specific type of nucleus.

\begin{figure}[htp]
\centering
\includegraphics[height=4.8cm,width=5.9cm]{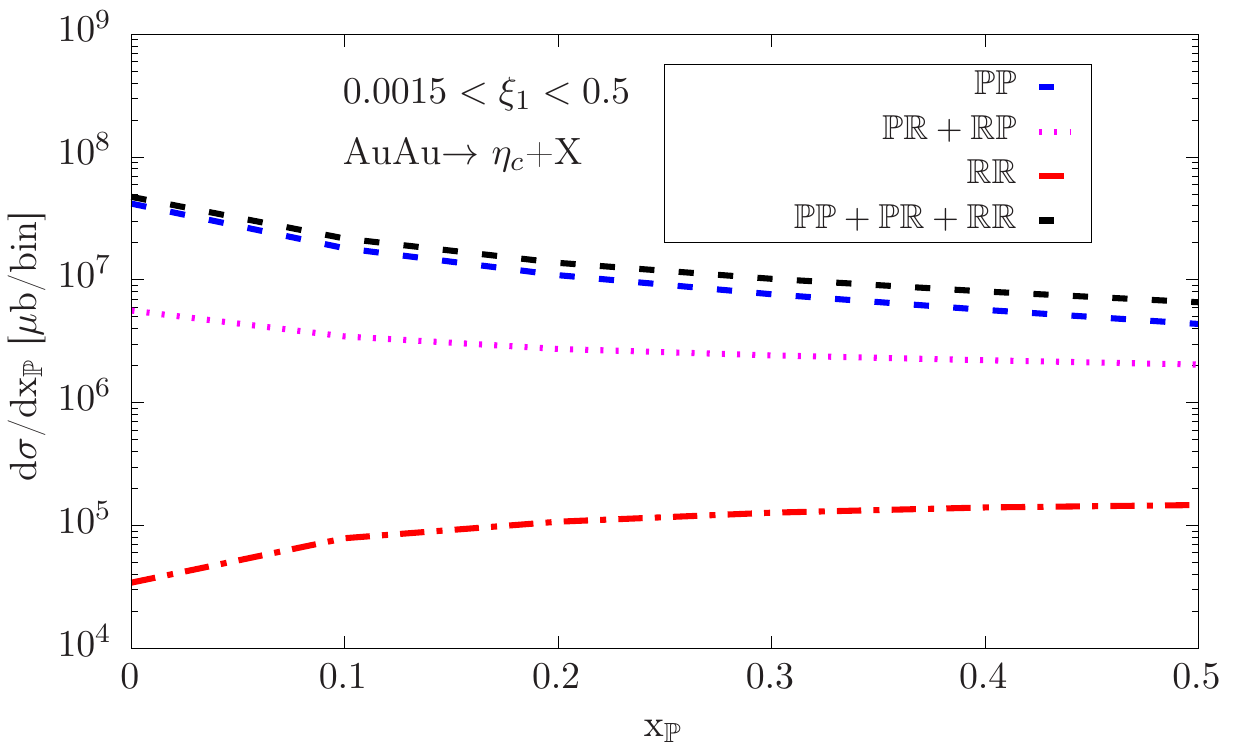}
\includegraphics[height=4.8cm,width=5.9cm]{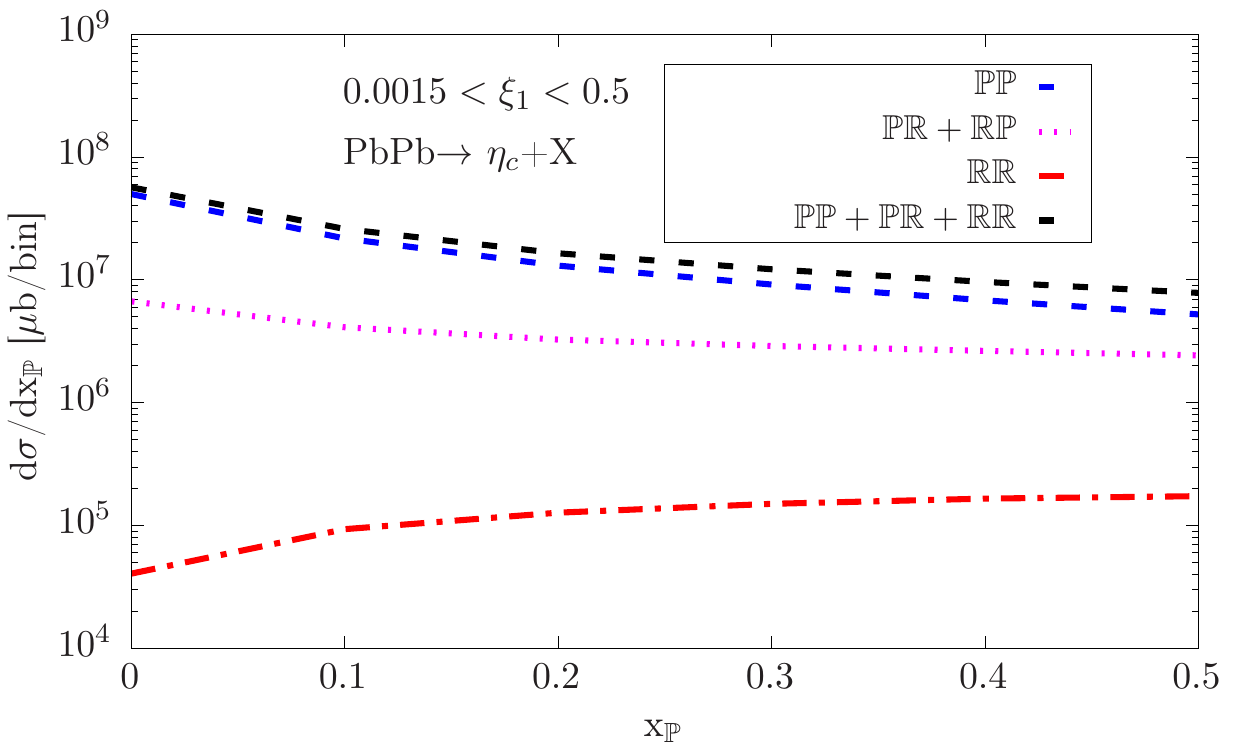}
\includegraphics[height=4.8cm,width=5.9cm]{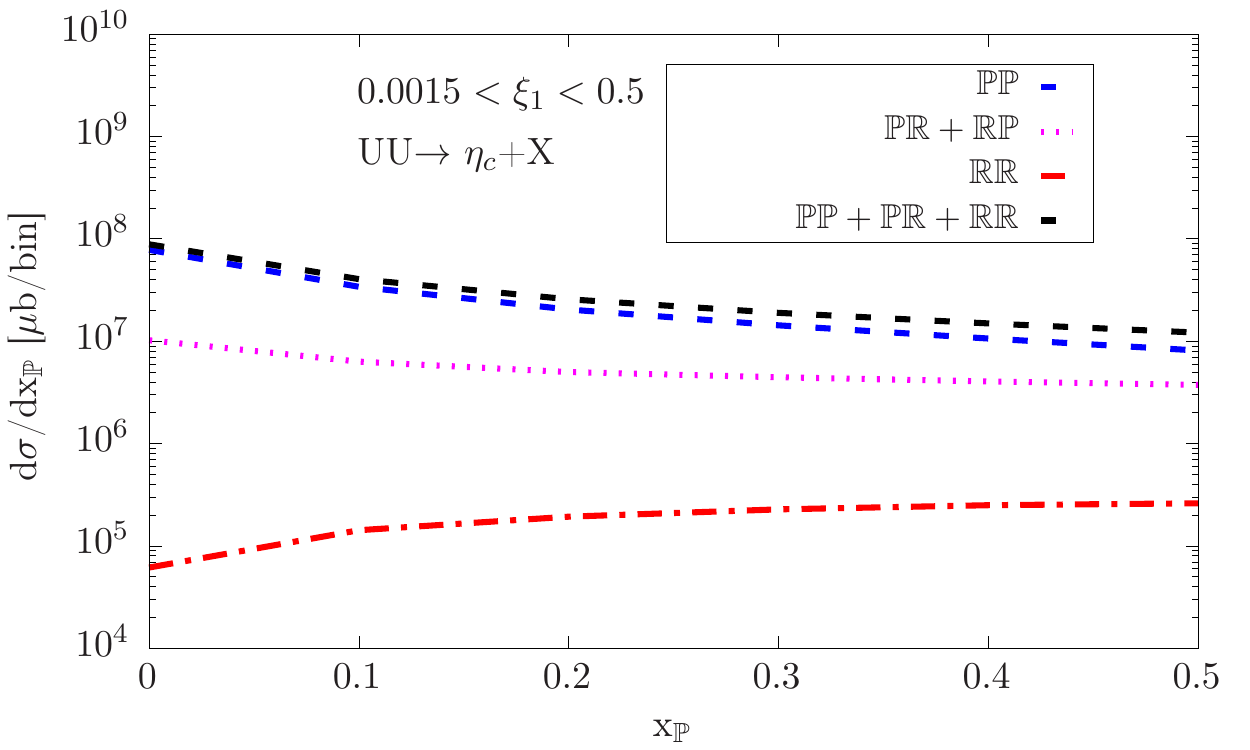}
\caption{ \normalsize (color online)
The $x_{\mathbb{P}}$ for the $\mathbb{PP}$ (blue dashed line), $\mathbb{ PR+RP}$ (purple dash dotted line), $\mathbb{ RR}$ (red dash dotted line) and $\mathbb{PP+PR+RP+RR}$ (black dotted line) in DD processes for AA collisions.}
\label{fig4:limits}
\end{figure}
The Fig.\ref{fig4:limits} shows $x_{\mathbb{P}}$ distributions of the nucleus-nucleus collision for heavy nuclei. Light and medium nuclei distributions have also the similar  profiles as heavy nuclei and consequently they are not displayed here. we have chosen the heavy nuclei ones because of their large cross sections. We have remarked that $\mathbb{PP}$ and $\mathbb{ PR+RP}$ contributions fall down for small $x_{\mathbb{P}}$ and turn out to be flat for large $x_{\mathbb{P}}$ whereas $\mathbb{RR}$ contributions enlarges for small $x_{\mathbb{P}}$ and also become flat for large $x_{\mathbb{P}}$. The dominant distribution hails from the $\mathbb{PP}$ while the minor contribution originates from
$\mathbb{RR}$. This nucleus longitudinal momentum fraction is important to investigate the $\mathbb{PP}$ contribution at the LHC. 

\begin{figure}[htp]
\centering
\includegraphics[height=4.8cm,width=5.9cm]{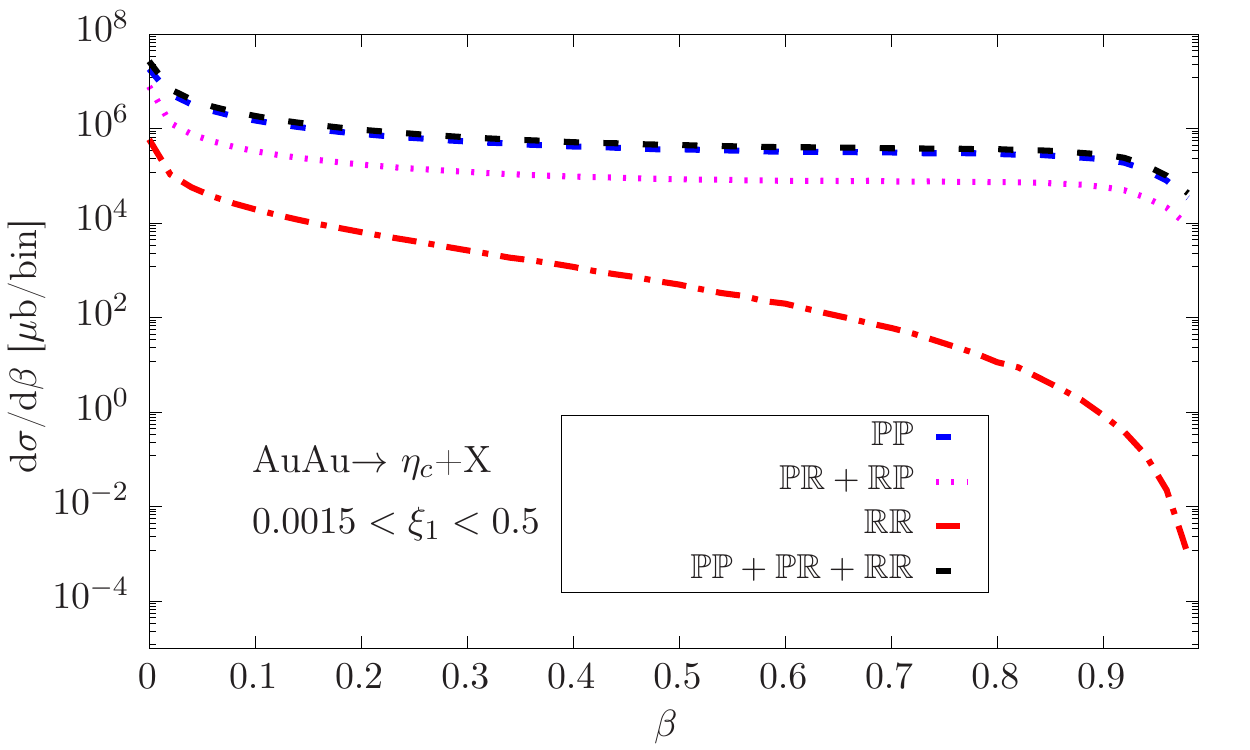}
\includegraphics[height=4.8cm,width=5.9cm]{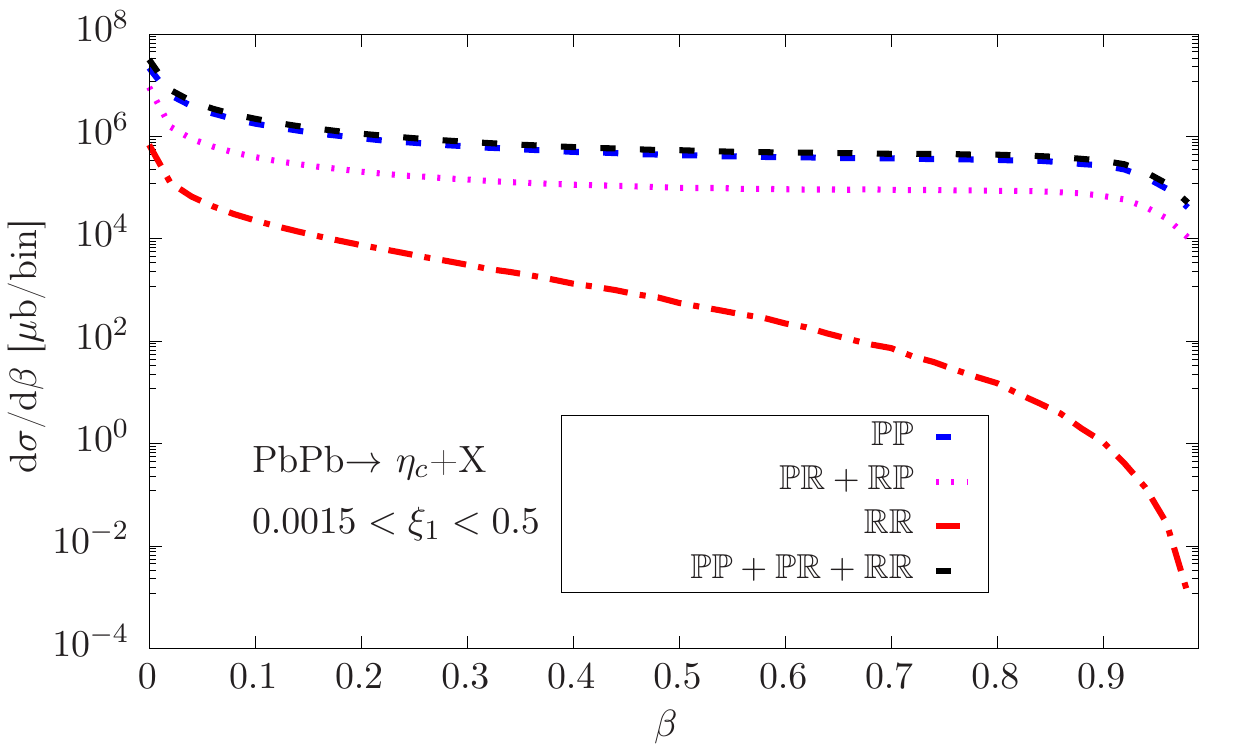}
\includegraphics[height=4.8cm,width=5.9cm]{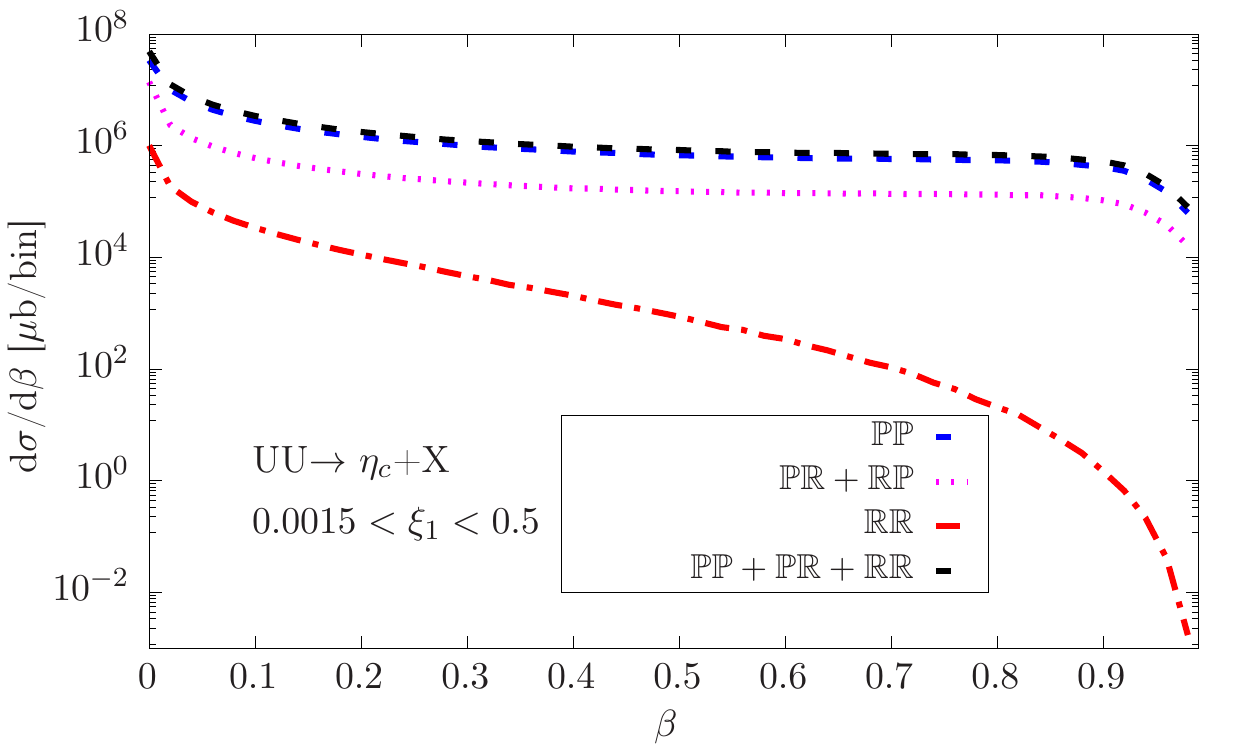}
\caption{ \normalsize (color online)
The $\beta$ distributions for the $\mathbb{PP}$ (blue dashed line), $\mathbb{ PR+RP}$ (purple dash dotted line), $\mathbb{ RR}$ (red dash dotted line) and $\mathbb{PP+PR+RP+RR}$ (black dotted line) in DD processes for AA collisions.}
\label{fig5:limits}
\end{figure}
We have also noted that light, medium and heavy nuclei possess  similar profiles in their $\beta$ distributions. According to this similarity, we have only plotted the $\beta$ distributions for the heavy nuclei in Fig.\ref{fig5:limits}. The $\mathbb{PP}$ and $\mathbb{ PR+RP}$ contributions reduce for small $\beta$, become flat for in-between values of $\beta$ and lessen again for large $\beta$. Nonetheless, the $\mathbb{ RR}$ decreases together with the enlargement of $\beta$. The concavity and the convexity are observed  for  $\mathbb{PP}$, $\mathbb{ PR+RP}$ and $\mathbb{ RR}$ distributions at their end points.

\subsubsection{pA collision}

The DD and ND cross section predictions of $\eta_{c}$ hadroproduction in $\rm pA$ mode for light, medium and heavy nuclei are arranged in Table.\ref{tab:4}. The cross sections of $\rm pA$ mode enlarge with the increase of the atomic number A from the light to heavy nuclei. The dominant contribution to the total cross section
 comes from the cross exchange $\mathbb{ PR+RP}$ in $\rm pA$ mode. But the lowest contribution  is still $\mathbb{ RR}$ one. 
 It has been noted that the total DD and $\mathbb{ PR+RP}$ cross sections are a same factor of magnitude, except for $\rm pBe$. The ND  and total DD cross sections are a same factor of magnitude for the heavy nuclei and $\rm p Cu$. However, the ND cross section should be larger than the total DD cross section by taking in account the survival gap probability. For the light nuclei, $\rm pCa$ and $\rm pAg$, the ND section is a factor $\gtrsim$ 10 than that of the total DD one.  The dominance of $\mathbb{PR+RP}$ and total DD cross sections  over ND cross section for heavy nuclei and $\rm pAg$ should be due to the non multiplication of the cross section by the gap survival probability. This absorption effect should decease the DD cross section which should remain smaller than the ND cross section as usual. The ND cross section in  $\rm AA$ mode is a factor $\gtrsim$ 10 larger than that of $\rm pA$ mode for the last medium and heavy nuclei. They are  a same factor of magnitude for the two first medium nuclei. The ND cross in $\rm AA$  mode is a factor $\gtrsim$ 10$^{2}$ larger than that of the total DD cross in $\rm pA$  mode.
The ND cross in $\rm pA$  mode is a factor $\gtrsim$ 10$^{2}$ larger than that of $\rm pp$ mode  in ND process. The increase of the ND cross section of $\rm pA$ mode with respect to $\rm pp$ mode can be accounted for by the non diffractive  gluon distribution from A , i.e., proportional to $\rm AR_{g}$. The total  DD cross in $\rm pA$ mode is a factor $\gtrsim$ 10 larger than that of pp mode in DD process. The enhancement of the total DD   cross section of pA mode with reference to $\rm pp$ mode can be explained by  the hard  diffractive gluon distribution from A  and the nuclear form factor, i.e., proportional to  $\rm A^{2}R_{g}F^{2}_{A}(t)$. The total DD cross section in $\rm AA$ mode is a factor 10 larger than that of $\rm pA$ mode for the heavy nuclei and the last medium nuclei.  They are  a same factor of magnitude for the two first medium nuclei and the two last light nuclei. The total DD cross section in $\rm AA$ mode is a factor 10 smaller than that of $\rm pA$ mode for the first light medium.

\begin{table*}[htbp]
	\begin{center}
		\begin{tabular}{|c |c | c | c |c |c |c |c |}\hline
			\multicolumn{3}{|c|}{ }&\multicolumn{4}{c|}{DD}& ND  \\ \hline	
			Nuclei   & pA      & $\rm \langle \left\vert S\right\vert ^{2}\rangle $             & $\mathbb{PP}$              & $\mathbb{PR+RP}$& $\mathbb{RR}$ & Total & gg             \\ \hline
		\multirow{3}{*}{Light nucleus}& pBe & 3.65$\times$10$^{-3}$  & 3.59$\times$10$^{1}$   &5.44$\times$10$^{1}$ &1.43$\times$10$^{1}$ &1.05$\times$10$^{2}$  &1.27$\times$10$^{4}$\\
		\multirow{3}{*}{}&pC  & 2.73$\times$10$^{-3}$   & 5.85$\times$10$^{1}$   &1.05$\times$10$^{2}$ &2.19$\times$10$^{1}$ &1.85$\times$10$^{2}$ & 1.67$\times$10$^{4}$\\
		& pO   &2.05$\times$10$^{-3}$    & 9.52$\times$10$^{1}$& 2.12$\times$10$^{2}$&3.36$\times$10$^{1}$ & 3.41$\times$10$^{2}$ &2.20$\times$10$^{4}$\\
		\cline{1-8}
		\multirow{3}{*}{Medium nucleus}& pCa&8.26$\times$10$^{-4}$   & 4.43$\times$10$^{2}$   & 2.58$\times$10$^{3}$ &1.30$\times$10$^{2}$ & 3.15$\times$10$^{3}$ & 5.28$\times$10$^{4}$\\
		\multirow{3}{*}{}&pCu  &5.14$\times$10$^{-4}$   &9.71$\times$10$^{2}$    &1.03$\times$10$^{4}$ &2.59$\times$10$^{2}$ & 1.15$\times$10$^{4}$ & 8.25$\times$10$^{4}$\\
		&pAg  & 3.04$\times$10$^{-6}$   &2.31$\times$10$^{3}$ & 5.03$\times$10$^{4}$ &5.55$\times$10$^{2}$ &5.32$\times$10$^{4}$ &1.35$\times$10$^{5}$\\
		\cline{1-8}
		\multirow{3}{*}{Heavy nucleus}& pAu  & 1.67$\times$10$^{-4}$ & 6.26$\times$10$^{3}$  & 3.17$\times$10$^{5}$   &1.33$\times$10$^{3}$ &3.24$\times$10$^{5}$ & 2.38$\times$10$^{5}$\\
		\multirow{3}{*}{}&pPb &1.58$\times$10$^{-4}$   & 6.84$\times$10$^{3}$ & 3.75$\times$10$^{5}$&1.44$\times$10$^{3}$ &3.83$\times$10$^{5}$ &2.50$\times$10$^{5}$\\
		& pU  &1.38$\times$10$^{-4}$  &8.55$\times$10$^{3}$ &5.67$\times$10$^{5}$ &1.75$\times$10$^{3}$ &5.77$\times$10$^{5}$ &2.84$\times$10$^{5}$\\
			\hline
		\end{tabular}
	\end{center}\caption{\label{tab:4}
		The DD and ND cross sections in $\mu$b  for the $\eta_{c}$ hadron-nucleus production  in $\rm pA$ mode at LHC with forward detector acceptance, $0.0015<\xi<0.5$.}
\end{table*}

\begin{figure}[htp]
\centering
\includegraphics[height=4.8cm,width=5.9cm]{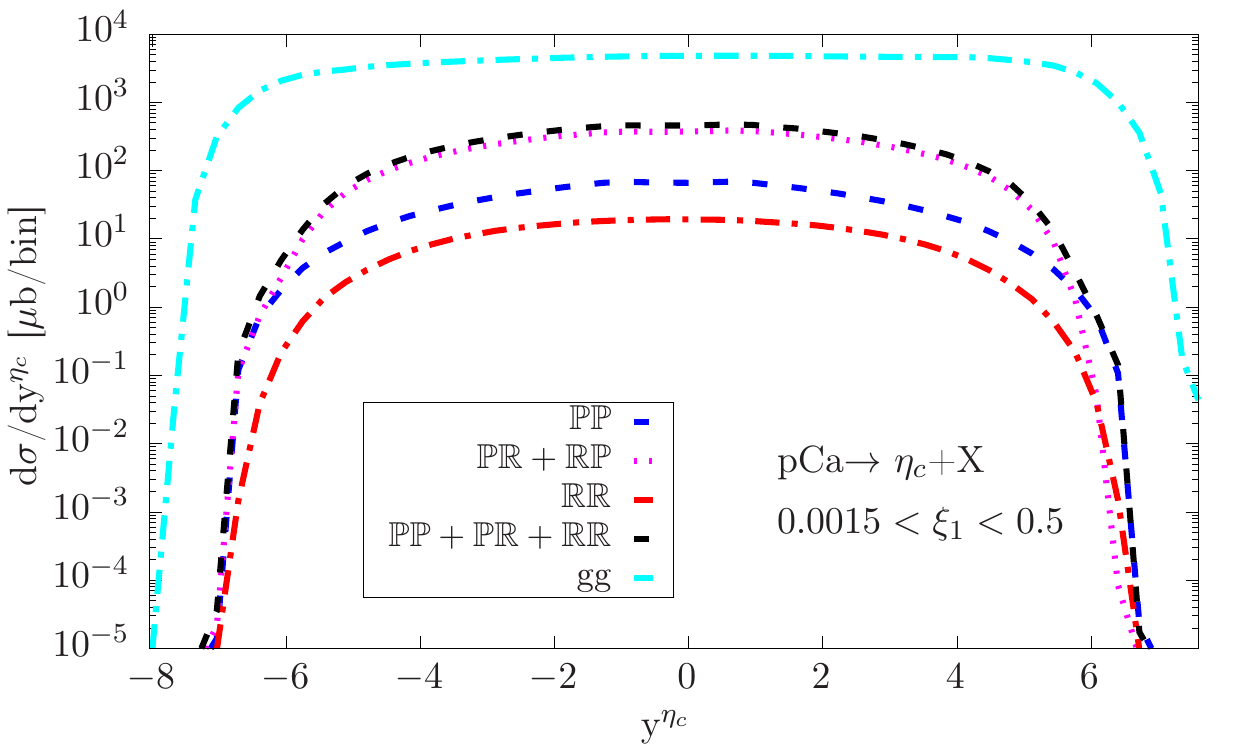}
\includegraphics[height=4.8cm,width=5.9cm]{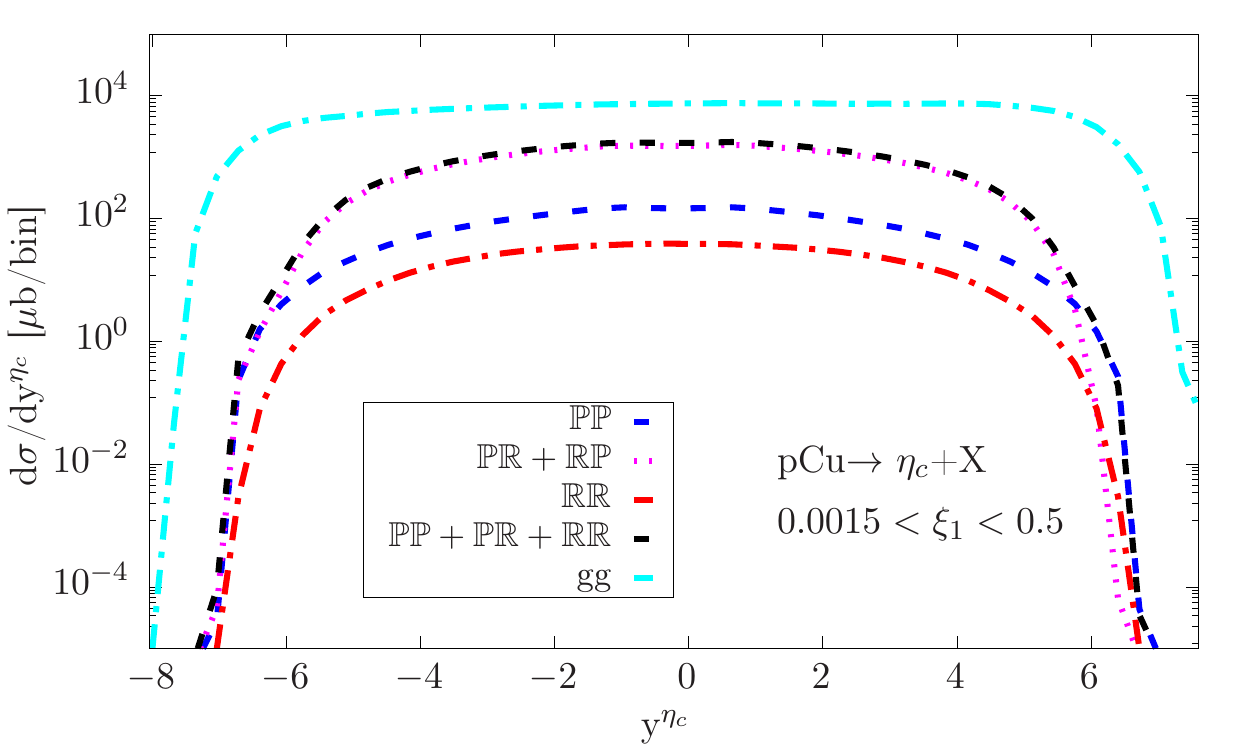}
\includegraphics[height=4.8cm,width=5.9cm]{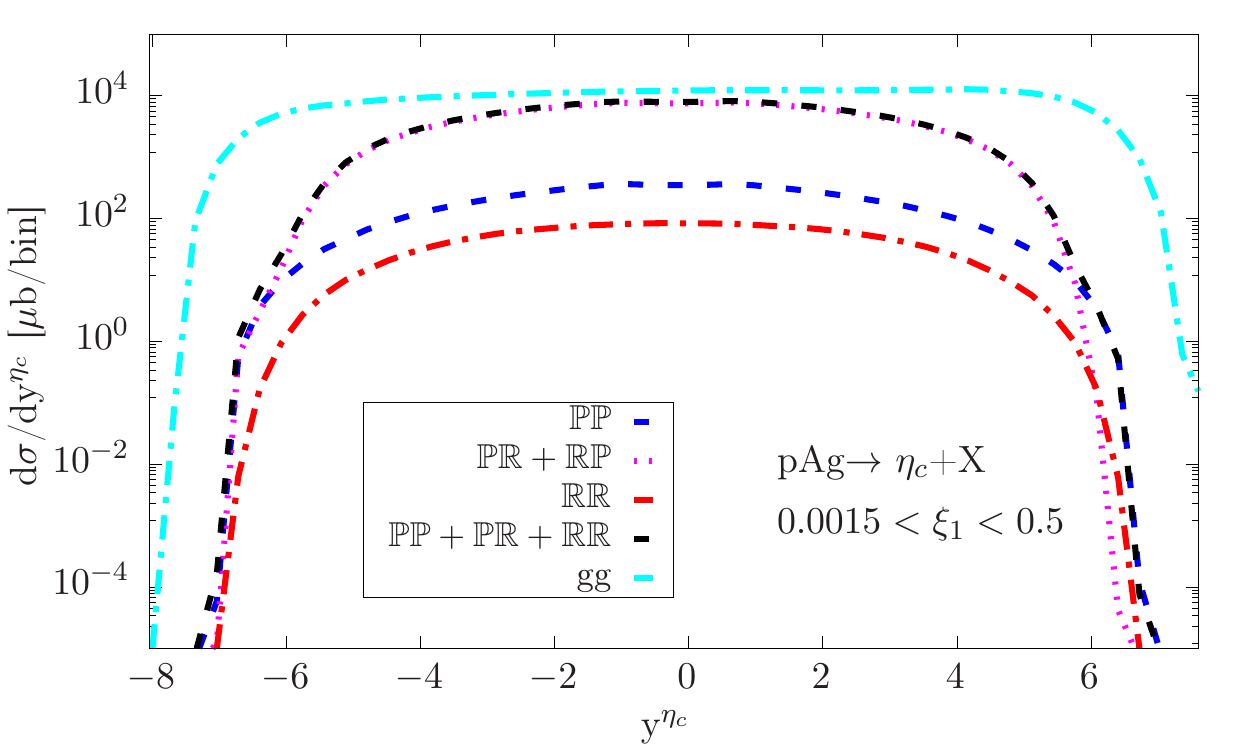}\\
\includegraphics[height=4.8cm,width=5.9cm]{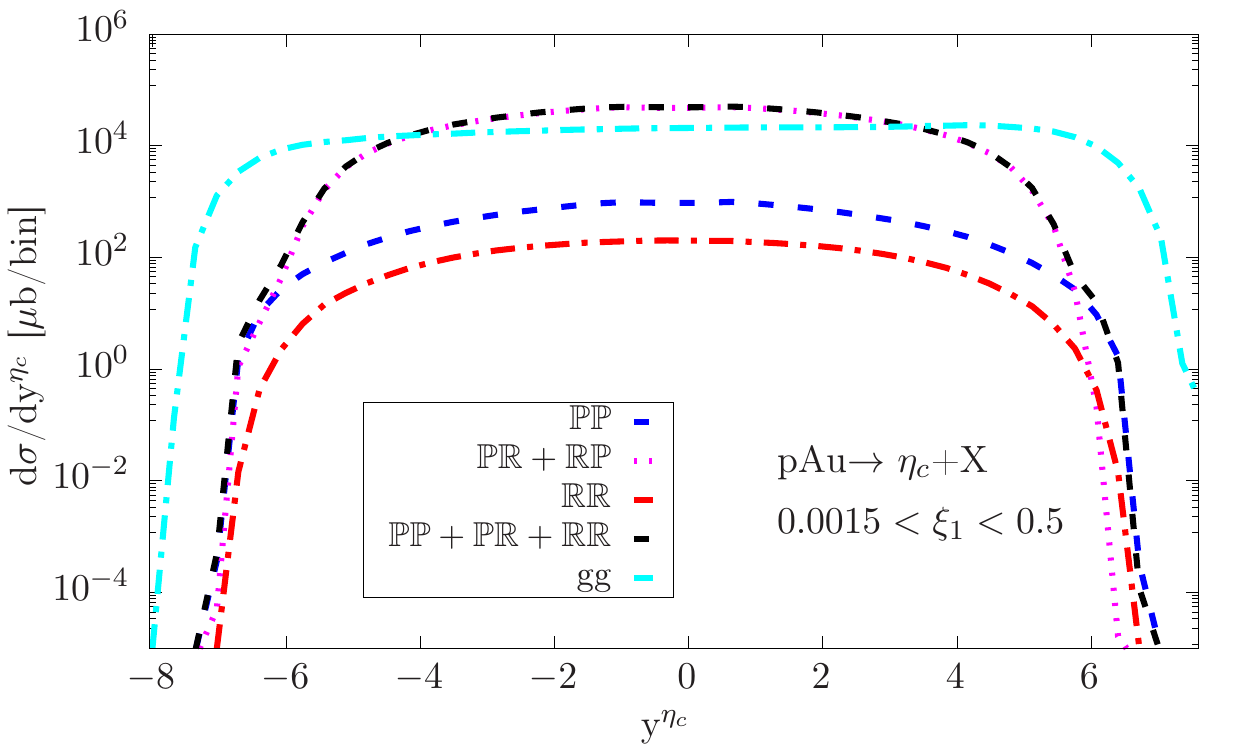}
\includegraphics[height=4.8cm,width=5.9cm]{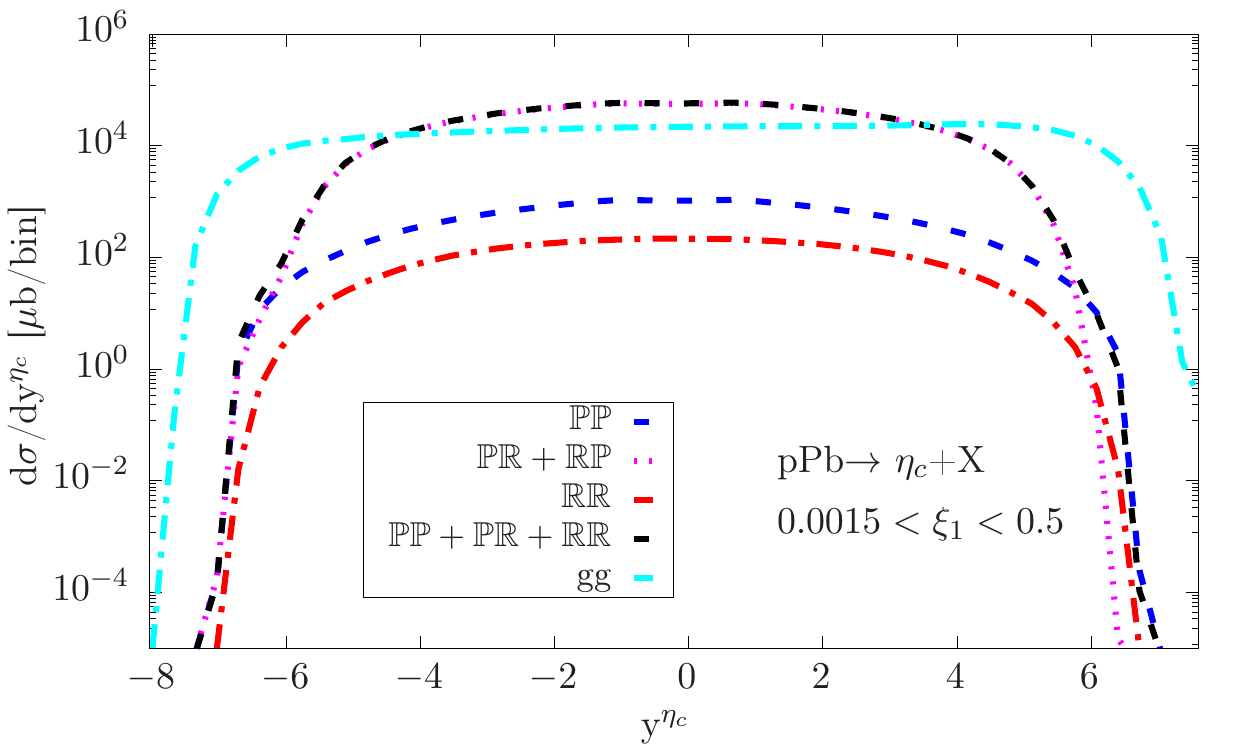}
\includegraphics[height=4.8cm,width=5.9cm]{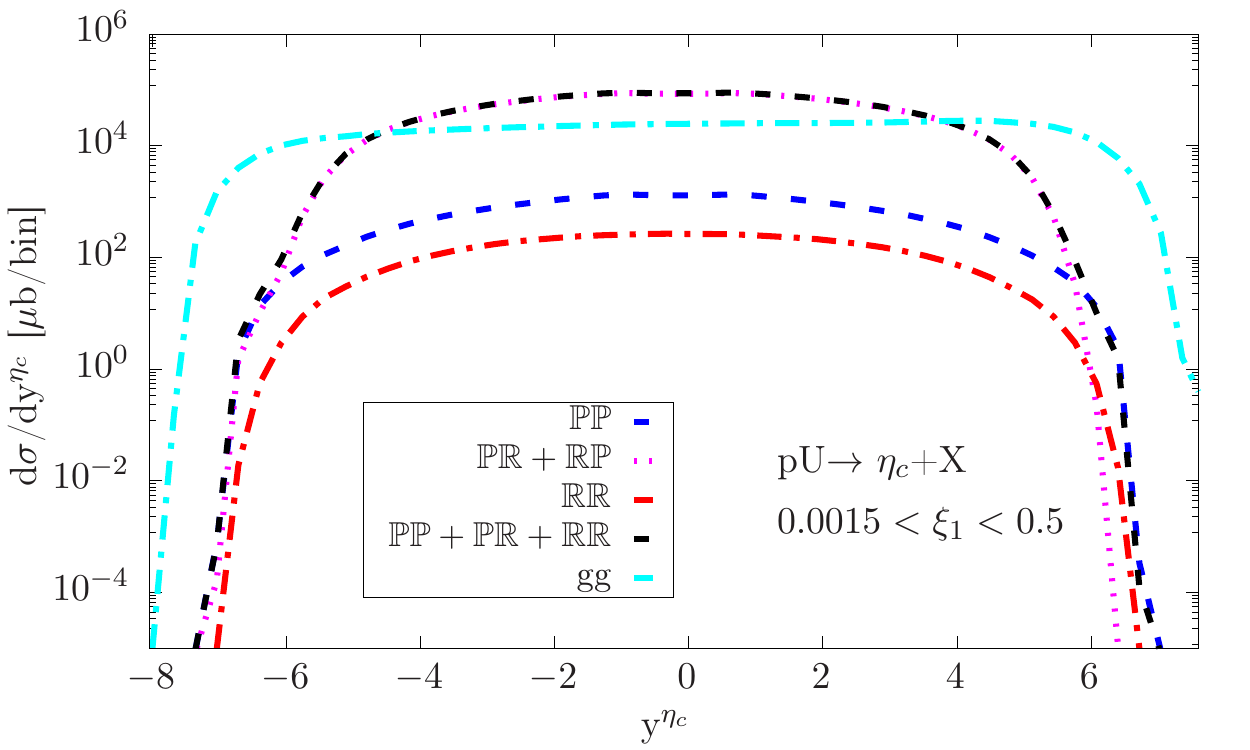}\\
\caption{\normalsize (color online) The $\rm y^{\eta_{c}}$ distributions for the $\mathbb{PP}$ (blue dashed line), $\mathbb{ PR+RP}$ (purple dash dotted line), $\mathbb{ RR}$(red dash dotted line), $\mathbb{PP+PR+RP+RR}$ (black dotted line) and $\rm gg$ (cyan dotted line) in DD processes for pA collisions.}
\label{fig6:limits}
\end{figure}
The rapidity distribution in $\rm pA$ mode for $\eta_{c}$ hadroproduction in DD processes is shown in Fig.\ref{fig6:limits}. In this mode, the incoming particles are dissimilar. One of the two incoming colliding particles, say, proton emits the hard diffractive gluon through the Pomeron or Reggeon while the second of the two incoming colliding particles, say, nucleus also radiates the hard diffractive gluon through the Pomeron or Reggeon. The swap of the two incoming particles $\rm pA$ to $\rm Ap$ leads to the same contribution. While the proton and nucleus emit the colorless Pomeron and Reggeon, they change into intact or early intact proton and nucleus which are detected by the forward tagging proton and nucleus detectors. The two hard diffractive gluons collide each other to produce $\eta_{c}$ charmonium state. The Pomeron and Reggeon originating from the incoming proton and incoming nucleus radiate not only the hard diffractive gluons, but also the remnants. The remnant and the  $\eta_{c}$ go to the central detector.  In ND processes, the proton and nucleus radiate hard non diffractive gluons and remnants. Then, the two hard non diffractive gluons interact to generate the $\eta_{c}$ meson. There are no emission of Pomeron and Reggeon. The central detector capture the remnant and $\eta_{c}$ meson. There is no use of the forward detector in ND process for the hadroproduction. We have observed that the light and medium nuclei present the similar distribution profile and behavior which are different from  heavy nuclei distribution profile. The medium nuclei distribution is chosen in favor of light nuclei because of its valuable cross section, and the heavy nuclei distribution as well. The medium nuclei distributions in top panel reveals that $\mathbb{PR+RP}$ and total DD distributions are almost equal, but the non diffractive distribution is still dominant. In the bottom panel, the heavy nuclei distribution brings to light that $\mathbb{PR+RP}$ and total DD distributions are still early equal and dominant over the non diffractive distribution. The dominance of $\mathbb{PR+RP}$ and total DD distributions over ND distribution should be reduced by including the  effect of the soft inelastic interactions which should generate new secondary particles (survival gap probability). Then, the ND distribution should be larger than the diffractive cross sections.The $\mathbb{PR+RP}$ distribution is  significant over the other diffractive distributions. The $\mathbb{RR}$ distribution remain the smallest one. The examination of $\mathbb{PR+RP}$ contribution in $\rm pA$ mode can not be neglected for $\xi$ at the LHC.

\begin{figure}[htp]
\centering
\includegraphics[height=4.8cm,width=5.9cm]{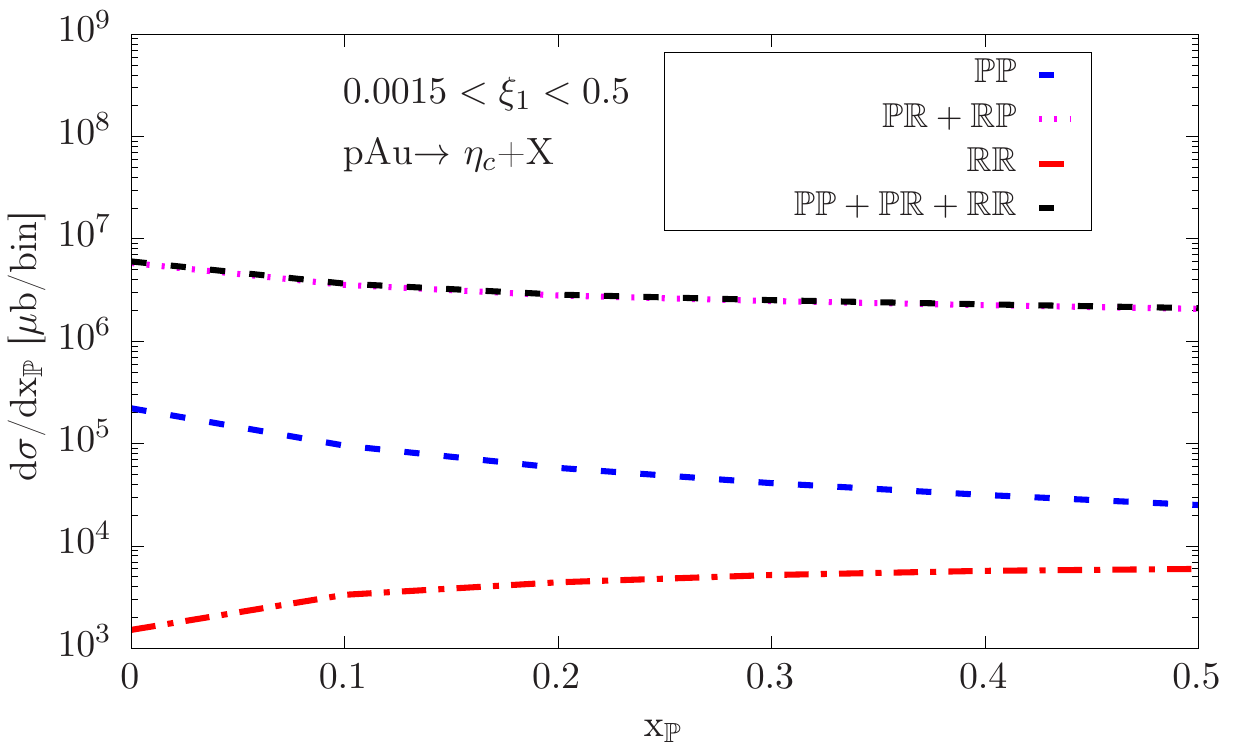}
\includegraphics[height=4.8cm,width=5.9cm]{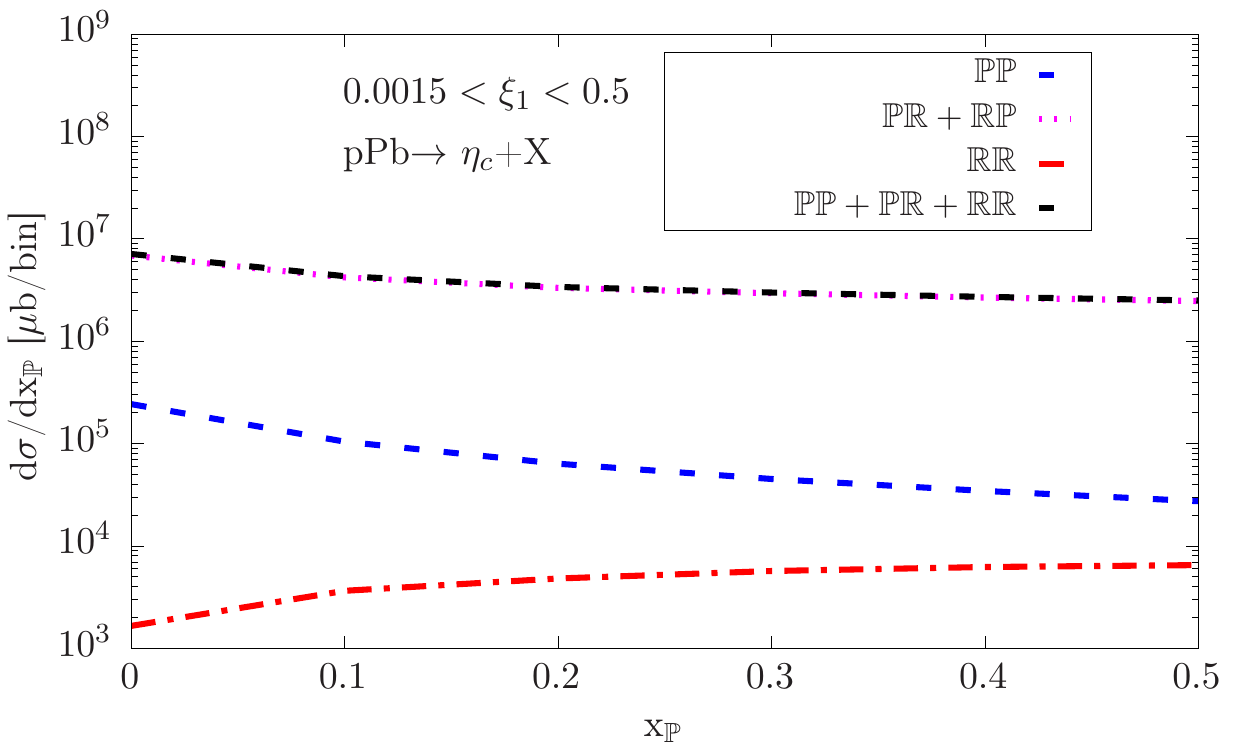}
\includegraphics[height=4.8cm,width=5.9cm]{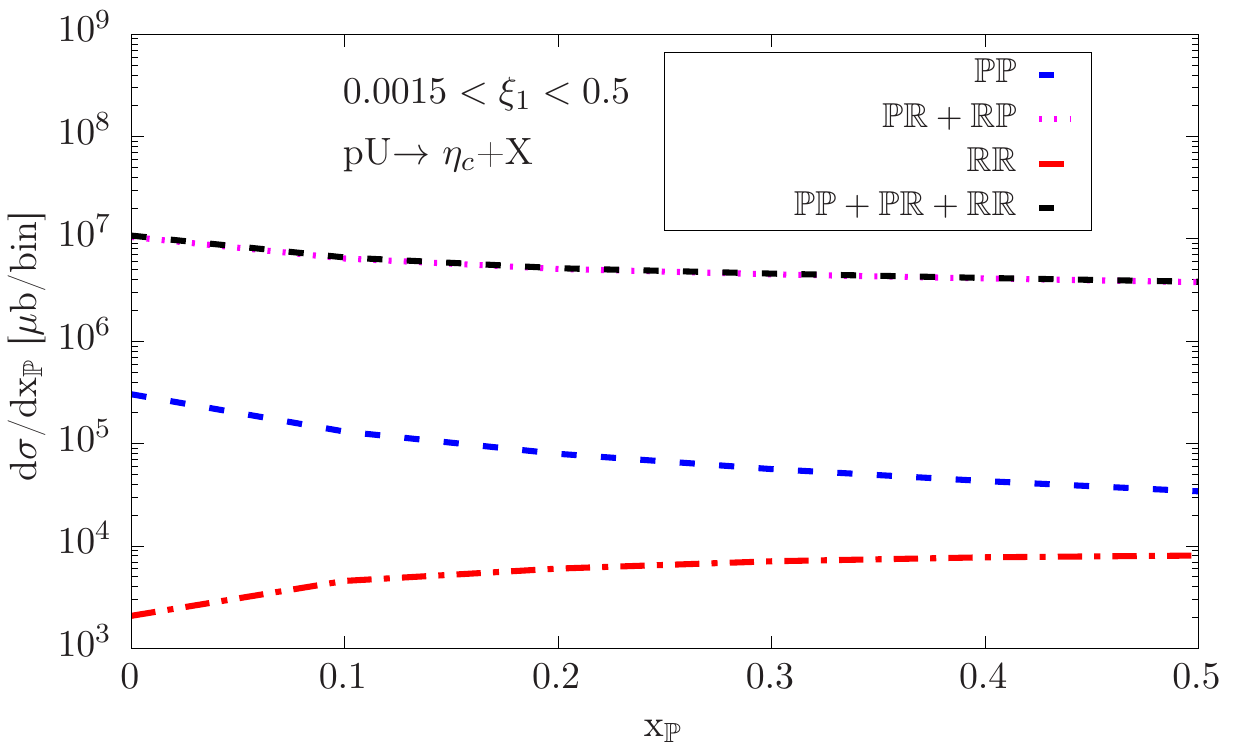}\\
\caption{ \normalsize (color online)
The $x_{\mathbb{P}}$ for the $\mathbb{PP}$ (blue dashed line), $\mathbb{ PR+RP}$ (purple dash dotted line), $\mathbb{ RR}$ (red dash dotted line) and $\mathbb{PP+PR+RP+RR}$ (black dotted line) in DD processes for pA collisions.}
\label{fig7:limits}
\end{figure}
The proton or nucleus longitudinal momentum fraction distributions for $\rm pA$ mode are given in Fig.\ref{fig7:limits}. The behavior and profile distributions of $x_{\mathbb{P}}$ show that the $\mathbb{ RR}$ distribution increases for small $x_{\mathbb{P}}$ and turns out to be flat for large  small $x_{\mathbb{P}}$. The $\mathbb{PP}$ and $\mathbb{ PR+RP}$ distributions decrease for small $x_{\mathbb{P}}$ and come to be flat for for large  $x_{\mathbb{P}}$.  The  $\mathbb{ PR+RP}$ distribution keeps the largest one and the $\mathbb{ RR}$  distribution remains the smallest one.

\begin{figure}[htp]
\centering
\includegraphics[height=4.8cm,width=5.9cm]{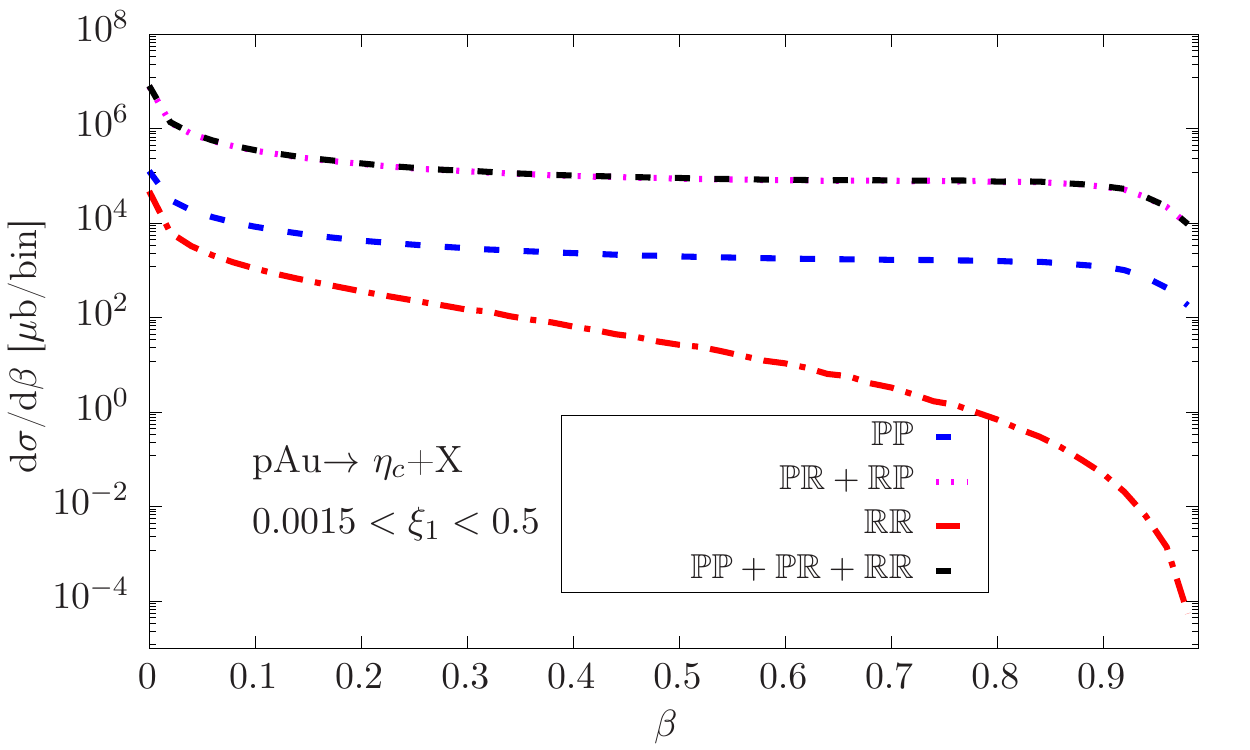}
\includegraphics[height=4.8cm,width=5.9cm]{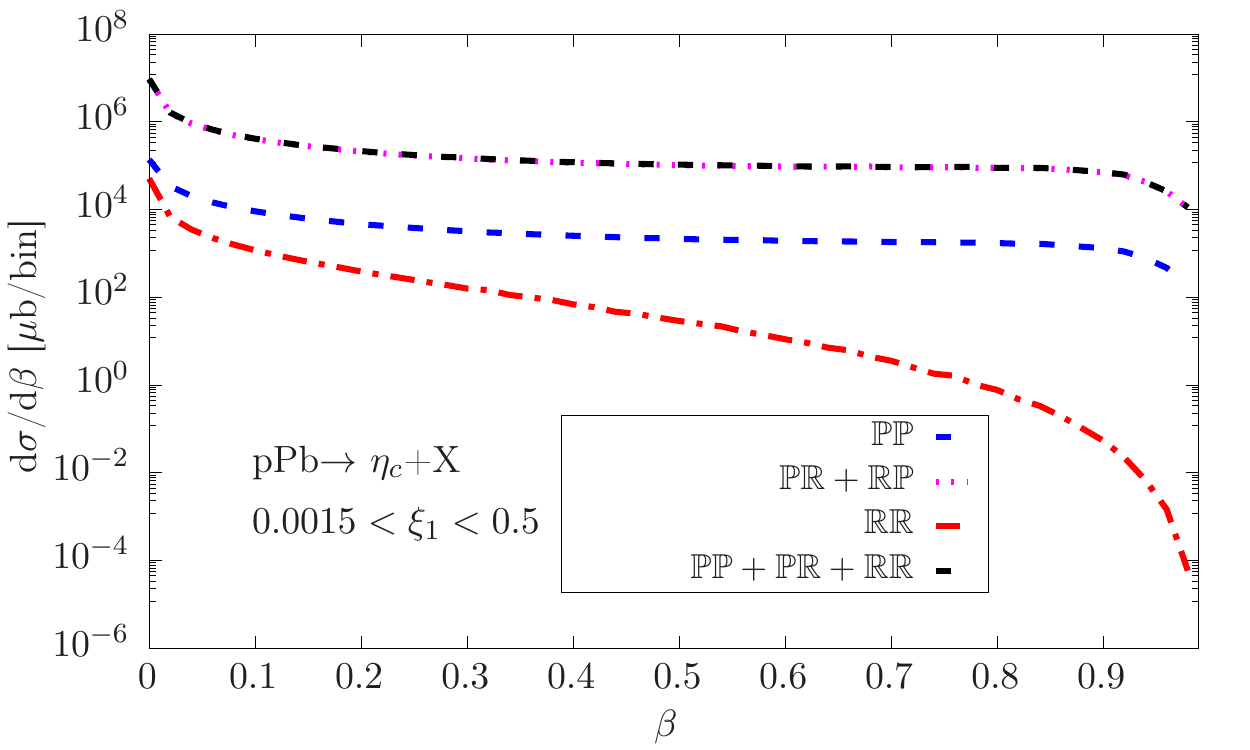}
\includegraphics[height=4.8cm,width=5.9cm]{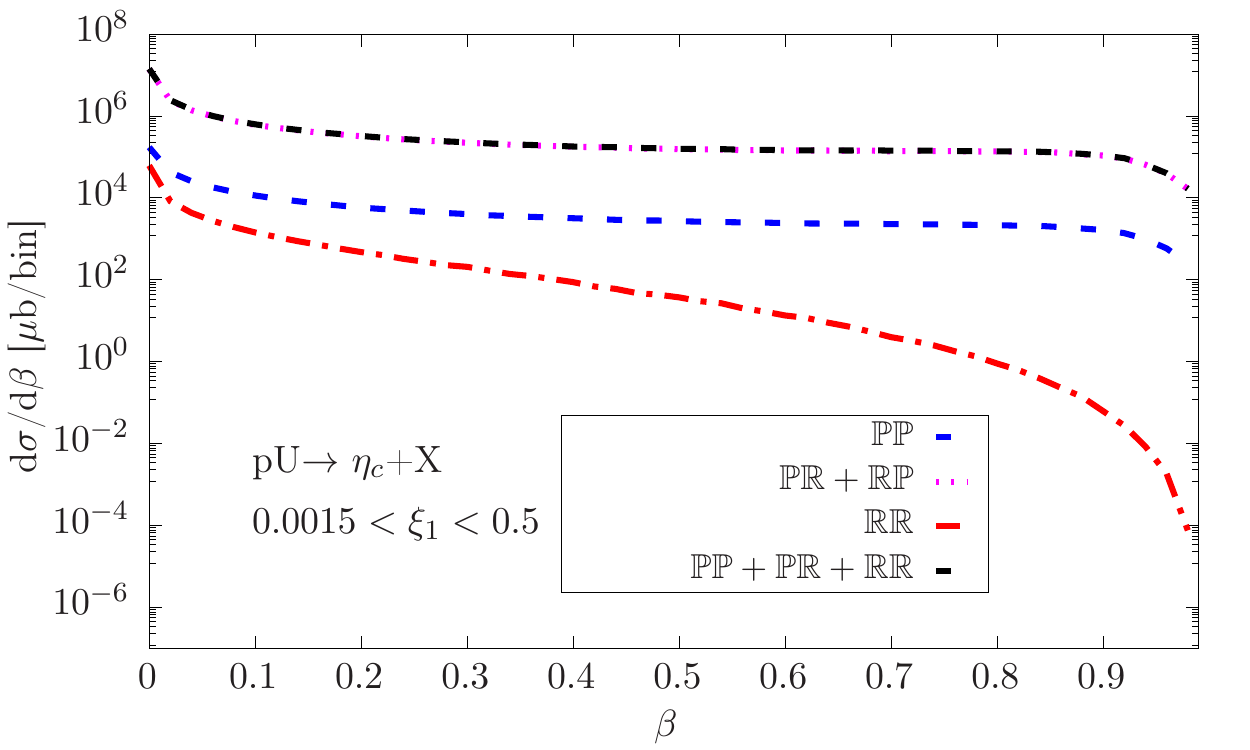}\\
\caption{ \normalsize (color online)
The $\beta$ distributions for the $\mathbb{PP}$ (blue dashed line), $\mathbb{ PR+RP}$ (purple dash dotted line), $\mathbb{ RR}$ (red dash dotted line) and $\mathbb{PP+PR+RP+RR}$ (black dotted line) in DD processes for pA collisions.}
\label{fig8:limits}
\end{figure}
The distribution of gluon longitudinal momentum fraction with respect to the exchanged Pomeron or Reggeon is provided in Fig.\ref{fig8:limits}. The convexity and the concavity are shown in the $\mathbb{PP}$, $\mathbb{ PR+RP}$ and  $\mathbb{ RR}$ distributions. The  $\mathbb{PP}$ and  largest $\mathbb{ PR+RP}$ distributions decrease for small $\beta$, become flat for the in-between value of $\beta$  and decrease for large $\beta$. The lowest $\mathbb{ RR}$ distribution falls for small and large $\beta$. 
\subsection{Single diffractive production}

\subsubsection{pp collision}

The SD and ND cross section assessments of $\eta_{c}$ hadroproduction in pp mode are also tabulated in Table.\ref{tab:2} for different channels. The Reggeon cross section is still important than the Pomeron one. The total SD cross section has been increased by the $\mathbb{R}$p contribution. It has noted that the ND cross section is  a same factor of magnitude than SD one. The SD cross section is a factor 10 times larger than the total DD cross section. Our attention has been drawn due to the fact that the Reggeon cross section can not be ignored in $\rm pp$ mode. The total DD cross section of $\rm pA$ ($\rm AA$ ) mode is a same factor than that of SD cross section of  $\rm pp$ for the light nuclei (except for the Beryllium). The total DD cross section of $\rm pA$ and $\rm AA$  modes is a factor $\gtrsim$ 10 larger than that of SD cross section of  $\rm pp$ for the medium and heavy nuclei.
\begin{figure}[htp]
\centering
\includegraphics[height=4.8cm,width=5.9cm]{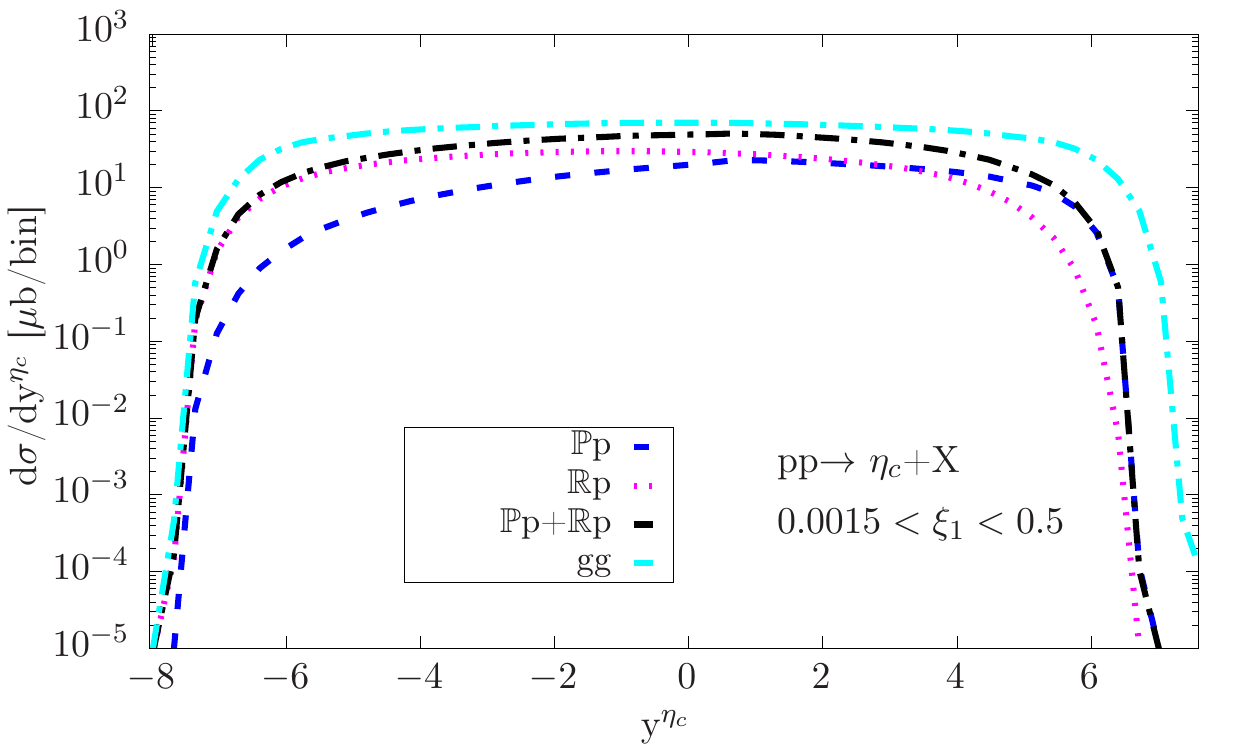}
\includegraphics[height=4.8cm,width=5.9cm]{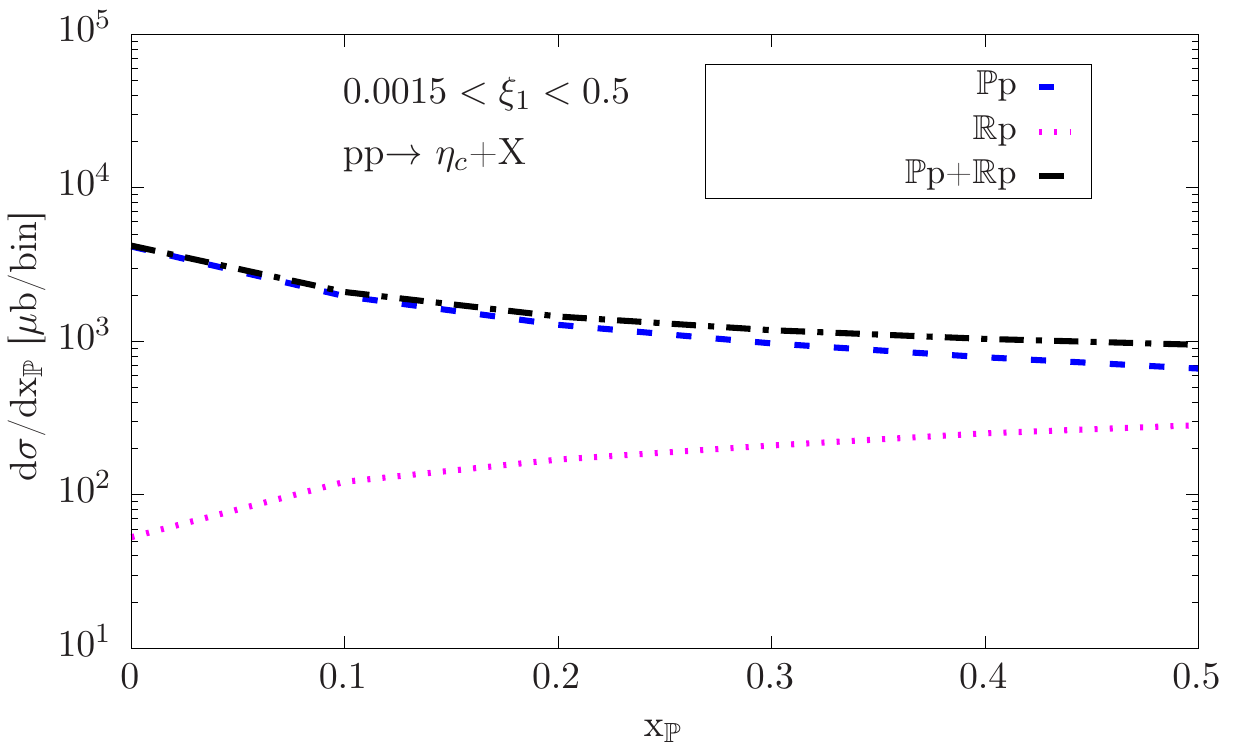}
\includegraphics[height=4.8cm,width=5.9cm]{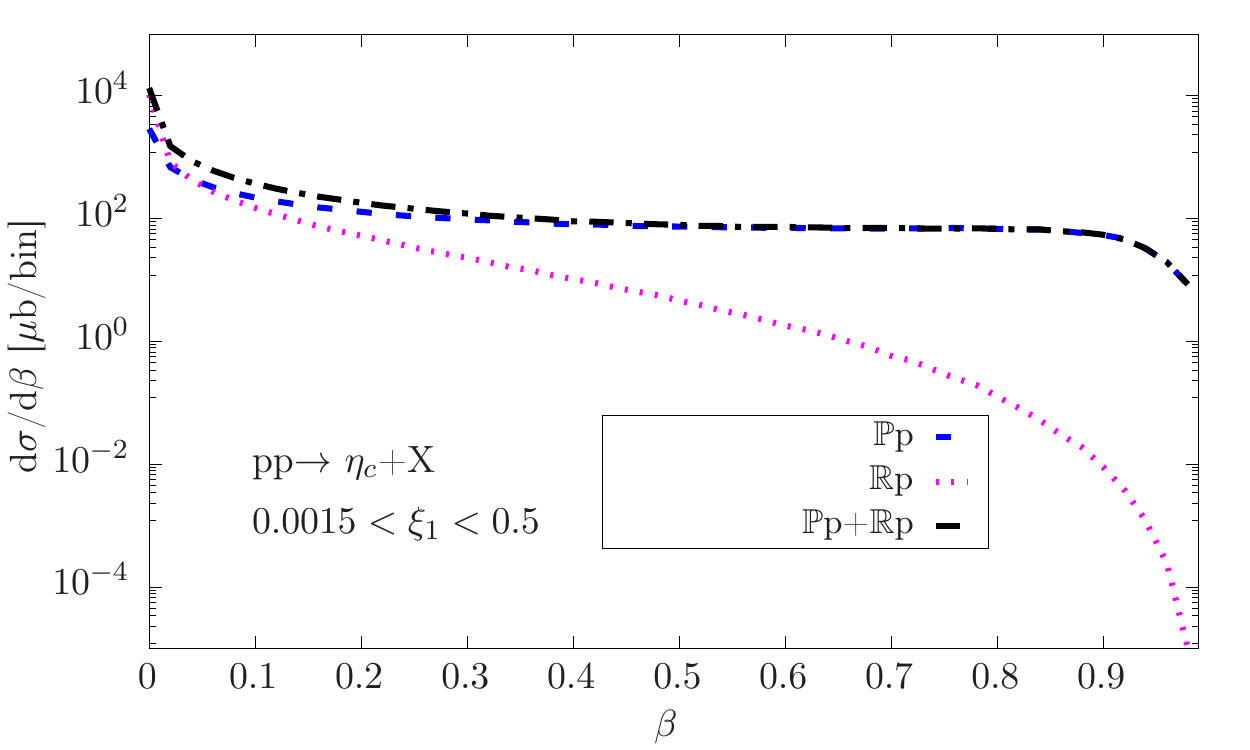}\\
\caption{\normalsize (color online) The $\rm y^{\eta_{c}}$, $x_{\mathbb{P}}$ and $\beta$ distributions for the $\mathbb{P}$p (blue dashed line), $\mathbb{R}$p (magenta dotted line), $\mathbb{P}$p+$\mathbb{R}$p (black dash dotted line) and $\rm gg$ (cyan dash dotted line) in DD processes for pp.}
\label{fig9:limits}
\end{figure}

The rapidity, gluon longitudinal momentum fraction and gluon longitudinal momentum fraction with respect to the exchanged Pomeron and Reggeon are displayed in Fig.\ref{fig9:limits} for pp mode in SD process. In this case, the two initial incoming particles are protons. The first proton emits  hard non diffractive gluon and turns into remnant system $\rm X$ while the second proton emits hard diffractive gluon via Pomeron or Reggeon and changes into intact or early intact proton. The hard process proceeds through the interaction of  hard non diffractive  gluon and hard diffractive gluon to produce the $\eta_{c}$ meson. The almost intact proton, remnant and  $\eta_{c}$ meson are observed in the final state. In ND proces, the two protons radiate hard non diffractive gluon to generate the $\eta_{c}$ meson, escorted by the remnant in the final state without intact protons. The  $\eta_{c}$ meson and the remnant are observed in the central detector in SD and SD processes whereas the intact proton is observed in forward tagging proton detector.  On the left side of the figure, the rapidity distribution of $\mathbb{P}$p and $\mathbb{R}$p are asymmetric due to unequal backward and forward distributions while the non diffractive are symmetric with respect to mid rapidity. Negative (positive) pseudorapidity  is referred to as left or forward (right or bacward) side of the detector for $\rm y < 0 $ ($\rm y > 0$). The dominant distribution is from the Reggeon distribution. The lowest $\mathbb{P}$p ( largest $\mathbb{R}$p) rapidity distributions for the SD dissociation have maximums shifted to forward (backward) rapidities with respect to the non diffractive case. Our SD results in $\rm pA$ mode at the LHC could be used to investigate the Reggeon contribution, since a kinematic window of dominance has been found which could be used experimentally to isolate and constrain it. On the middle side, proton longitudinal momentum fraction distribution shows that the large Pomeron distribution decreases for small $x_{\mathbb{P}}$ and becomes flat for large $x_{\mathbb{P}}$. The lower Reggeon distribution increases for small $x_{\mathbb{P}}$ and becomes flat for large $x_{\mathbb{P}}$.
 On the right panel, the gluon longitudinal momentum fraction with respect to the exchanged Pomeron or Reggeon distribution reveals that the Reggeon distribution is reducing for small and large $\beta$ whereas the Pomeron one is reducing  for small $\beta$, becoming flat and reducing again for large $\beta$. The convexity and the concavity are revealed in their end points. The constraint on Reggeon distribution at LHC should enhance the theoretical predictions for $\rm \eta_{c}$. 

\subsubsection{AA collision}
The SD and ND cross section evaluations in AA mode are tabularized in Table.\ref{tab:5} for the light, medium and heavy nuclei. The cross section enhances with the enlargement of atomic mass number, A. The cross section becomes considerable for heavy nuclei. As we can see, the Pomeron contribution to total SD cross section is larger than the Reggeon one. The Pomeron contribution is  a same factor than total SD cross section. The total SD cross sections  is a factor $\gtrsim$ 10 larger than that of the total DD one for the light and medium nuclei in $\rm AA$ mode. For the heavy nuclei, they keeps the same factor of magnitude. The SD cross in $\rm AA$ mode  is a factor $\gtrsim$ 10 larger than that of $\rm pp$ mode for SD and DD processes. The SD cross section  in $\rm AA$ mode  is a factor $\gtrsim$ 10 larger than that of $\rm pA$ mode in DD process. The increase of SD cross section in $\rm AA$ mode with respect to $\rm pp$ mode  can be explained by the  hard non diffractive gluon distribution from A, the hard diffractive gluon distribution from A and the nuclear form factor, i.e., proportional to $\rm A^{3}R^{2}_{g}F^{2}_{A}(t)$. The ND cross section in $\rm AA$ mode is a factor $\gtrsim$ 10 larger than that of SD process in the same mode.
\begin{table*}[htbp]
	\begin{center}
		\begin{tabular}{|c |c | c | c |c |c |c |}\hline
			\multicolumn{3}{|c|}{ }&\multicolumn{3}{c|}{SD}& ND  \\ \hline	
			Nuclei   & AA      & $\rm \langle \left\vert S\right\vert ^{2}\rangle $             & $\mathbb{P}$A              & $\mathbb{R}$A & Total & gg            \\ \hline
		\multirow{3}{*}{Light nucleus}& BeBe &  4.06$\times$10$^{-4}$ &1.29$\times$10$^{3}$    & 3.56$\times$10$^{2}$&1.65$\times$10$^{3}$ &5.45$\times$10$^{4}$  \\
		\multirow{3}{*}{}&CC  & 2.05$\times$10$^{-2}$  & 2.74$\times$10$^{3}$   &7.23$\times$10$^{2}$ & 3.46$\times$10$^{3}$& 9.45$\times$10$^{4}$ \\
		& OO   & 6.51$\times$10$^{-3}$   &5.85$\times$10$^{3}$ &1.46$\times$10$^{3}$ &7.31$\times$10$^{3}$  &1.64$\times$10$^{5}$\\
		\cline{1-7}
		\multirow{3}{*}{Medium nucleus}& CaCa   &1.67$\times$10$^{-4}$    &6.42$\times$10$^{4}$&1.37$\times$10$^{4}$ & 7.79$\times$10$^{4}$  &9.37$\times$10$^{5}$\\
		\multirow{3}{*}{}&CuCu  &2.54$\times$10$^{-5}$   &2.18$\times$10$^{5}$&4.31$\times$10$^{4}$ &2.61$\times$10$^{5}$  &2.28$\times$10$^{6}$\\
		&AgAg  &  2.82$\times$10$^{-6}$   & 8.42$\times$10$^{5}$& 1.53$\times$10$^{5}$ &9.95$\times$10$^{5}$ &6.10$\times$10$^{6}$\\
		\cline{1-7}
		\multirow{3}{*}{Heavy nucleus}& AuAu& 2.83$\times$10$^{-7}$  &3.95$\times$10$^{6}$&6.52$\times$10$^{5}$ & 4.60$\times$10$^{6}$ &1.87$\times$10$^{7}$\\
		\multirow{3}{*}{}&PbPb  &2.29$\times$10$^{-7}$   &4.54$\times$10$^{6}$  &7.43$\times$10$^{5}$ &5.28$\times$10$^{6}$  &2.07$\times$10$^{7}$\\
		& UU   &1.28$\times$10$^{-7}$  &6.42$\times$10$^{6}$   &1.03$\times$10$^{6}$ & 7.45$\times$10$^{6}$ &2.66$\times$10$^{7}$\\
			\hline
		\end{tabular}
	\end{center}\caption{\label{tab:5}
		The SD and ND cross sections in $\mu$b for the $\eta_{c}$ hadroproduction  $\rm AA$ mode  at LHC with forward detector acceptance, $0.0015<\xi<0.5$.}
\end{table*}

\begin{figure}[htp]
\centering
\includegraphics[height=4.8cm,width=5.9cm]{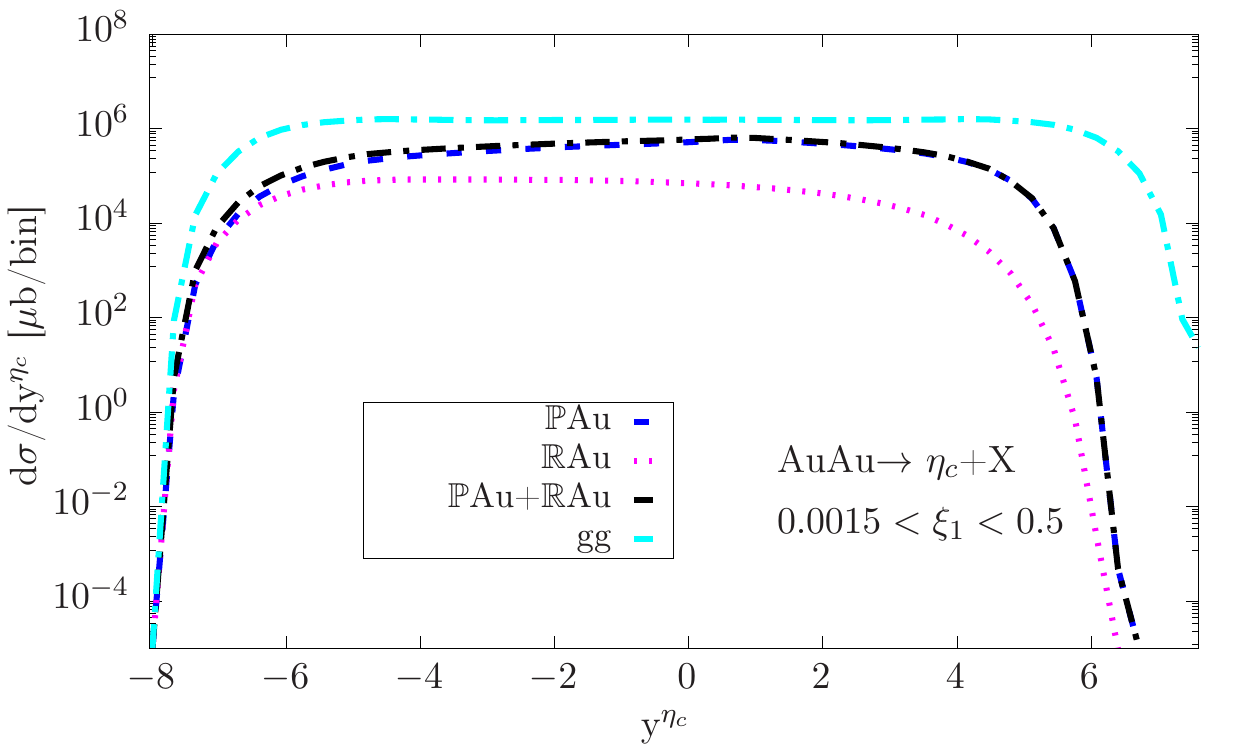}
\includegraphics[height=4.8cm,width=5.9cm]{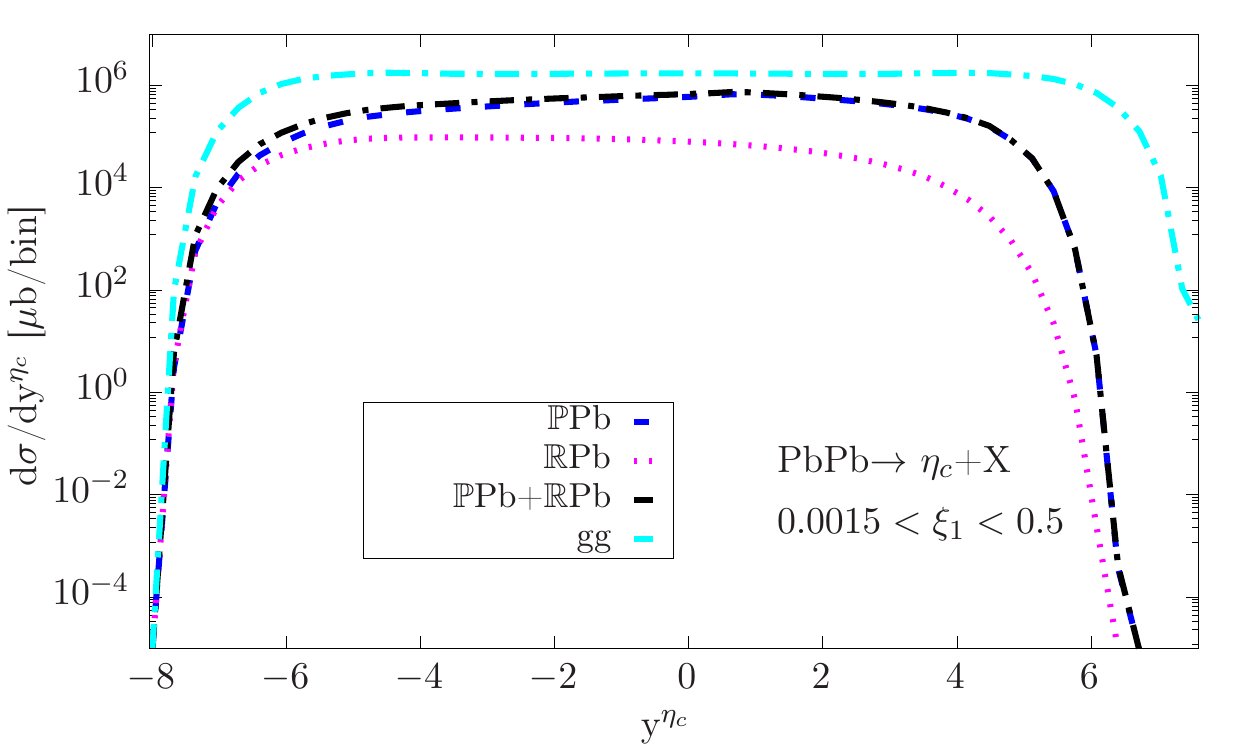}
\includegraphics[height=4.8cm,width=5.9cm]{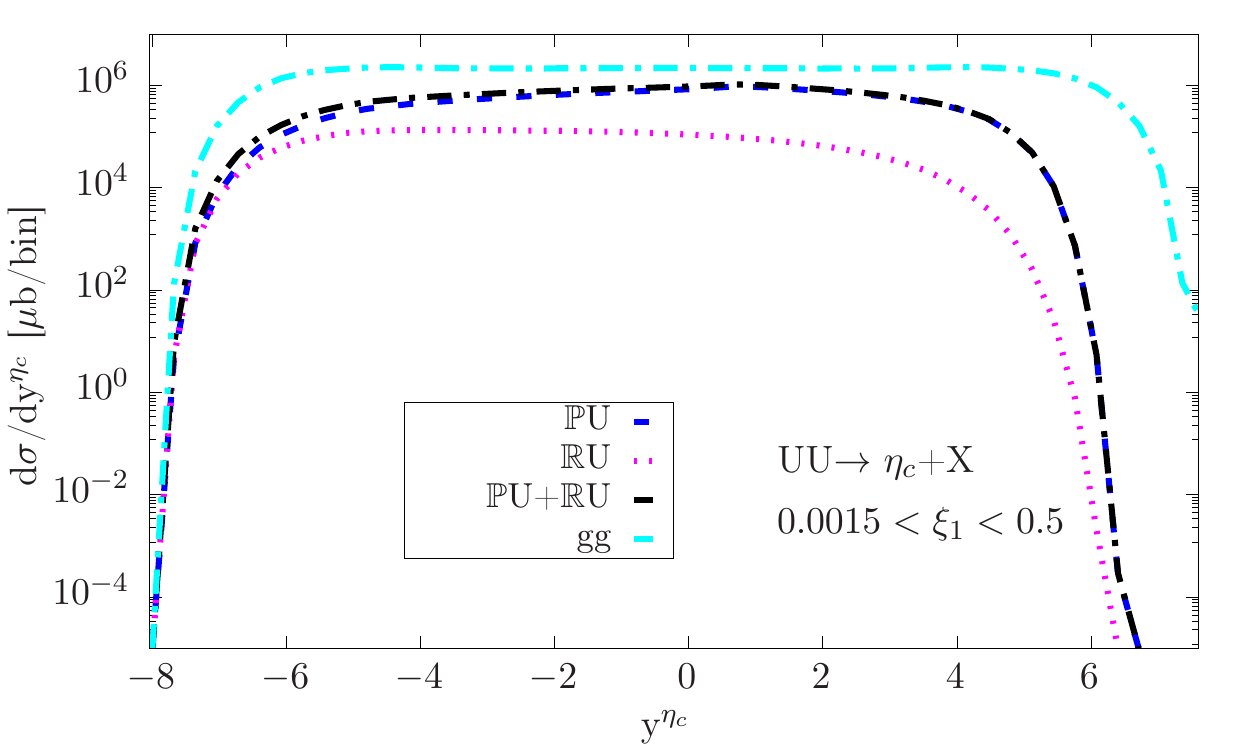}
\caption{\normalsize (color online) The $\rm y^{\eta_{c}}$   distributions for the the $\mathbb{P}$A (blue dashed line), $\mathbb{R}$A (magenta dotted line), $\mathbb{P}$A+$\mathbb{R}$A (black dash dotted line) and $\rm gg$ (cyan dash dotted line)  in SD processes for  AA collisions.}
\label{fig10:limits}
\end{figure}
The rapidity distributions in AA mode for SD process are given in Fig.\ref{fig10:limits} for three types of nucleus.
The two incoming colliding nuclei are identical. In this interaction, one of two incoming nuclei emit hard non diffractive gluon and changes into remnant system $\rm X$ while the second incoming nucleus radiate the hard diffractive gluon through Pomeron or Reggeon and turns in to intact or almost intact nucleus.The Pomeron or Reggeon
radiates not only the gluon, but also the remnant. The $\rm \eta_{c}$ meson is produced through the hard collision of the non diffractive gluon with the diffractive gluon, which is detected in the central detector with the remnant. The intact nucleus is captured by the forward tagging nucleus detector. The diffractive gluon from nucleus and the non diffractive gluon from nucleus  are differently parametrized in terms of their parton distribution functions. The SD distribution is asymmetric and ND distribution is symmetric. The larger $\mathbb{P}$A ( lower $\mathbb{R}$A) rapidity distributions for the single diffractive dissociation have maximums shifted to forward (backward) rapidities with respect to the non diffractive.

\begin{figure}[htp]
\centering
\includegraphics[height=4.8cm,width=5.9cm]{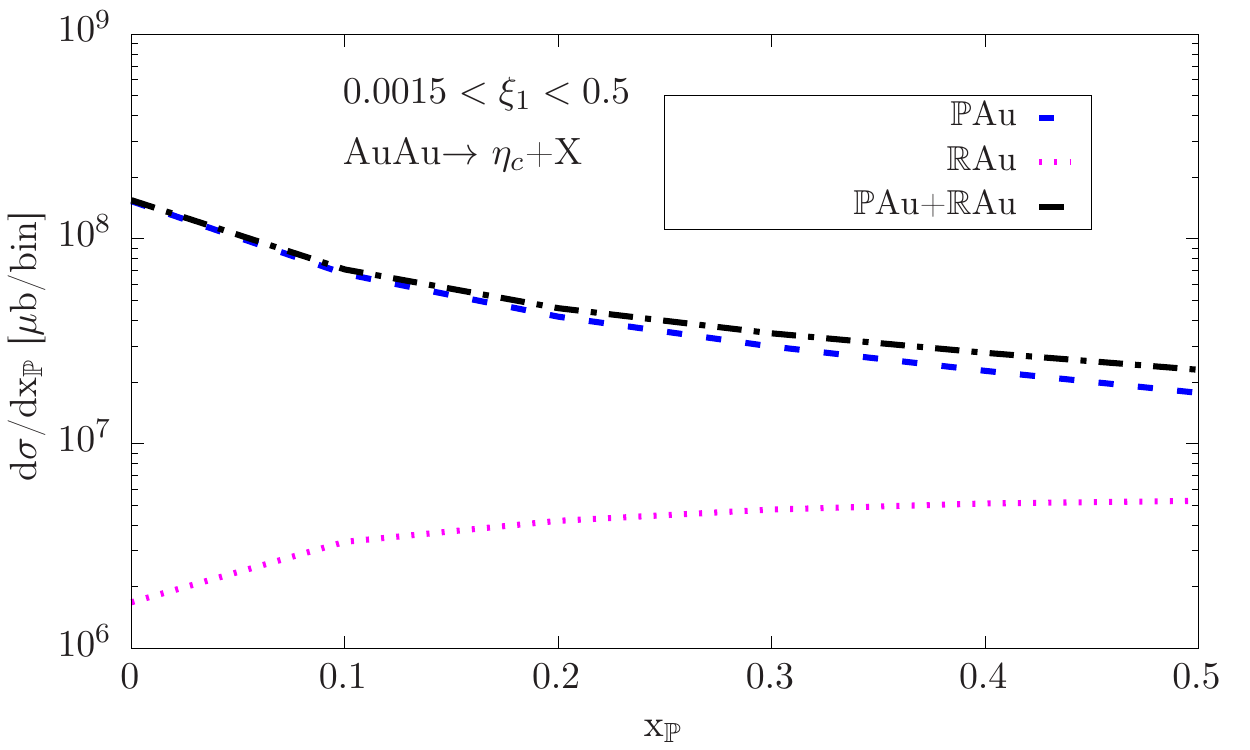}
\includegraphics[height=4.8cm,width=5.9cm]{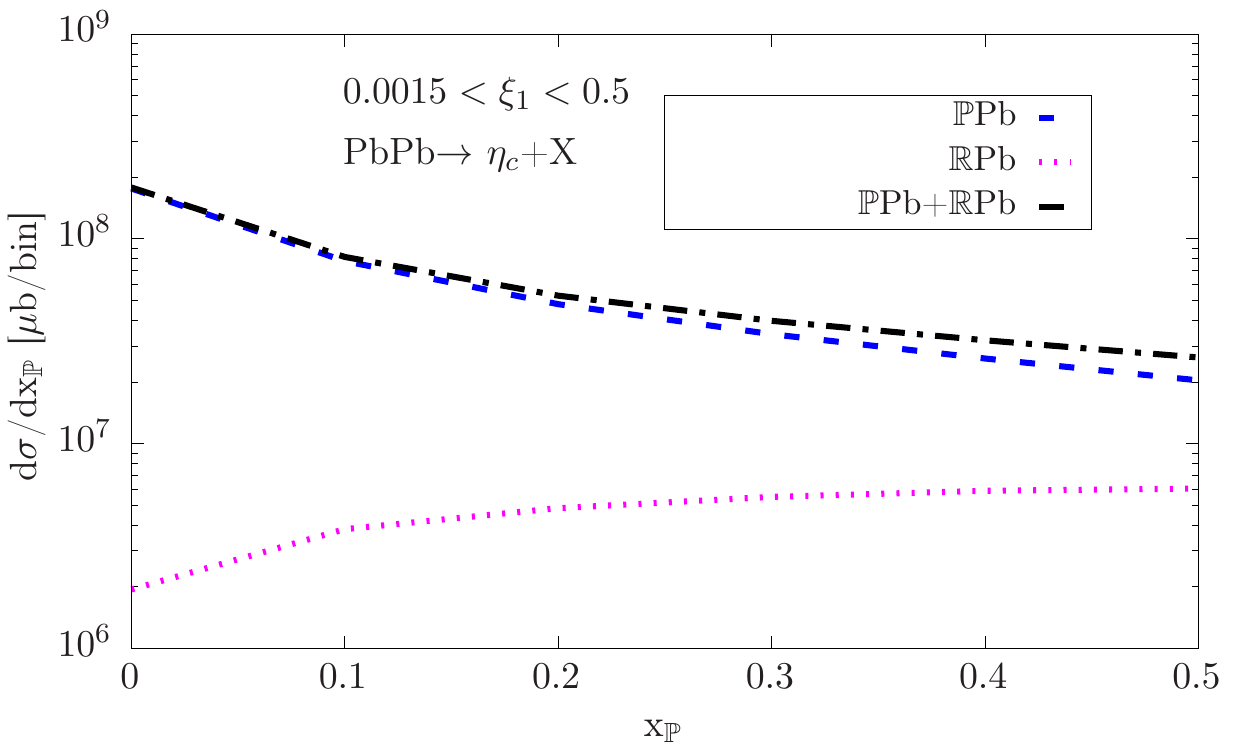}
\includegraphics[height=4.8cm,width=5.9cm]{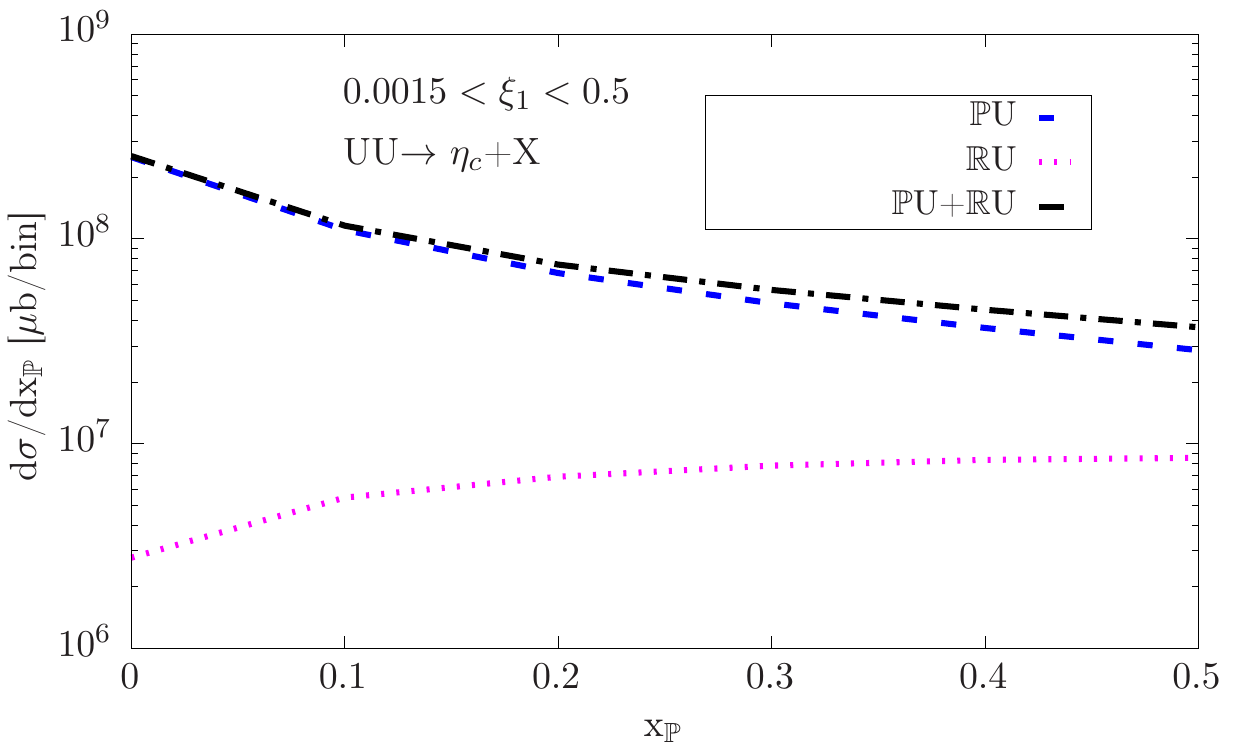}
\caption{ \normalsize (color online)
The $x_{\mathbb{P}}$ distributions for the $\mathbb{P}$A (blue dashed line), $\mathbb{R}$A (magenta dotted line) and $\mathbb{P+R}$A (black dash dotted line) in SD processes for AA collisions.}
\label{fig11:limits}
\end{figure}
The $x_{\mathbb{P}}$ distributions of $\eta_{c}$ for  heavy nuclei in AA mode are arranged in Fig.\ref{fig11:limits} for SD processes.
The distribution of $\mathbb{P}$A falls for small $x_{\mathbb{P}}$ and become flat for large $x_{\mathbb{P}}$. The distribution of $\mathbb{R}$A enlarges for small $x_{\mathbb{P}}$ and become flat for large $x_{\mathbb{P}}$.

\begin{figure}[htp]
\centering
\includegraphics[height=4.8cm,width=5.9cm]{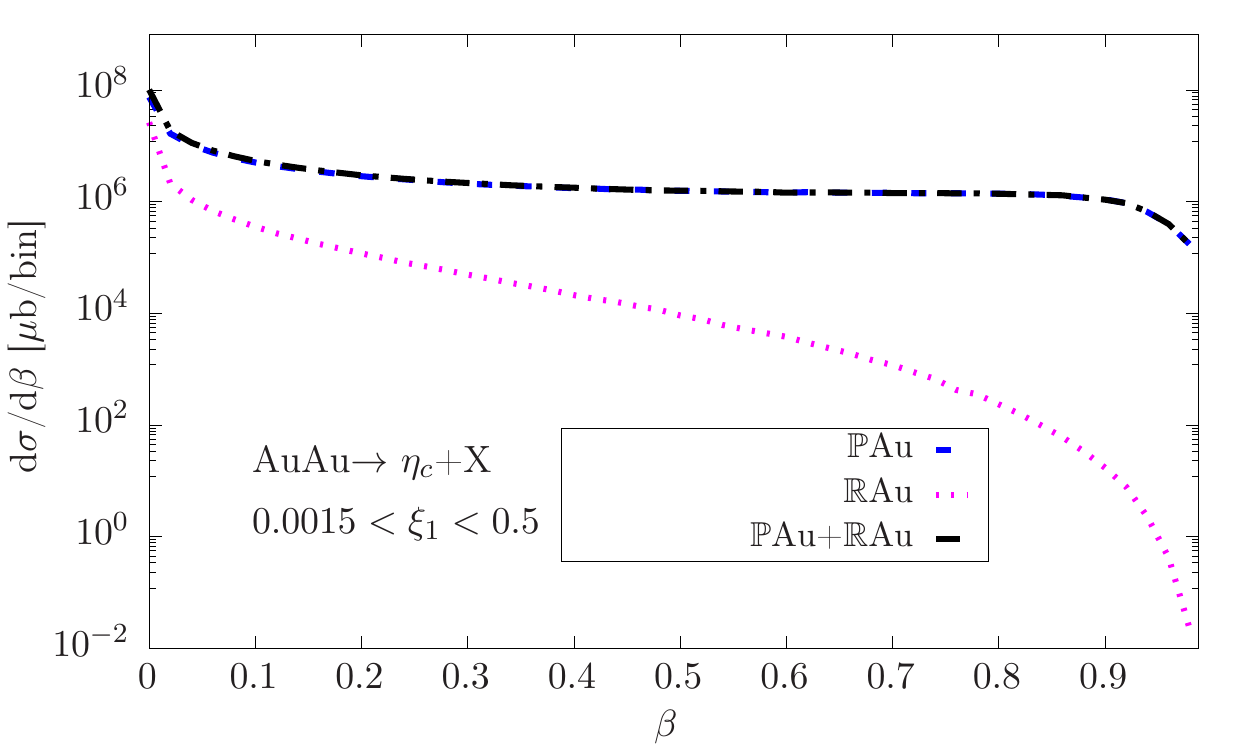}
\includegraphics[height=4.8cm,width=5.9cm]{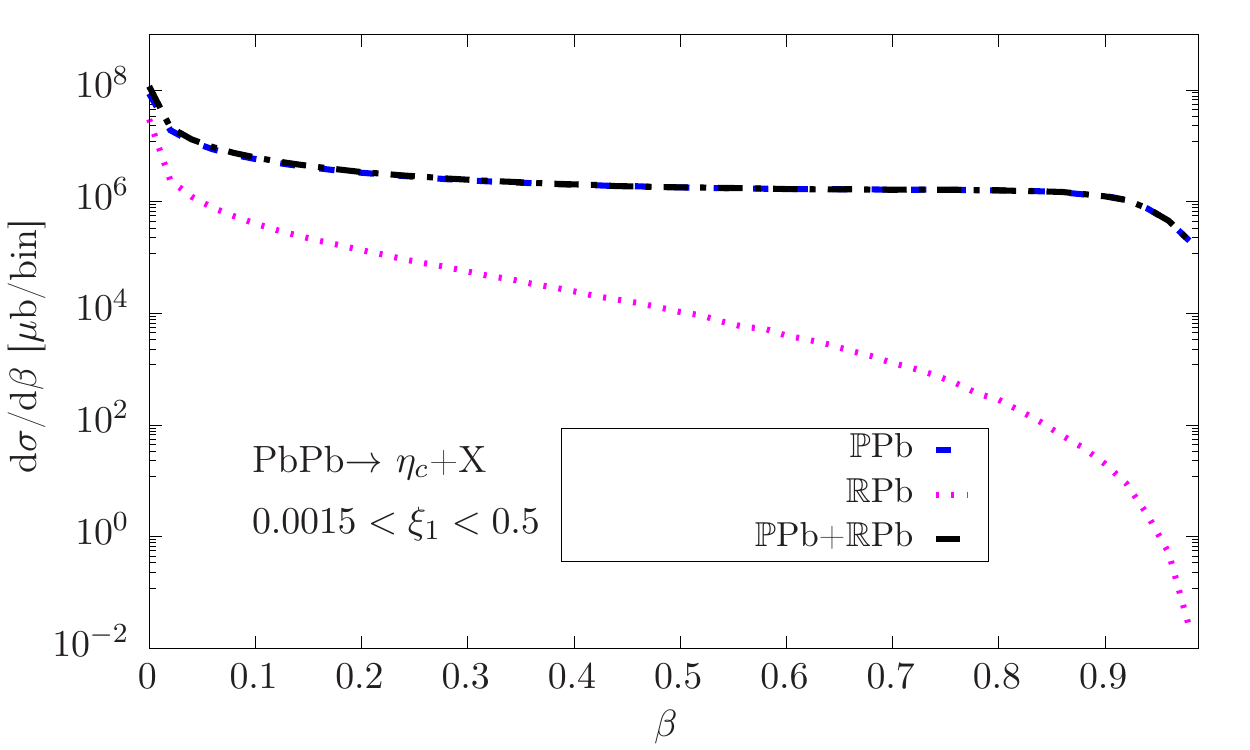}
\includegraphics[height=4.8cm,width=5.9cm]{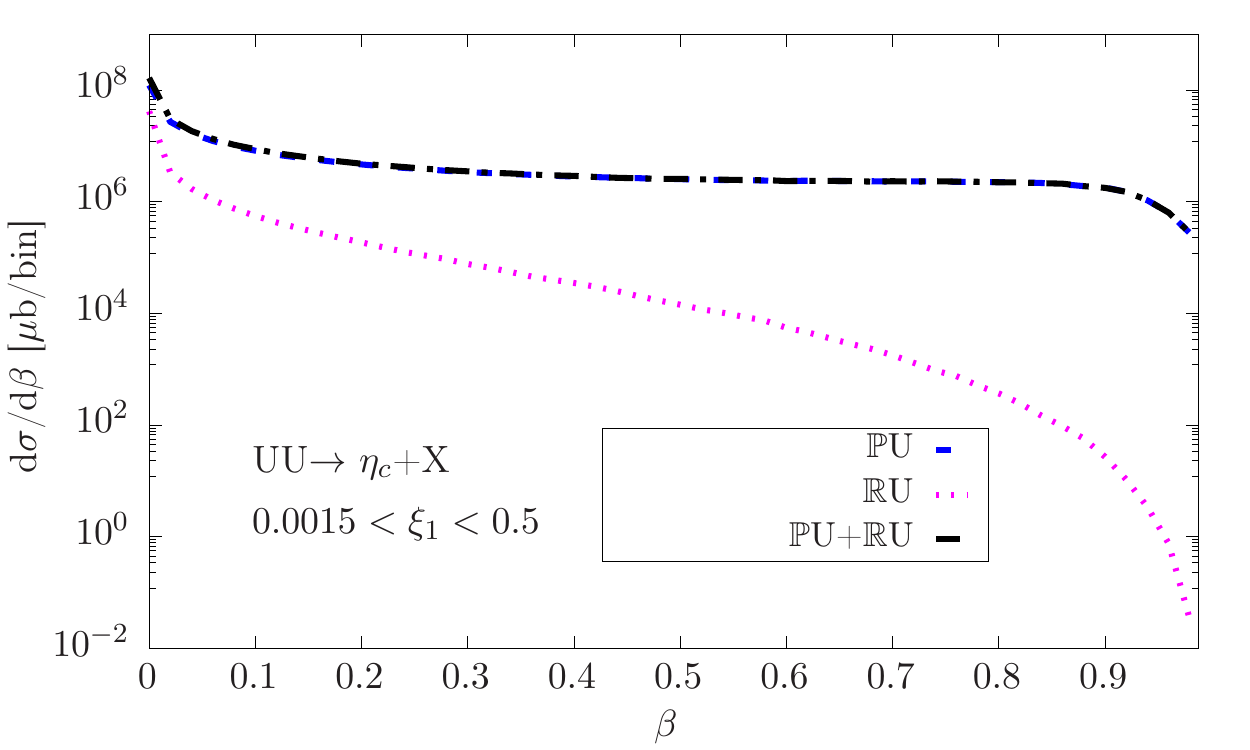}
\caption{ \normalsize (color online)
The $\beta$ distributions for the $\mathbb{P}$A (blue dashed line), $\mathbb{R}$A (magenta dotted line) and $\mathbb{ P+R}$A (black dash dotted line) in SD processes for AA collisions.}
\label{fig12:limits}
\end{figure}
The Fig.\ref{fig12:limits} shows the $\beta$ distributions of $\eta_{c}$ for heavy nuclei in $\rm AA$ mode. The $\beta$ distribution for the $\mathbb{P}$A decreases for small $\beta$, becomes flat and decreases again for large $\beta$. The $\beta$ distribution for the $\mathbb{R}$A drops along with  $\beta$

\subsubsection{pA collision}

The SD and ND cross section estimations in pA mode for $\eta_{c}$ hadroproduction is organized in Tab.\ref{tab:6}. In this table two situations are presented with two distinct incoming particles, $\rm A$ and $\rm p$. For the first situation, one of the two incoming particles, for example nucleus $\rm A$ emits the hard non diffractive gluon and the other incoming particle, for example proton radiates the hard  diffractive gluon via the Pomeron or Reggeon. The hard non diffractive gluon collides with the hard  diffractive gluon to give rise to $\eta_{c}$ meson in the final state. There are also presence of remnants emitted by the Pomeron or Reggeon as well as the intact proton in the final state. In this case, the Reggeon contribution to the total cross section overpasses the Pomeron contribution. The total SD cross section in $\rm AA$ mode  is a factor $\gtrsim$ 10 larger than that of  $\rm pA$ mode in SD process for the medium and heavy nuclei. The total SD cross section in $\rm AA$ mode  is  a same factor than that of $\rm pA$ mode in SD process for the light nuclei. The total SD cross section in $\rm pA$ mode  is a factor $\gtrsim$ 10 larger than that of $\rm pp$ mode in SD process. The ND cross section in $\rm pA$ mode is a factor $\gtrsim$ 10 larger than that of $\rm pA$ mode in SD process and they are a same factor for the middle medium nucleus. The total DD (SD) cross section in  $\rm pA$ mode is a factor 10 times larger than that of the total SD (DD) cross section in  $\rm pA$ for the heavy nuclei (light nuclei). They are of a same order for the medium nuclei. The total DD cross section in  $\rm AA$ mode is a factor 10$^{2}$ times larger than that of the total SD  cross section in  $\rm pA$ for the heavy nuclei. The total SD cross section in $\rm pA$ mode  is a factor $\gtrsim$ 10 larger than that of $\rm AA$ mode in DD process for the light nuclei. The total DD cross section in  $\rm AA$ mode is  a factor 10 times larger than that of the total SD  cross section in  $\rm pA$ for the last medium nuclei while they are a same factor for the two first medium nuclei. The total SD cross section in $\rm pA$ mode  is a factor $\gtrsim$ 10$^{2}$ larger than that of $\rm pp$ mode in DD process. The boost of $\rm pA$ mode cross section in SD processes pertaining to $\rm pp$ mode can be elucidated by the hard non diffractive gluon distribution from A, i.e., proportional to $\rm AR_{g}$. 

For the second situation, one of the two incoming particle $\rm A$ radiates the  hard diffractive gluon and the other incoming particle proton emits the hard non diffractive gluon. The $\eta_{c}$ meson is obtained for the hard collision of the different gluons. The final sate is composed of $\eta_{c}$ meson, remnant and intact nucleus. The  Pomeron contribution to the cross section surpasses the Reggeon contribution. The ND cross section in $\rm pA$ mode is a factor $\gtrsim$ 10$^{2}$ larger than that of $\rm pA$ mode in SD process. The SD cross section in $\rm pA$ mode  is a factor $\gtrsim$ 10 larger than that of $\rm pp$ mode in SD process for the medium and heavy nuclei and they are a same factor for light nuclei. The total SD cross section in $\rm AA$ mode  is a factor $\gtrsim$ 10 larger than that of $\rm pA$ mode in SD process.
The total DD cross section in $\rm AA$ mode is a factor 10$^{2}$ times larger than that of the total SD  cross section in  $\rm pA$ for the heavy nuclei. The total DD cross section in $\rm AA$ mode and the total SD  cross section in  $\rm pA$ are of a same order of magnitude for the two last light and the first medium nuclei. The total DD (SD) cross section in $\rm AA$ ($\rm pA$) mode is  a factor 10 times larger than that of the total SD (DD)  cross section in  $\rm pA$ ($\rm AA$) mode for the last medium nuclei. The DD  cross section in $\rm pA$ mode is  a factor 10 times larger than that of the total SD cross section in  $\rm pA$  mode for the heavy nuclei. The total DD  cross section in $\rm pA$ mode is a same factor than that of the total SD cross section in  $\rm pA$  mode for the light nuclei. The total DD  cross section in $\rm pA$ mode is a same factor than that of the total SD cross section in  $\rm pA$  mode for the first and the last medium nuclei, while  for the middle medium nucleus, the total DD  cross section in $\rm pA$ mode is a factor 10 times larger than that of the total SD cross section in  $\rm pA$  mode. The total SD cross section in $\rm pA$ mode  is a factor $\gtrsim$ 10 larger than that of $\rm pp$ mode in DD process. The improvement of p$\rm A$ mode cross sections in SD processes with regard to $\rm pp$ mode can be clarified by the hard diffractive gluon distribution from A and nuclear form factor, i.e., proportional to $\rm A^{2}R_{g}F^{2}_{A}(t)$. The exception is made for the light nuclei. The Pomeron and Reggeon contributions to the SD cross section should be measured separately because of their opposite contribution in pA mode at the LHC.
\begin{table*}[htbp]
	\begin{center}
		\begin{tabular}{|c |c | c | c |c |c |c |c |c |c |}\hline
			\multicolumn{3}{|c|}{ }&\multicolumn{6}{c|}{SD}& ND \\ \hline	
			Nuclei   & pA      & $\rm \langle \left\vert S\right\vert ^{2}\rangle $             & $\mathbb{P}$A              & $\mathbb{R}$A & Total1 & $\mathbb{P}$p & $\mathbb{R}$p & Total2 & gg           \\ \hline
		\multirow{3}{*}{Light nucleus}& pBe &3.65$\times$10$^{-3}$  & 7.28$\times$10$^{2}$ & 1.12$\times$10$^{3}$ & 1.84$\times$10$^{3}$&1.51$\times$10$^{2}$ & 4.11$\times$10$^{1}$  &1.92$\times$10$^{2}$ & 1.27$\times$10$^{4}$  \\
		\multirow{3}{*}{}&pC  &2.73$\times$10$^{-3}$  & 9.67$\times$10$^{2}$   & 1.48$\times$10$^{3}$ & 2.45$\times$10$^{3}$ & 2.46$\times$10$^{2}$&6.32$\times$10$^{1}$ &3.09$\times$10$^{2}$ & 1.67$\times$10$^{4}$\\
		& pO   & 2.05$\times$10$^{-3}$  & 1.28$\times$10$^{3}$&1.96$\times$10$^{3}$ &3.24$\times$10$^{3}$ &3.99$\times$10$^{2}$ & 9.70$\times$10$^{1}$&4.96$\times$10$^{2}$ &2.20$\times$10$^{4}$  \\
		\cline{1-10}
		\multirow{3}{*}{Medium nucleus}& pCa &8.26$\times$10$^{-4}$  &3.16$\times$10$^{3}$   & 4.82$\times$10$^{3}$    &7.98$\times$10$^{3}$ &1.84$\times$10$^{3}$ & 3.76$\times$10$^{2}$ & 2.22$\times$10$^{3}$ &5.28$\times$10$^{4}$   \\
		\multirow{3}{*}{}&pCu  &5.14$\times$10$^{-4}$    & 5.01$\times$10$^{3}$ &7.63$\times$10$^{3}$ &1.26$\times$10$^{4}$ &4.01$\times$10$^{3}$ & 7.50$\times$10$^{2}$ &4.76$\times$10$^{3}$ &8.25$\times$10$^{4}$ \\
		&pAg  & 3.04$\times$10$^{-6}$    &8.36$\times$10$^{3}$ &1.27$\times$10$^{4}$  & 2.11$\times$10$^{4}$ &9.52$\times$10$^{3}$  & 1.61$\times$10$^{3}$ &1.11$\times$10$^{4}$  &1.35$\times$10$^{5}$\\
		\cline{1-10}
		\multirow{3}{*}{Heavy nucleus}& pAu  &1.67$\times$10$^{-4}$   & 1.50$\times$10$^{4}$   &2.28$\times$10$^{4}$ &3.78$\times$10$^{4}$ &2.56$\times$10$^{4}$  & 3.88$\times$10$^{3}$ & 2.95$\times$10$^{4}$&2.38$\times$10$^{5}$ \\
		\multirow{3}{*}{}&pPb  &1.58$\times$10$^{-4}$  & 1.58$\times$10$^{4}$&2.40$\times$10$^{4}$ &3.98$\times$10$^{4}$ &2.80$\times$10$^{4}$ &4.20$\times$10$^{3}$ &3.22$\times$10$^{4}$ &2.50$\times$10$^{5}$   \\
		& pU  &1.38$\times$10$^{-4}$   &  1.81$\times$10$^{4}$&2.74$\times$10$^{4}$ &4.55$\times$10$^{4}$ & 3.49$\times$10$^{4}$& 5.10$\times$10$^{3}$ & 4.00$\times$10$^{4}$ &2.84$\times$10$^{5}$  \\
		\hline	
		\end{tabular}
	\end{center}\caption{\label{tab:6}
		The SD and ND cross sections in $\mu$b for the $\eta_{c}$ proton-nucleus productions at LHC with forward detector acceptance, $0.0015<\xi_{1}<0.5$.}
\end{table*}

\begin{figure}[htp]
\centering
\includegraphics[height=4.8cm,width=5.9cm]{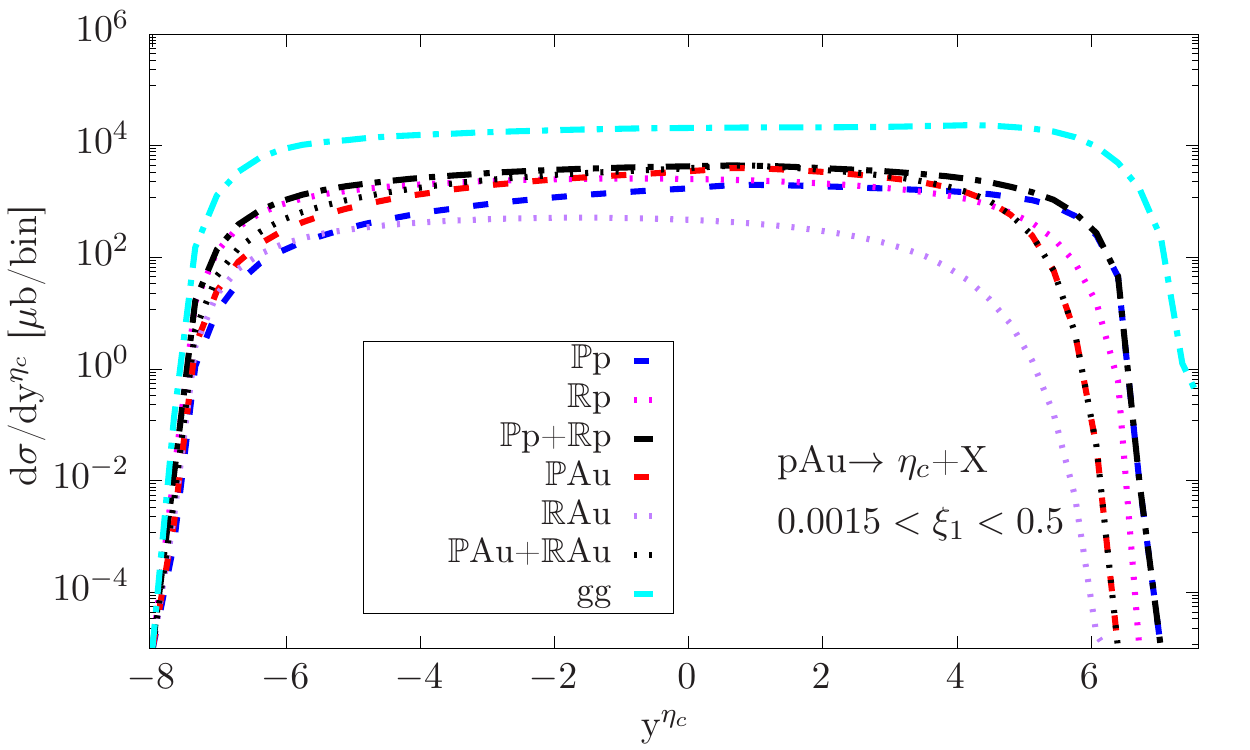}
\includegraphics[height=4.8cm,width=5.9cm]{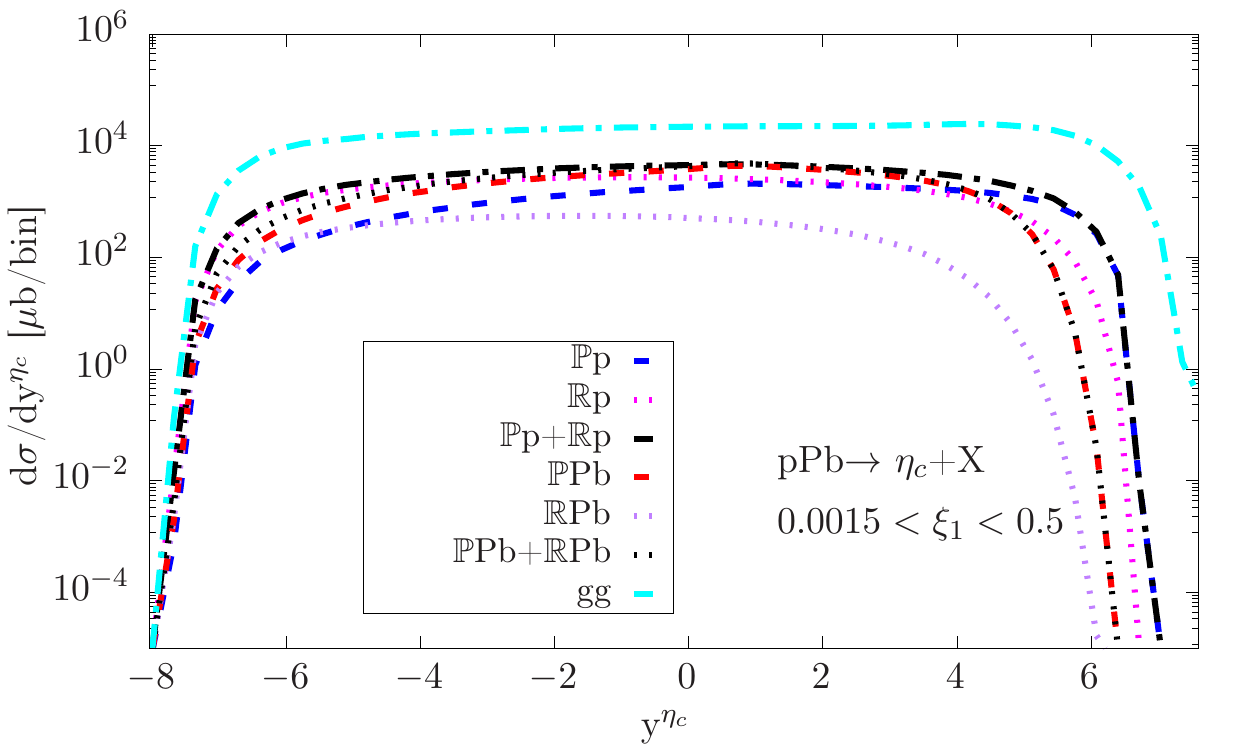}
\includegraphics[height=4.8cm,width=5.9cm]{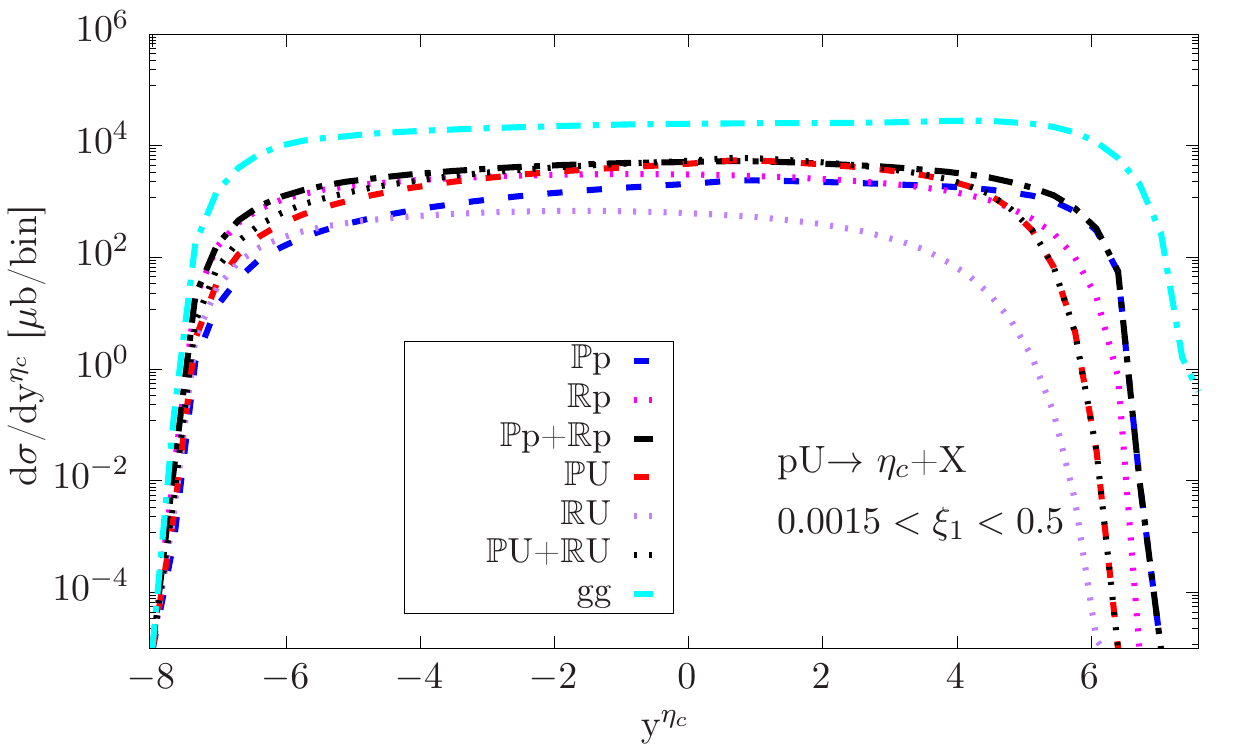}\\
\caption{\normalsize (color online) The $\rm y^{\eta_{c}}$ distributions for the $\mathbb{P}$p (blue dashed line), $\mathbb{R}$p (magenta dotted line), $\mathbb{P}$p+$\mathbb{R}$p (black  dash dotted line), $\mathbb{P}$A (red  dashed  line),
$\mathbb{R}$A (purple dotted line), $\mathbb{P}$A+$\mathbb{R}$A (black dotted line)
) and $\rm gg$ (cyan dash dotted line) in SD processes for pA collisions.}
\label{fig13:limits}
\end{figure}
The $\eta_{c}$ rapidity distributions in SD processes for two circumstances in pA mode are plotted in Fig.\ref{fig13:limits} for heavy nuclei. In the first circumstance, the Reggeon distribution dominant over the Pomeron distribution. In the second circumstance, the  Pomeron distribution dominant over  the Reggeon distribution. The diffractive and non diffractive distribution are asymmetric. The rapidity distributions for the single diffractive dissociation and non diffractive process have maximums shifted to forward or backward  rapidities with respect to the mid rapidity. The  rapidity distribution for the first circumstance dominates over the  rapidity distribution for the second circumstance for the light nuclei and Copper. They are of a same order of magnitude for the heavy nuclei, Calcium and Silver.

\begin{figure}[htp]
\centering
\includegraphics[height=4.8cm,width=5.9cm]{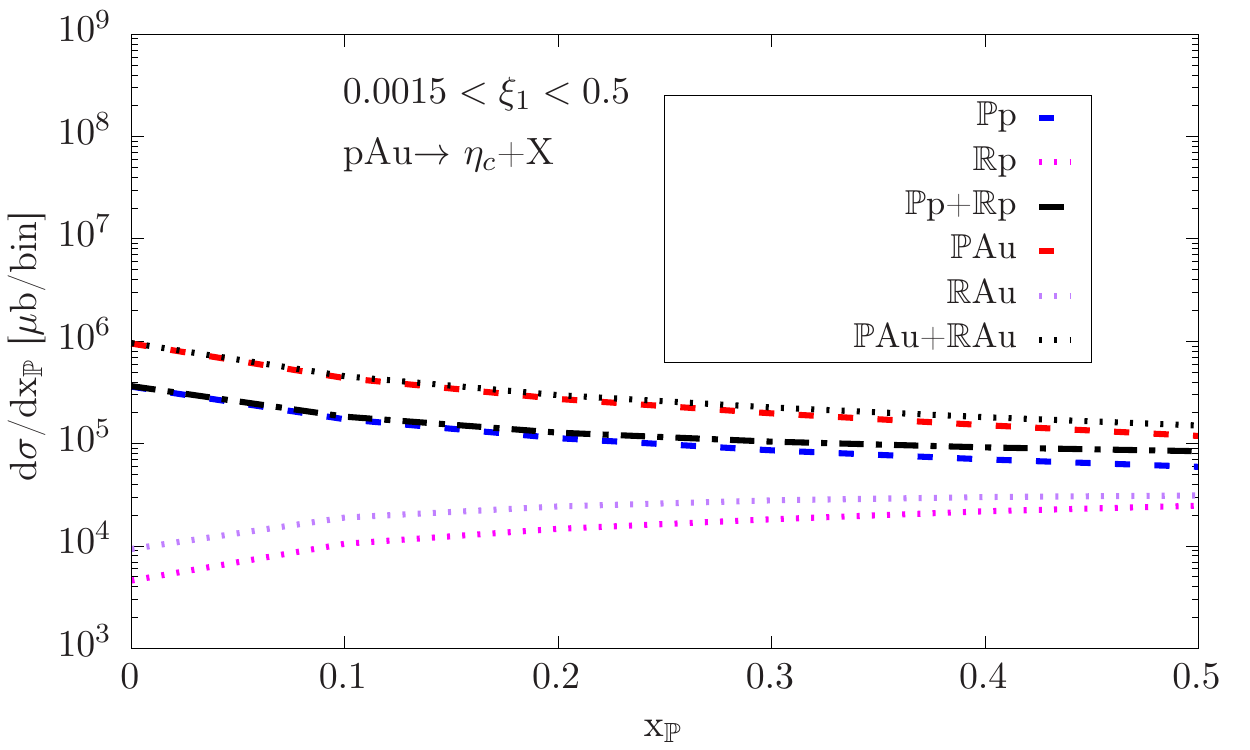}
\includegraphics[height=4.8cm,width=5.9cm]{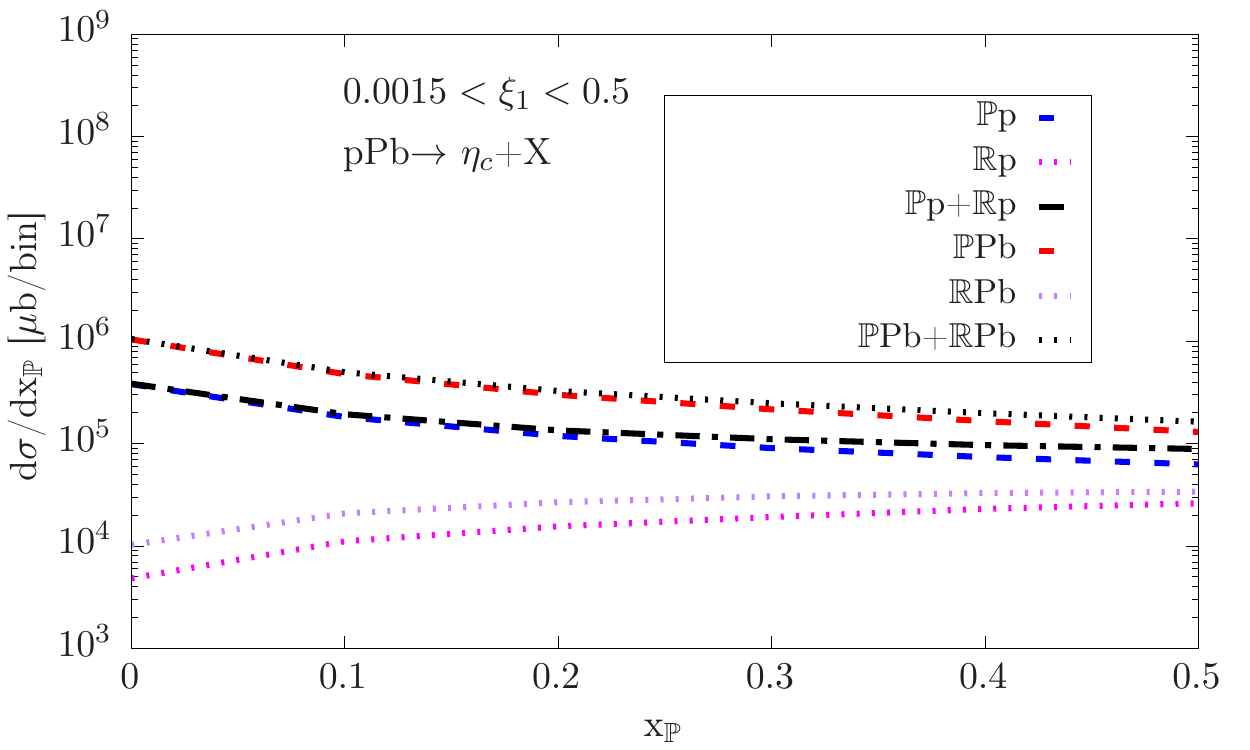}
\includegraphics[height=4.8cm,width=5.9cm]{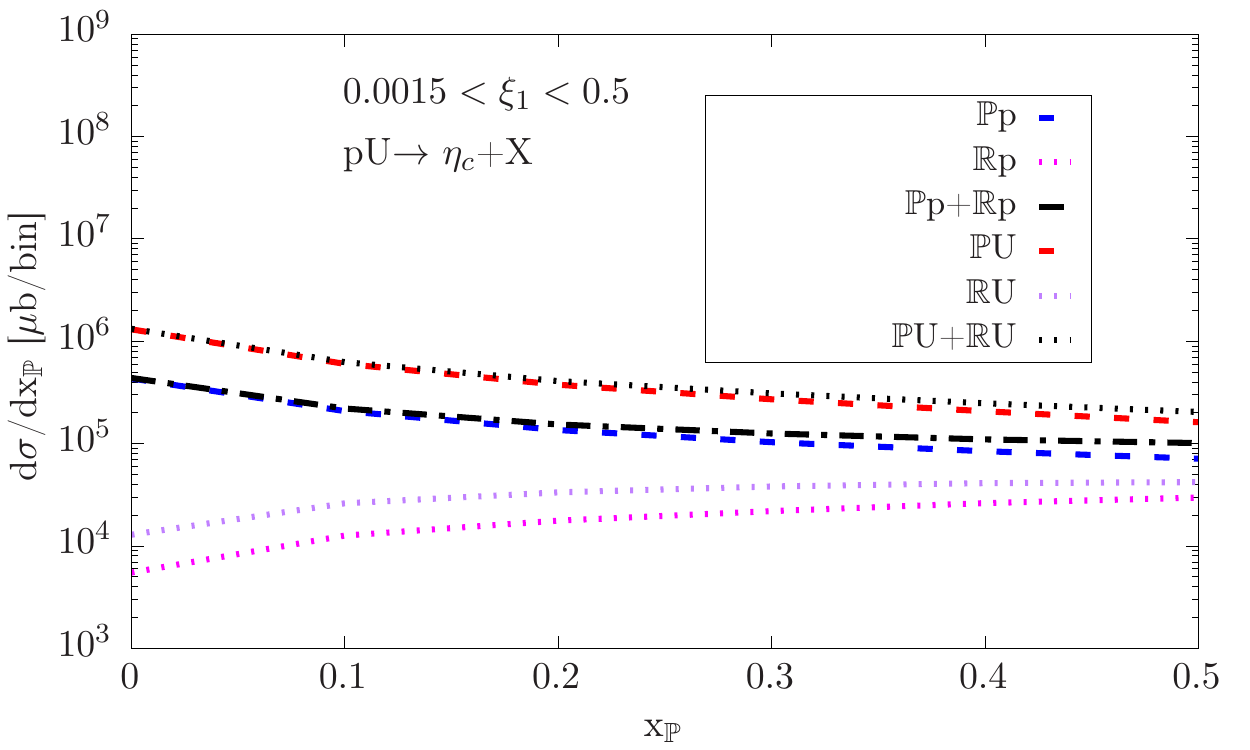}\\
\caption{ \normalsize (color online)
The $x_{\mathbb{P}}$ distributions for the $\mathbb{P}$p (blue dashed line), $\mathbb{R}$p (magenta dotted line), $\mathbb{P}$p+$\mathbb{R}$p (black  dash dotted line), $\mathbb{P}$A (red dashed line), $\mathbb{R}$A (purple dotted line) and $\mathbb{P}$A+$\mathbb{R}$A (black dotted line) in DD processes for pA collisions.}
\label{fig14:limits}
\end{figure}
The proton or nucleus longitudinal momentum fraction distributions of $\eta_{c}$ for two situations in $\rm pA$ mode via SD processes are given in Fig.\ref{fig14:limits} for heavy nuclei. The lower $x_{\mathbb{P}}$ of the Pomeron distribution for the first situation and the lower $x_{\mathbb{P}}$ of the Reggeon distribution for the second situation increase for the small $x_{\mathbb{P}}$ and become flat for the large $x_{\mathbb{P}}$.  The larger $x_{\mathbb{P}}$ of the Reggeon  distribution for the first situation and the larger $x_{\mathbb{P}}$ of the Pomeron  distribution for the second situation
decrease for the small $x_{\mathbb{P}}$ and become flat for the large $x_{\mathbb{P}}$

\begin{figure}[htp]
\centering
\includegraphics[height=4.8cm,width=5.9cm]{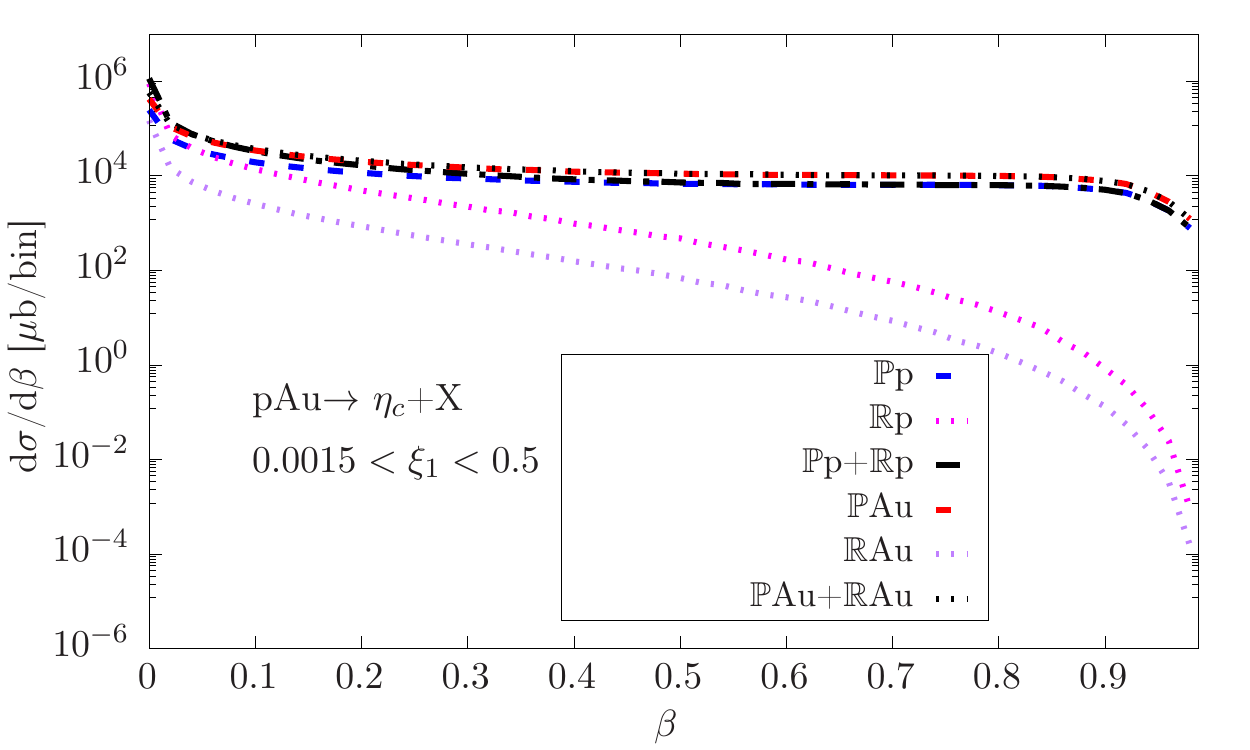}
\includegraphics[height=4.8cm,width=5.9cm]{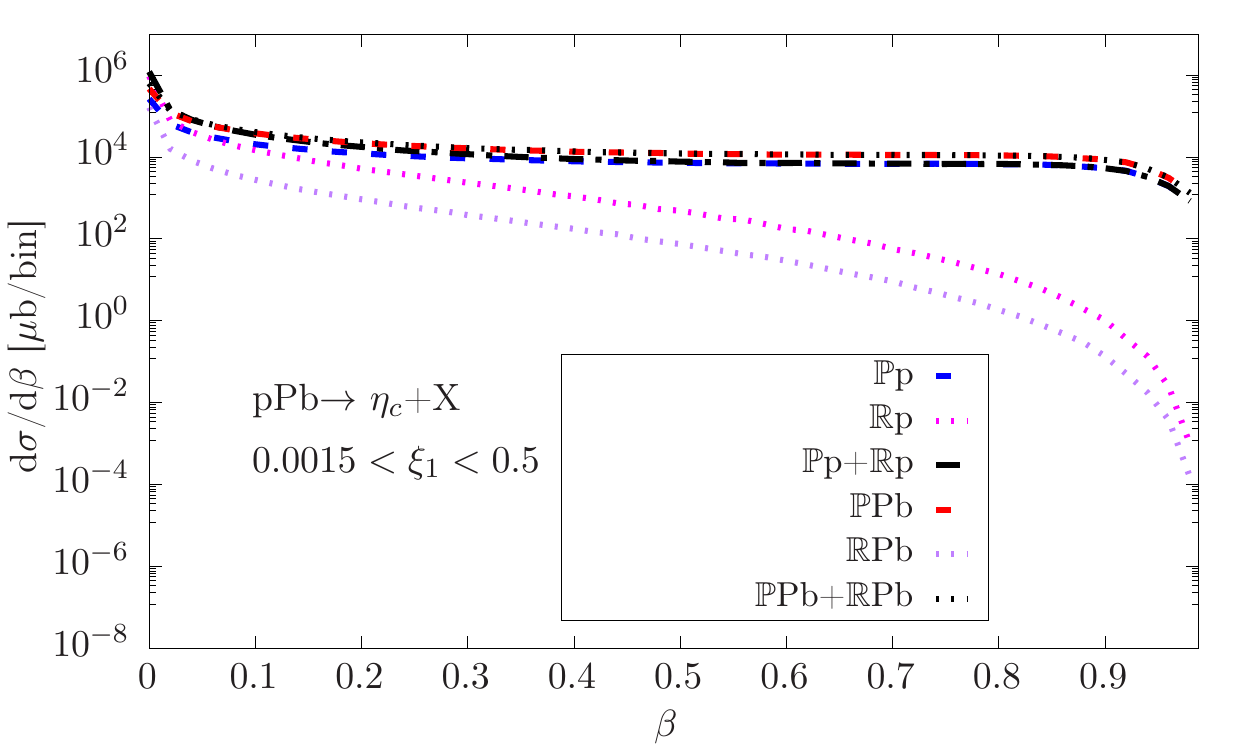}
\includegraphics[height=4.8cm,width=5.9cm]{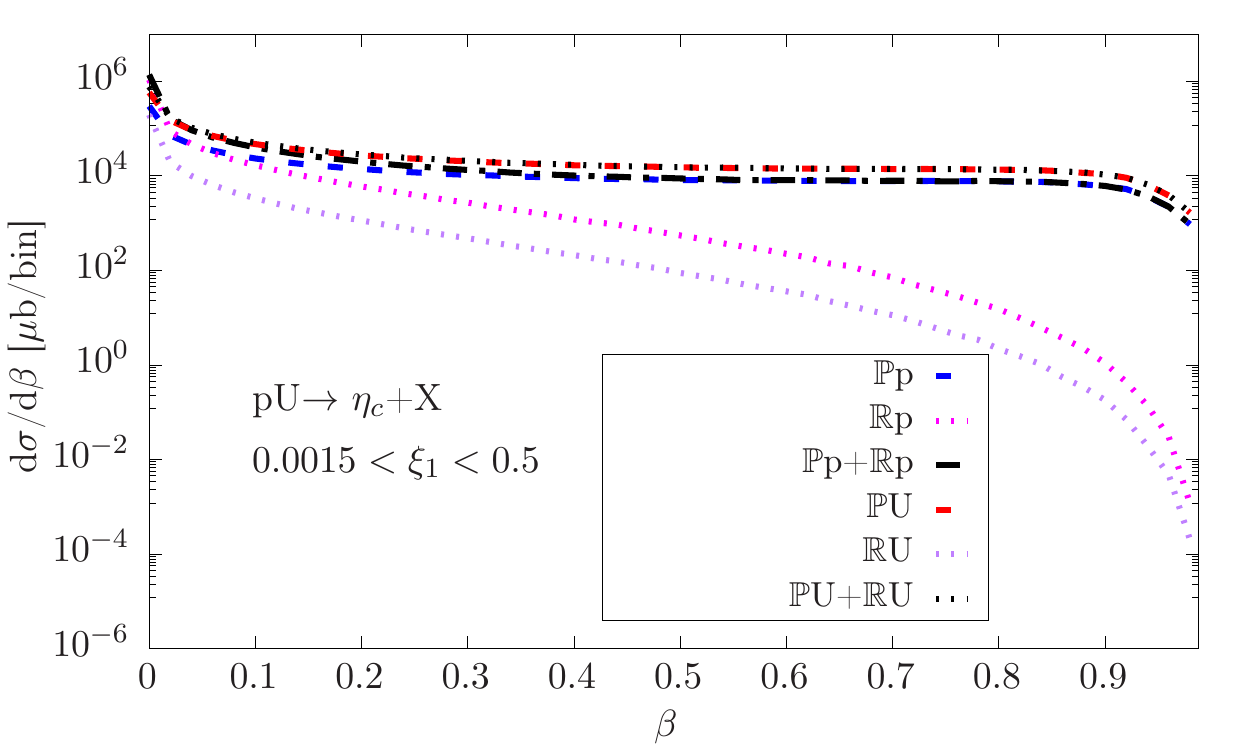}\\
\caption{ \normalsize (color online)
The $\beta$ distributions for the $\mathbb{P}$p (blue dashed line), $\mathbb{R}$p (magenta dotted line), $\mathbb{P}$p+$\mathbb{R}$p (black dash dotted line), $\mathbb{P}$A (red dashed line), $\mathbb{R}$A (purple dotted line) and $\mathbb{P}$A+$\mathbb{R}$A (black dotted line) in SD processes for pA collisions.}
\label{fig15:limits}
\end{figure}
The gluon longitudinal momentum fraction with respect to the exchanged Pomeron or Reggeon distributions of $\eta_{c}$ for two circumstances in pA mode through SD processes are plotted in Fig.\ref{fig15:limits} for heavy nuclei. The lower $\beta$ of the Pomeron distribution for the first situation and the lower $\beta$ of the Reggeon distribution for the second situation decrease for the small and large $\beta$. 
The larger $\beta$ of the Reggeon  distribution for the first situation and the larger $\beta$ of the Pomeron  distribution for the second situation decrease for the small $\beta$ and become flat for the large $\beta$. The $\beta$ distribution for the two situation presents the convexity and concavity and their end points.

\subsection{Diffractive photon-nucleus production}

\subsubsection{pp collision}
The $\rm \eta_{c}$ cross section prediction in $\rm pp$ mode for  photon-Pomeron and -Reggeon induced interactions is arranged in Table.\ref{tab:7}. The exclusive photon-photon interaction is not included in this section due to the fact that we have only focused on the inclusive production of $\rm \eta_{c}$ meson
 which serves as its important background. The incoming particles are similar where the photon is emitted by one of the incoming particle $\rm p$ and the hard diffractive gluon through Pomeron or Reggeon is radiated by the other incoming particle $\rm p$. The both emitting protons remain intact or early intact after the collision. The hard process of photon and hard diffractive gluon produces $\rm \eta_{c}$ charmonium state accompanied with remnants. The intact or early intact proton, remnants and $\rm \eta_{c}$ are observed in the final states.  As we can see, the Pomeron contribution to the total cross section  is a factor $10$ larger than the Reggeon contribution. The total gluon-Pomeron and-Reggeon cross section in SD (DD) process in $\rm pp$ mode is dominant of a factor $10^{6}$($10^{5}$) over the total photon-Pomeron and -Reggeon cross section in the same mode. The DD cross section for gluon-Pomeron and-Reggeon process in $\rm AA$ ($\rm pA$) mode is a factor $\gtrsim$ $10^{5}$ ($10^{6}$) larger than that of $\rm pp$ mode in photon-Pomeron and -Reggeon induced interactions. The SD cross section for gluon-Pomeron and-Reggeon process in $\rm AA$  mode is a factor $\gtrsim$ $10^{7}$  larger than that of $\rm pp$ mode in photon-Pomeron and -Reggeon induced interactions. For the both situations, the SD cross section for gluon-Pomeron and-Reggeon process in $\rm pA$ mode is a factor $\gtrsim$ $10^{7}$ ($10^{6}$)larger than that of $\rm pp$ mode in photon-Pomeron and -Reggeon induced interactions.
\begin{table*}[htbp]
	\begin{center}
		\begin{tabular}{|c |c | c | c |c |}\hline
			Hadron       & $\rm \langle \left\vert S\right\vert ^{2}\rangle $             &$\mathbb{P}$p &$\mathbb{R}$p &Total             \\ \hline
			pp    & 1 &   3.92$\times$10$^{-5}$&1.36$\times$10$^{-4}$  & 1.75$\times$10$^{-4}$             \\ \hline	
		\end{tabular}
	\end{center}\caption{\label{tab:7}
		The photon-Pomeron and -Reggeon cross sections in $\mu$b for the $\eta_{c}$ production at LHC with forward detector acceptance, $0.0015<\xi <0.5$, in pp collisions. }
\end{table*}

\begin{figure}[htp]
	\centering
	\includegraphics[height=4.8cm,width=5.9cm]{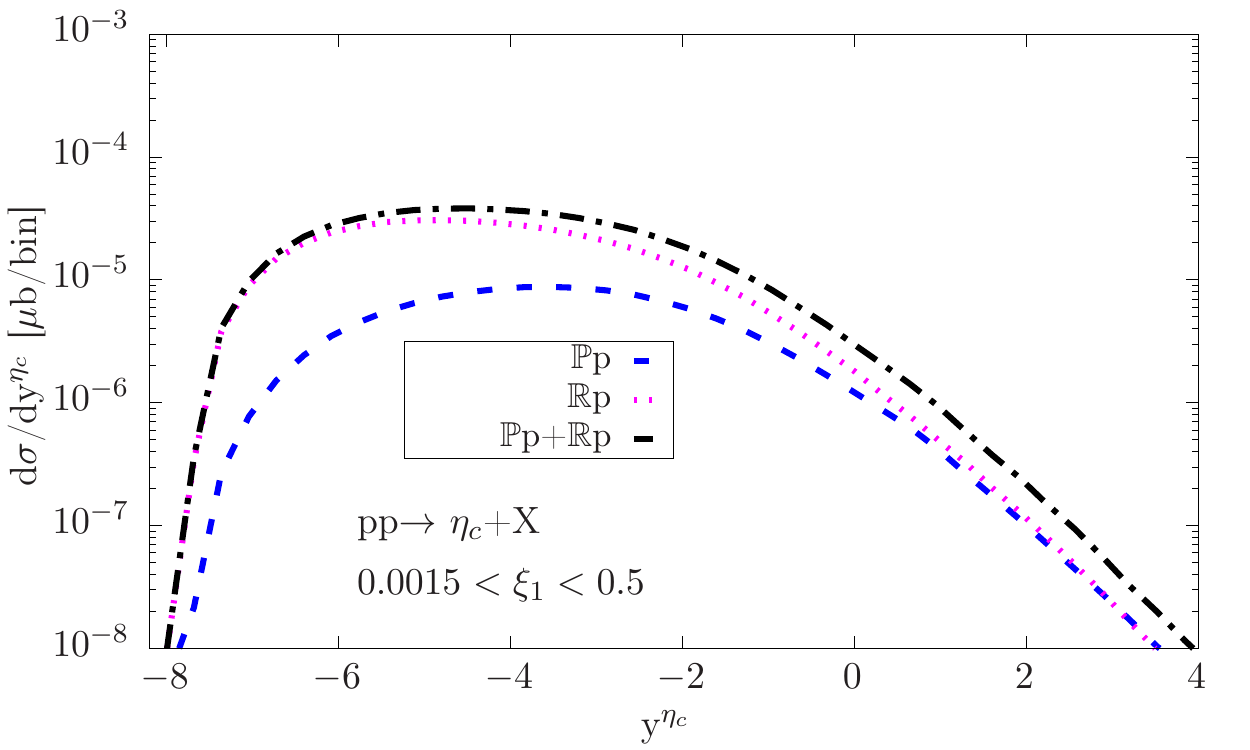}
	\includegraphics[height=4.8cm,width=5.9cm]{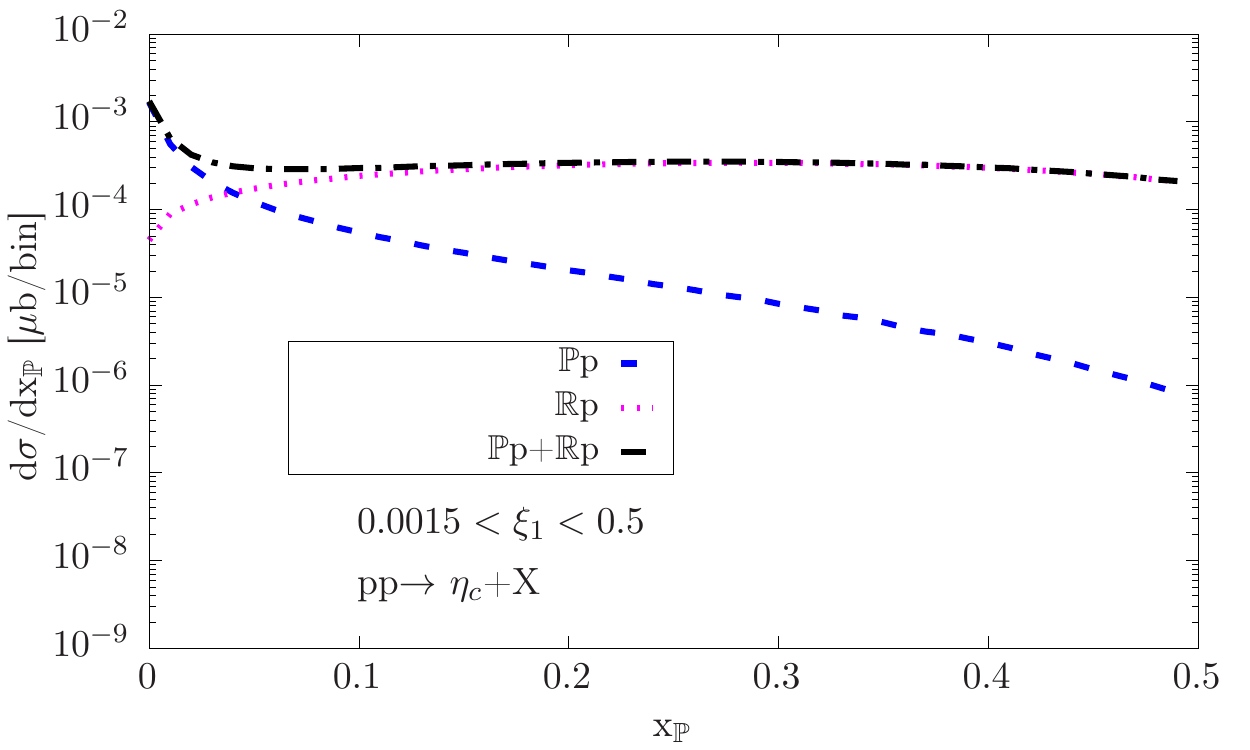}
	\includegraphics[height=4.8cm,width=5.9cm]{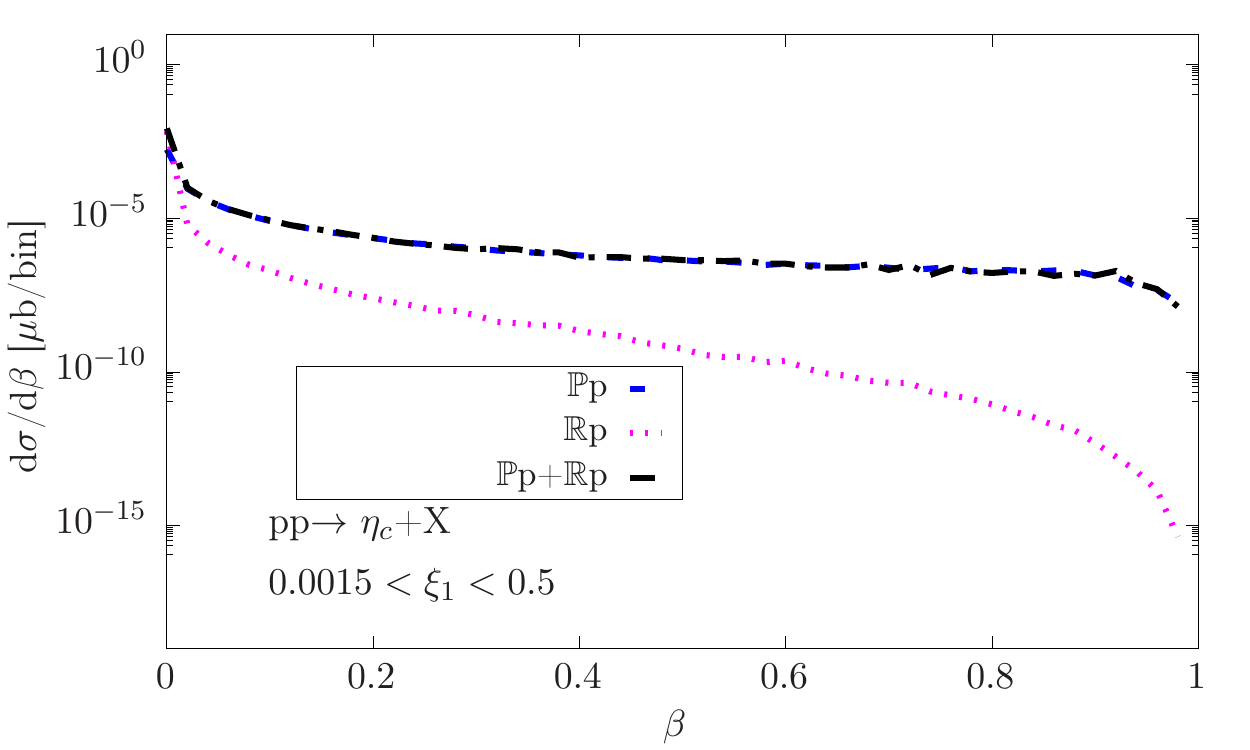}\\
	\caption{\normalsize (color online) The $\rm y^{\eta_{c}}$, $x_{\mathbb{P}}$ and $\beta$ distributions for the $\mathbb{P}$p (blue dashed line), $\mathbb{R}$p (magenta dotted line)and $\mathbb{P}$p+$\mathbb{R}$p (black dash dotted line) in  photon-Pomeron and -Reggeon induced processes for pp.}
	\label{fig16:limits}
\end{figure}

The Fig.\ref{fig16:limits} exhibits the rapidity, gluon longitudinal momentum fraction and gluon longitudinal momentum fraction with respect to the exchanged Pomeron and Reggeon  for $\rm pp$ mode in photon-Pomeron and -Reggeon induced interactions. On the left panel of the figure, the rapidity distributions  are asymmetric due to unequal backward and forward contributions which is expected since the photon and Pomeron or Reggeon fluxes from the both protons  are different. The Reggeon distribution is significant over Pomeron distribution. The rapidity distributions have maximums displaced to backward rapidities with respect to the mid rapidity. At the middle panel, proton longitudinal momentum fraction distribution reveals that  Reggeon contribution increases for small and becomes flat for large $x_{\mathbb{P}}$ while  Pomeron contribution decreases for small and large $x_{\mathbb{P}}$. On the right panel, the gluon longitudinal momentum fraction with respect to the exchanged Pomeron or Reggeon distribution shows that the Reggeon contribution is lessening for small and large $\beta$ whereas the Pomeron one is lowering for small $\beta$, becoming flat and lowering again for large $\beta$. The convexity and concavity are  visible at their end points. The Reggeon contribution is still important in photon-Pomeron and -Reggeon  induced interactions and should not be neglected.
\subsubsection{AA collision}
Table.\ref{tab:8}  provides the $\rm \eta_{c}$ cross section estimation in $\rm AA$ mode for the light, medium and heavy nuclei in photon-Pomeron and -Reggeon induced processes . The  cross section is evaluated where the both incoming nucleus are similar. The first incoming nucleus radiates the photon which collides with the hard diffractive gluon coming from Pomeron or Reggeon of the second incoming nucleus. Notice that the total cross section increases from light to heavy nuclei with the increase of $\rm Z$
and $\rm A$. On the one hand, the Pomeron and Reggeon contribution to the total cross section are a same factor for certain nucleus for example the light nucleus, Calcium, Copper and Gold. One the other hand, the Pomeron  contribution is a factor $10$ (Uranium and Lead) or $10^{2}$ (Silver) greater than the Reggeon  contribution. The total gluon-Pomeron and-Reggeon cross section in SD (DD) process in $\rm pp$ mode is  a factor $\gtrsim$ 10 greater than the total photon-Pomeron and -Reggeon cross section in $\rm AA$ (for light and medium nuclei). The total gluon-Pomeron and-Reggeon cross section in DD process in $\rm pp$ mode is  a same factor than the total photon-Pomeron and -Reggeon cross section in $\rm AA$ for heavy nuclei. The DD cross section for gluon-Pomeron and-Reggeon process in $\rm AA$ ($\rm pA$) mode is a factor $\gtrsim$ $10^{5}$ ($10^{4}$)  larger than that of $\rm AA$ mode in photon-Pomeron and -Reggeon induced interactions.The SD cross section for gluon-Pomeron and-Reggeon process in $\rm AA$  mode is a factor $\gtrsim$ $10^{5}$  larger than that of $\rm AA$  mode in photon-Pomeron and -Reggeon induced interactions. For the both scenarios, the SD cross section for gluon-Pomeron and-Reggeon process in $\rm pA$ mode is a factor $\gtrsim$ $10^{3}$ larger than that of $\rm AA$ mode in photon-Pomeron and -Reggeon induced interactions. The total  $\rm AA$ cross section is a factor $\gtrsim$ 10  larger than that of $\rm pp$ cross section in  photon-Pomeron and -Reggeon induced processes due  to 
 the diffractive gluon distribution from A and the nuclear form factor, i.e., proportional to $\rm A^{2}R_{g}F^{2}_{A}(t)$ as well as  the photon distribution from A, i.e., proportional to $\rm Z^{2}$. The exception is made for the first light nucleus where $\rm pp$ and $\rm AA$ modes are  a same factor in  photon-Pomeron and -Reggeon induced process. 
	\begin{table*}[htbp]
		\begin{center}
			\begin{tabular}{|c |c | c | c |c |c |}\hline	
				Nuclei   & AA      & $\rm \langle \left\vert S\right\vert ^{2}\rangle $             & $\mathbb{P}$A              & $\mathbb{R}$A & Total             \\ \hline
				\multirow{3}{*}{Light nucleus}& BeBe &  1 &2.84$\times$10$^{-4}$    &
				1.34$\times$10$^{-4}$& 4.18$\times$10$^{-4}$   \\
				\multirow{3}{*}{}&CC  & 1 & 9.87$\times$10$^{-4}$   &4.34$\times$10$^{-4}$ & 1.42$\times$10$^{-3}$ \\
				& OO   & 1   &2.70$\times$10$^{-3}$ &1.11$\times$10$^{-3}$ &3.81$\times$10$^{-3}$  \\
				\cline{1-6}
				\multirow{3}{*}{Medium nucleus}& CaCa   &1    &6.52$\times$10$^{-2}$&2.16$\times$10$^{-2}$ & 8.68$\times$10$^{-2}$  \\
				\multirow{3}{*}{}&CuCu  &1  &27.13$\times$10$^{-2}$&8.06$\times$10$^{-2}$ &35.19$\times$10$^{-2}$  \\
				&AgAg  &  1  & 1.51$\times$10$^{0}$& 39.74$\times$10$^{-2}$ &1.91$\times$10$^{0}$ \\
				\cline{1-6}
				\multirow{3}{*}{Heavy nucleus}& AuAu& 1  
				&9.96$\times$10$^{0}$&2.29$\times$10$^{0}$ & 1.22$\times$10$^{1}$ \\
				\multirow{3}{*}{}&PbPb  &1 &1.16$\times$10$^{1}$  &2.62$\times$10$^{0}$ &1.42$\times$10$^{1}$  \\
				& UU   &1 
				&1.76$\times$10$^{1}$   &3.87$\times$10$^{0}$ & 2.15$\times$10$^{1}$ \\
				\hline
			\end{tabular}
		\end{center}\caption{\label{tab:8}
			The photon-Pomeron and -Reggeon induced cross sections in $\mu$b for the $\eta_{c}$ nucleus-nucleus productions at LHC with forward detector acceptance, $0.0015<\xi<0.5$.}
	\end{table*}

\begin{figure}[htp]
		\centering
		\includegraphics[height=4.8cm,width=5.9cm]{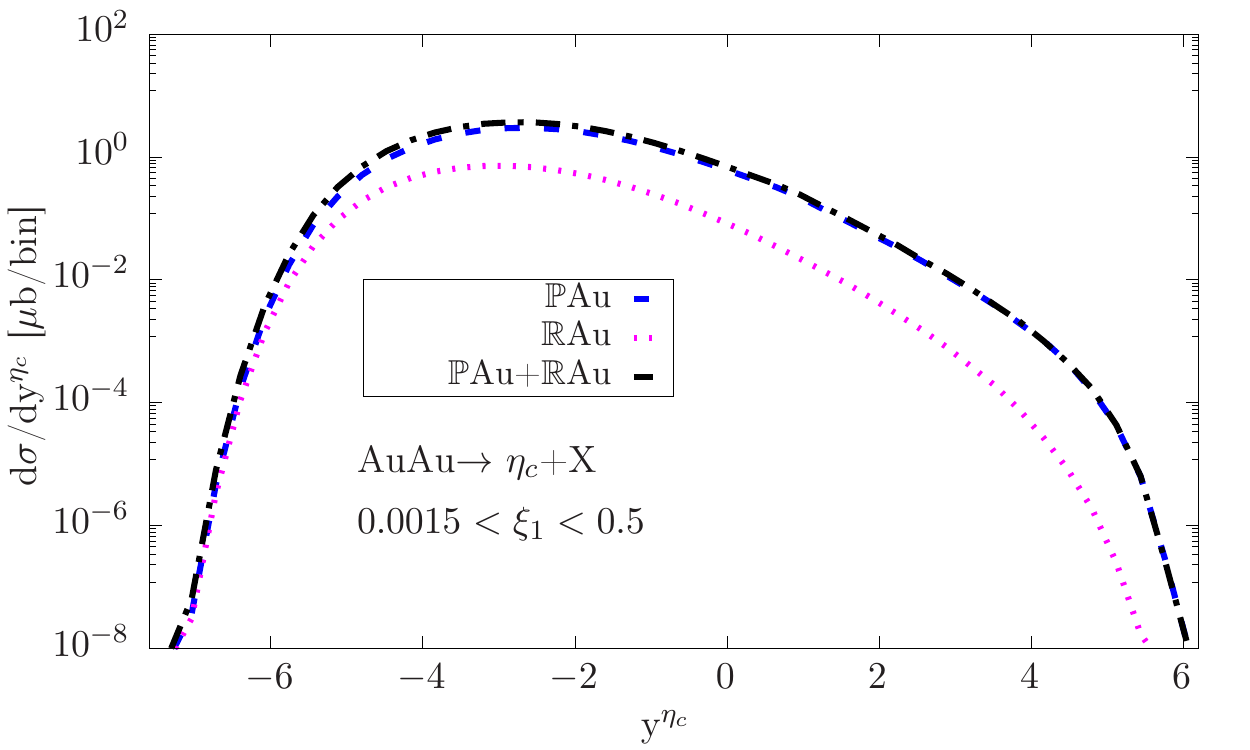}
		\includegraphics[height=4.8cm,width=5.9cm]{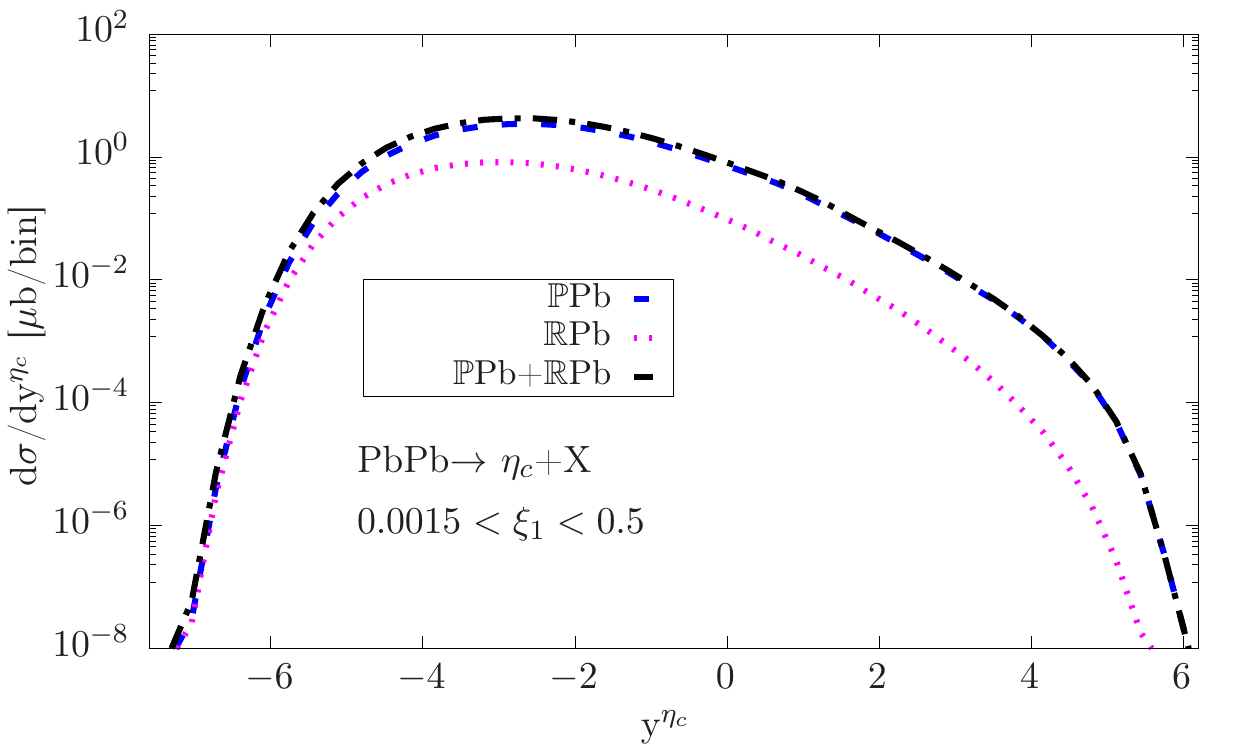}
		\includegraphics[height=4.8cm,width=5.9cm]{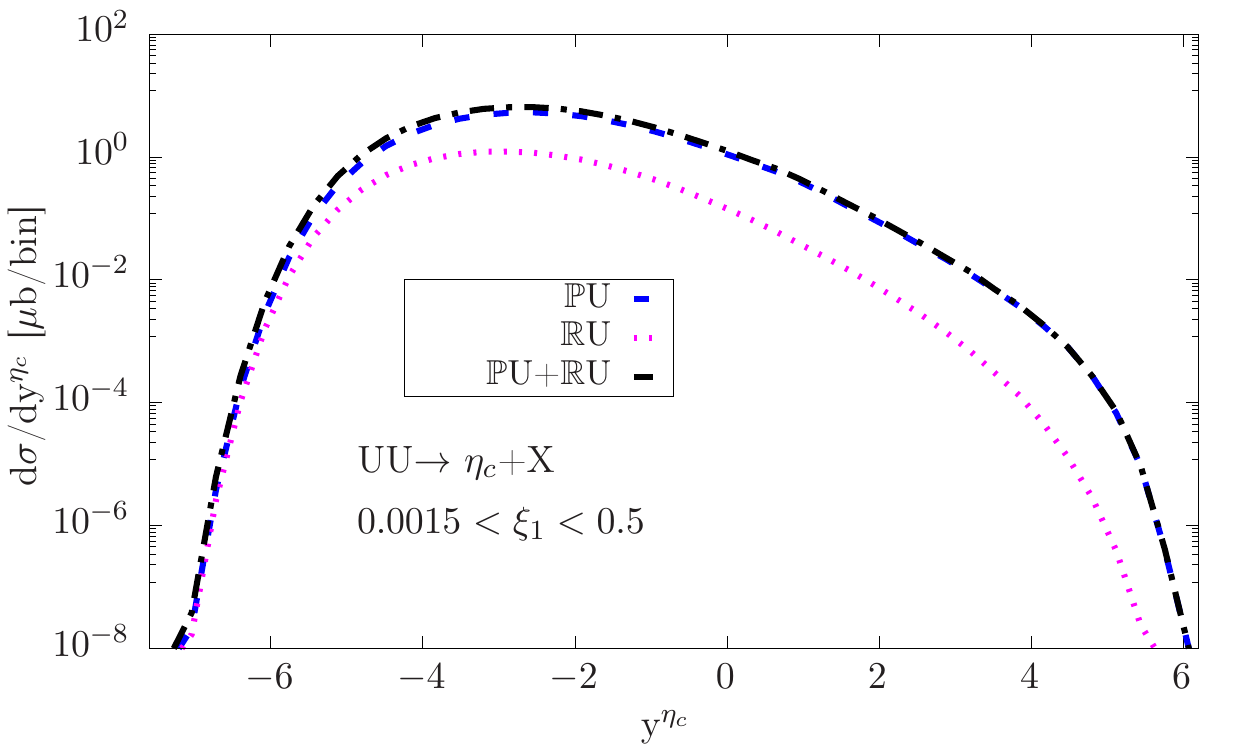}
		\caption{\normalsize (color online) The $\rm y^{\eta_{c}}$   distributions for the the $\mathbb{P}$A (blue dashed line), $\mathbb{R}$A (magenta dotted line), $\mathbb{P}$A+$\mathbb{R}$A (black dash dotted line) and $\rm gg$ (cyan dash dotted line)  in  photon-Pomeron and-Reggeon induced processes  for  AA collisions.}
		\label{fig17:limits}
	\end{figure}

In Fig.\ref{fig17:limits} we show the rapidity distributions    for the $\rm AA$ mode in photon-Pomeron and- Reggeon induced interactions for the heavy nuclei. The rapidity distributions  are asymmetric due to unequal backward and forward contributions elucidated by the photon and pomeron or Reggeon flux differences from both nuclei. The rapidity distributions present maximums moved to backward rapidities with respect to the mid rapidity. The Pomeron contribution to the total cross section keeps the dominant part.	
	
\begin{figure}[htp]
		\centering
		\includegraphics[height=4.8cm,width=5.9cm]{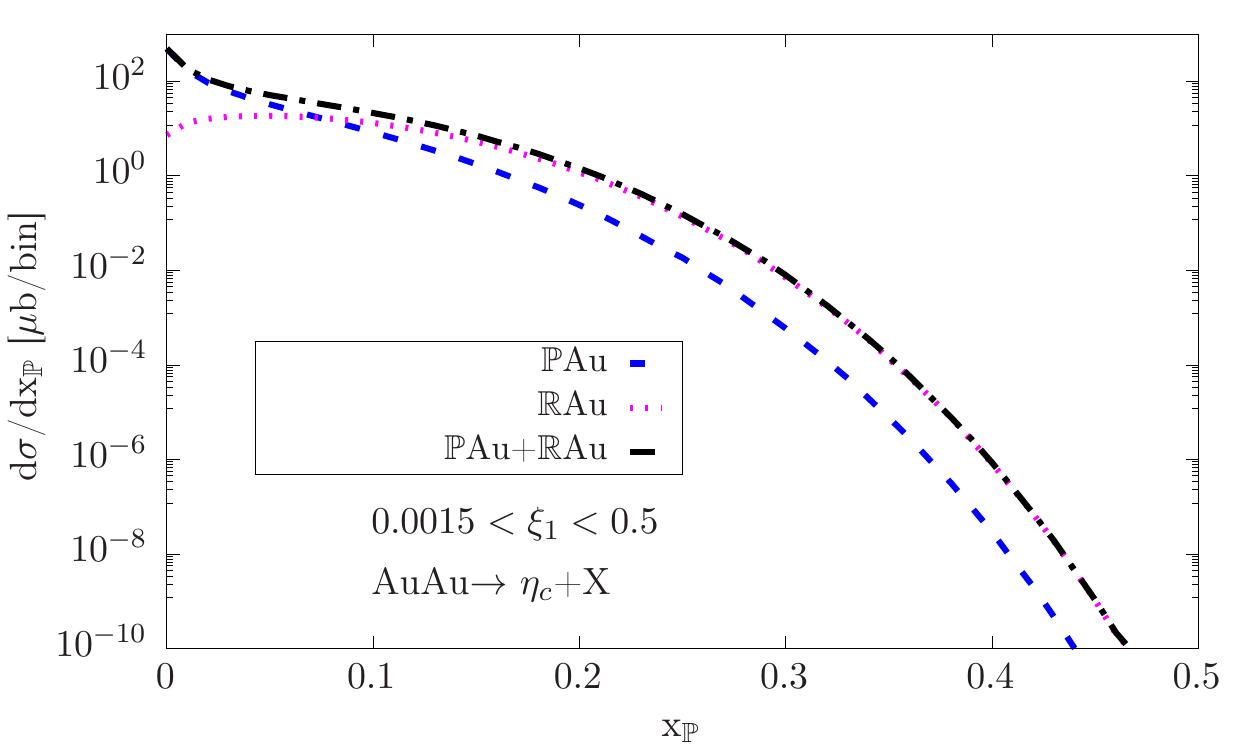}
		\includegraphics[height=4.8cm,width=5.9cm]{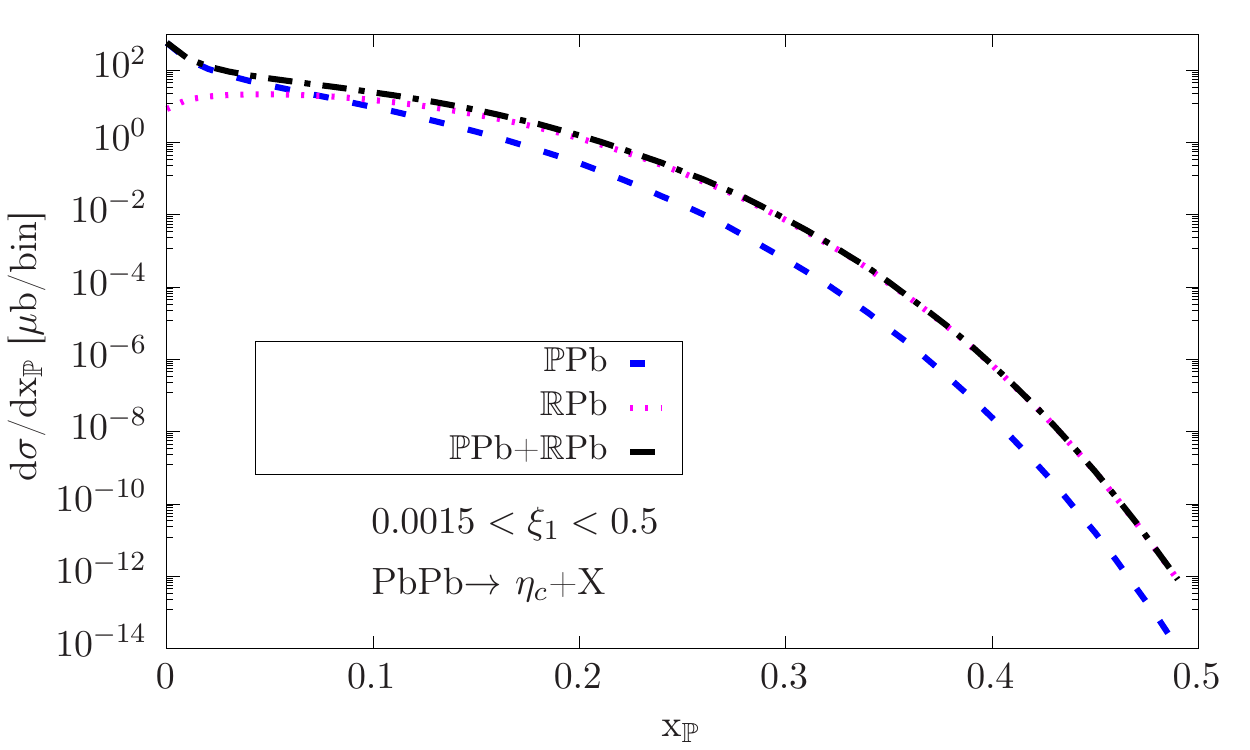}
		\includegraphics[height=4.8cm,width=5.9cm]{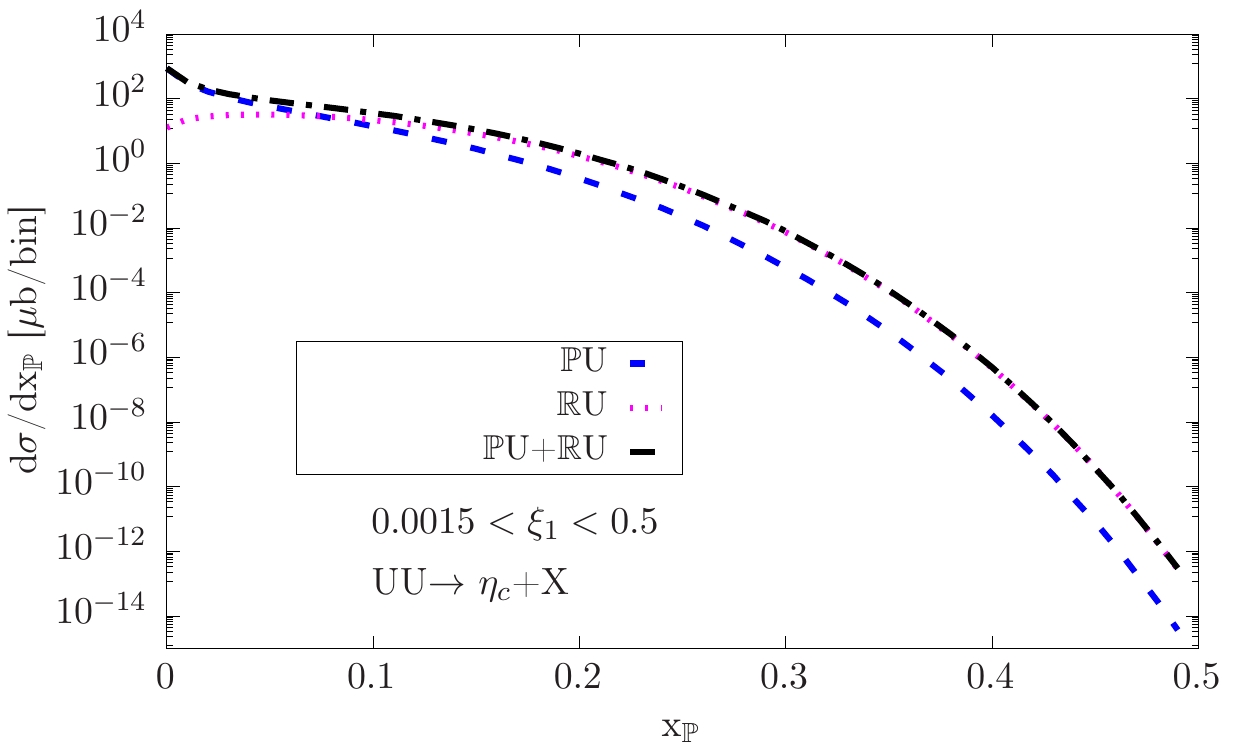}
		\caption{ \normalsize (color online)
			The $x_{\mathbb{P}}$ distributions for the $\mathbb{P}$A (blue dashed line), $\mathbb{R}$A (magenta dotted line) and $\mathbb{P+R}$A (black dash dotted line) in  photon-Pomeron and-Reggeon induced processes  for  AA collisions.}
		\label{fig18:limits}
	\end{figure}

The Fig.\ref{fig18:limits} shows $x_{\mathbb{P}}$ distributions of the nucleus-nucleus collision for heavy nuclei. It has been pointed out that the $x_{\mathbb{P}}$ distribution of Pomeron  falls for small and large $x_{\mathbb{P}}$  while $x_{\mathbb{P}}$ distribution of Reggeon  lightly increases for small $x_{\mathbb{P}}$ and considerably decreases  for large $x_{\mathbb{P}}$. The Pomeron distribution is important for  small $x_{\mathbb{P}}$ whereas the Reggeon contribution is considerable for large $x_{\mathbb{P}}$.	

\begin{figure}[htp]
		\centering
		\includegraphics[height=4.8cm,width=5.9cm]{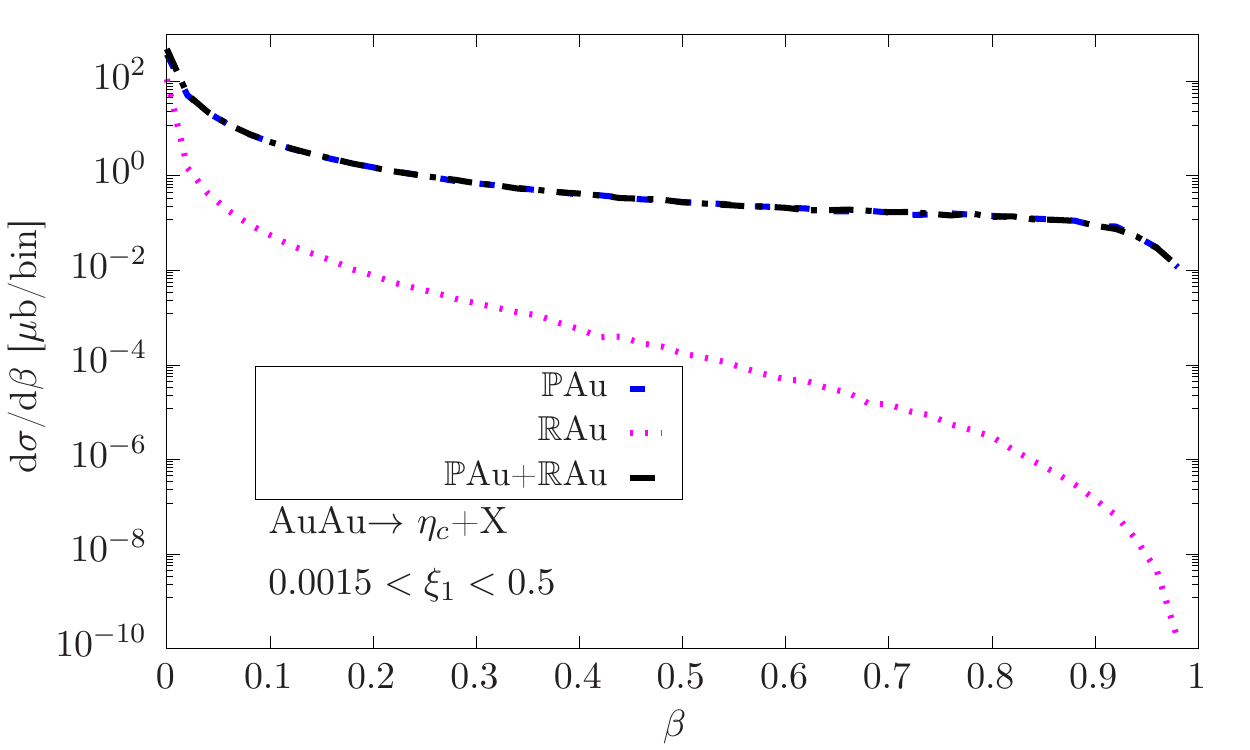}
		\includegraphics[height=4.8cm,width=5.9cm]{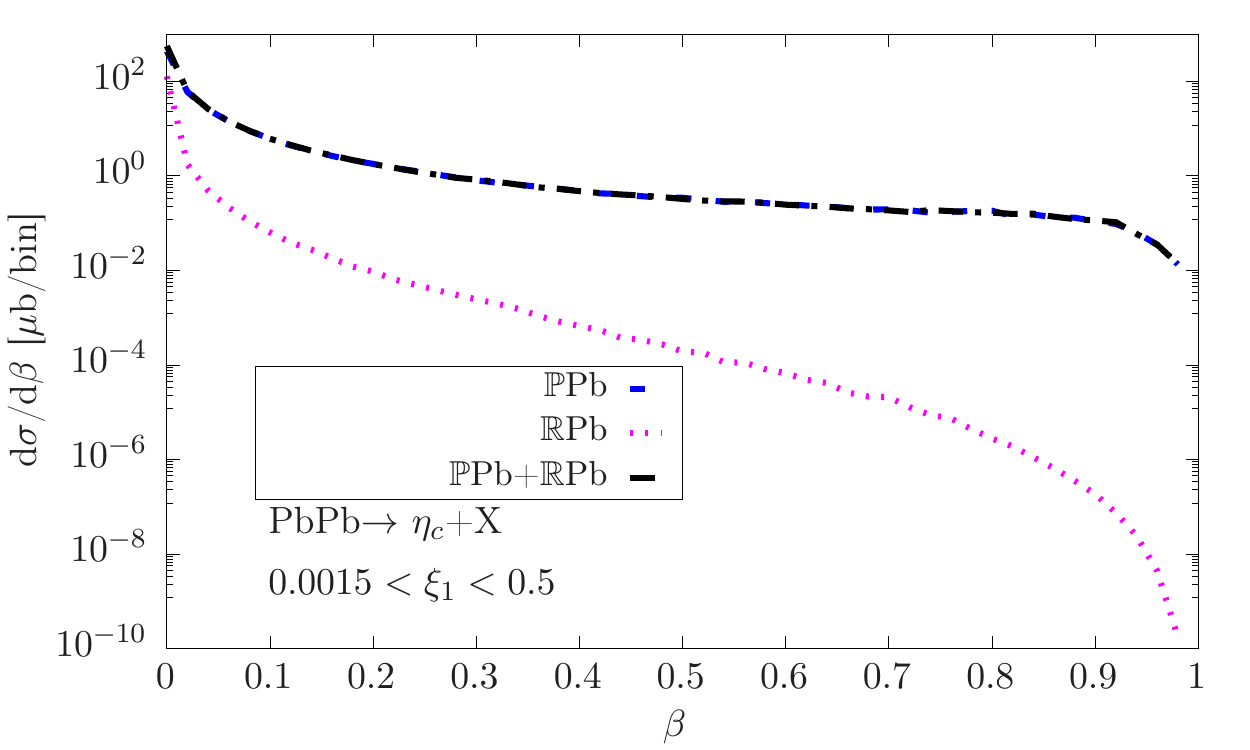}
		\includegraphics[height=4.8cm,width=5.9cm]{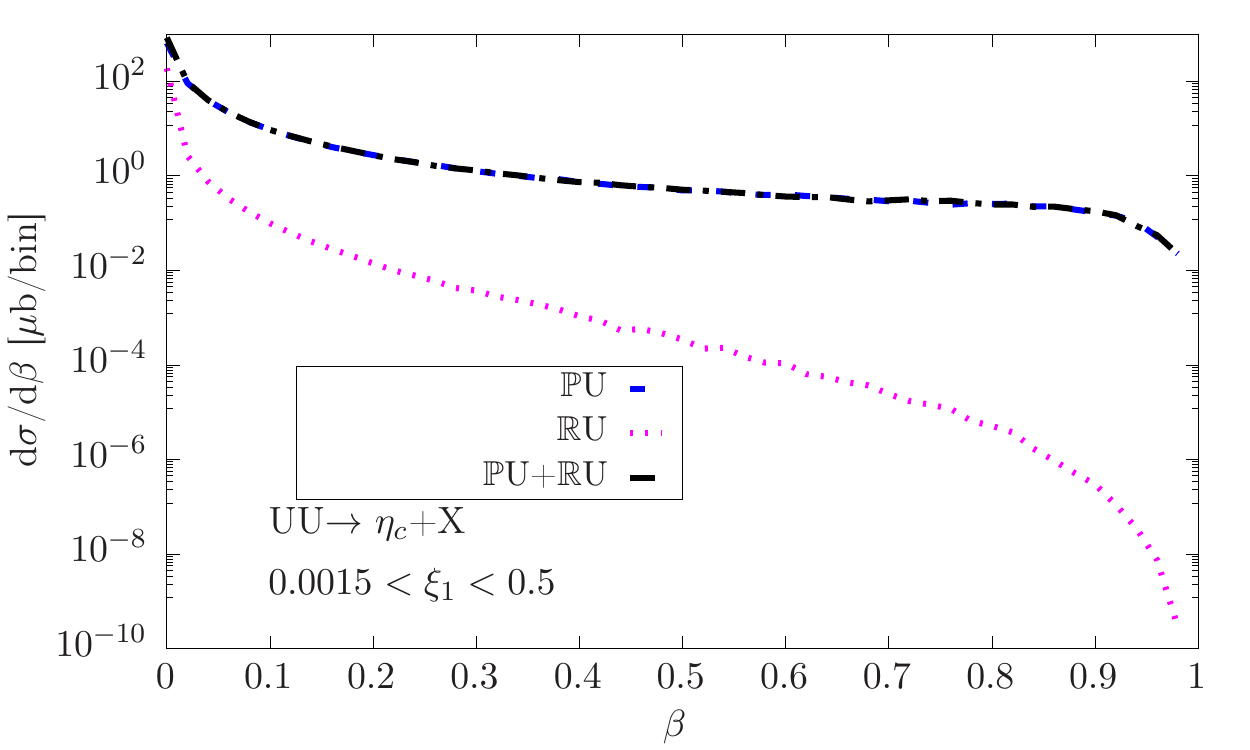}
		\caption{ \normalsize (color online)
			The $\beta$ distributions for the $\mathbb{P}$A (blue dashed line), $\mathbb{R}$A (magenta dotted line) and $\mathbb{ P+R}$A (black dash dotted line) in   photon-Pomeron and-Reggeon induced processes  for  AA collisions.}
		\label{fig19:limits}
	\end{figure}

In Fig.\ref{fig19:limits} we plot $\beta$ distributions for the heavy nuclei. The $\beta$ Pomeron and Reggeon distributions present the concavity and convexity at the end point of the distributions. The Reggeon contribution decrease with the increase of $\beta$. The Pomeron contribution drops for small $\beta$, comes to be flat and falls again for large $\beta$. There are  concavity and convexity at the end points. 	
\subsubsection{pA collision}
The  photon-Pomeron and-Reggeon induced cross section results in $\rm pA$ mode for  $\eta_{c}$ production is organized in Table.\ref{tab:9}. Two different events are considered in this case. The incoming colliding particles are dissimilar. 
For the first event ($\mathbb{P}$A and $\mathbb{R}$A), the   incoming nucleus $\rm A$ radiates photon
which collides with the hard diffractive gluon hailing from Pomeron or Reggeon of the incoming proton $\rm p$. The $\eta_{c}$, undissociated proton and nucleus along with remnants are observed in the final state.
We have noticed that the  Reggeon contribution to the total cross section is significant than that of the Pomeron.
The total DD cross section for  gluon-Pomeron and-Reggeon in $\rm pp$ ($\rm AA$) ($\rm pA$) mode is a factor $\gtrsim$ $10^{2}$ ($10^{5}$) ($10^{5}$) larger than that of $\rm pA$ mode in photon-Pomeron and -Reggeon induced interactions. The total SD cross section for  gluon-Pomeron and-Reggeon in $\rm pp$ ($\rm AA$) ($\rm pA$) mode is a factor $\gtrsim$ $10^{3}$ ($10^{6}$) ($10^{5}$) larger than that of $\rm pA$ mode in photon-Pomeron and -Reggeon induced interactions. The total cross section in $\rm pA$ mode
is a factor $\gtrsim$ 10  larger than that of $\rm pp$ mode in photon-Pomeron and -Reggeon induced interactions which is due the photon distribution from $\rm A$, i.e., proportional to $\rm Z^{2}$, except for the first light nucleus where $\rm pA$ and $\rm pp$ modes are of a same order of magnitude. The total cross section for  photon-Pomeron and -Reggeon induced interactions in $\rm AA$  mode is a factor $10^{3}$ ($10^{2}$)(10) larger than that of $\rm pA$ mode for the two first heavy nuclei(last heavy and medium nuclei)(first medium nuclei). They are of a same order of magnitude for the light a nuclei and the middle medium nucleus. 

For the second event ($\mathbb{P}$p and $\mathbb{R}$p), the incoming proton $\rm p$ radiates photon
which interacts with the hard diffractive gluon stemming from Pomeron or Reggeon of the  incoming nucleus $\rm A$. The $\eta_{c}$ is formed from the hard interaction of photon with the hard diffractive gluon, plus remnants. The intact or early intact proton and nucleus are also detected in the final state. We observe that the Pomeron contribution to the total cross section is substantial than that of the Reggeon. The total DD cross section for  gluon-Pomeron and-Reggeon in $\rm pp$ ($\rm AA$) ($\rm pA$) mode is a factor $\gtrsim$ $10^{3}$ ($10^{6}$) ($10^{6}$) larger than that of $\rm pA$ mode in photon-Pomeron and -Reggeon induced interactions. The total SD cross section for  gluon-Pomeron and-Reggeon in $\rm pp$ ($\rm AA$) ($\rm pA$) mode is a factor $\gtrsim$ $10^{4}$ ($10^{7}$) ($10^{6}$) larger than that of $\rm pA$ mode in photon-Pomeron and -Reggeon induced interactions. The total cross section in $\rm pA$  is a factor $\gtrsim$ 10  larger than that of $\rm pp$ mode in photon-Pomeron and -Reggeon induced interactions which is due to the photon distribution from $\rm p$, i.e., proportional to $\rm Z^{2}$, the hard diffractive gluon distribution from $\rm A$ and the nuclear form factor, i.e., proportional toto $\rm A^{2}R_{g}F^{2}_{A}(t)$, except for the two last light nuclei where $\rm pA$ and $\rm pp$ modes are of a same order of magnitude. The total cross section for  photon-Pomeron and -Reggeon induced interactions in $\rm AA$  mode is a factor $\gtrsim$ 10 larger than that of $\rm pA$ mode in the same process.
 The Pomeron and Reggeon contributions to the total cross section should be measured separately in two situations because of their opposite contribution  at the LHC for $\rm pA$ mode.
	\begin{table*}[htbp]
		\begin{center}
			\begin{tabular}{|c |c | c | c |c |c |c |c |c |}\hline	
				Nuclei   & pA      & $\rm \langle \left\vert S\right\vert ^{2}\rangle $             & $\mathbb{P}$A              & $\mathbb{R}$A & Total1 & $\mathbb{P}$p & $\mathbb{R}$p & Total2           \\ \hline
				\multirow{3}{*}{Light nucleus}& pBe &1 &1.38$\times$10$^{-4}$ & 3.80$\times$10$^{-4}$  &5.18$\times$10$^{-4}$ & 4.06$\times$10$^{-5}$ & 2.36$\times$10$^{-5}$ &6.42$\times$10$^{-5}$\\
				\multirow{3}{*}{}&pC  &1 &2.94$\times$10$^{-4}$& 7.98$\times$10$^{-4}$ &1.09$\times$10$^{-3}$
				& 6.64$\times$10$^{-5}$  & 3.64$\times$10$^{-5}$ &  1.03$\times$10$^{-4}$ \\
				& pO   & 1 & 4.93$\times$10$^{-4}$ & 1.32$\times$10$^{-3}$&1.81$\times$10$^{-3}$
				& 1.08$\times$10$^{-4}$&5.61$\times$10$^{-5}$ &1.64$\times$10$^{-4}$ \\
				\cline{1-9}
				\multirow{3}{*}{Medium nucleus}& pCa &1 &	2.54$\times$10$^{-3}$ & 6.52$\times$10$^{-3}$ & 9.06$\times$10$^{-3}$ &5.11$\times$10$^{-4}$   & 2.21$\times$10$^{-4}$    &7.32$\times$10$^{-4}$   \\
				\multirow{3}{*}{}&pCu  &1&4.81$\times$10$^{-3}$ & 1.21$\times$10$^{-2}$ &1.69$\times$10$^{-2}$ 
				& 1.12$\times$10$^{-3}$ &4.44$\times$10$^{-4}$ &1.57$\times$10$^{-3}$
				 \\
				&pAg  & 1 &1.12$\times$10$^{-2}$  & 2.74$\times$10$^{-2}$ & 3.86$\times$10$^{-2}$ 	
				&2.70$\times$10$^{-3}$ &9.64$\times$10$^{-4}$  & 3.66$\times$10$^{-3}$\\
				\cline{1-9}
				\multirow{3}{*}{Heavy nucleus}& pAu  &1 & 2.72$\times$10$^{-2}$  & 6.50$\times$10$^{-2}$ & 9.22$\times$10$^{-2}$& 7.37$\times$10$^{-3}$   &2.34$\times$10$^{-3}$ &9.70$\times$10$^{-3}$  \\
				\multirow{3}{*}{}&pPb  &1&2.89$\times$10$^{-2}$ &6.89$\times$10$^{-2}$ &9.78$\times$10$^{-2}$  
				& 8.06$\times$10$^{-3}$&2.54$\times$10$^{-3}$ &1.06$\times$10$^{-2}$
				   \\
	
				& pU  &1  &  3.51$\times$10$^{-2}$& 8.32$\times$10$^{-2}$ & 1.18$\times$10$^{-1}$
				&  1.01$\times$10$^{-2}$&3.09$\times$10$^{-3}$ &1.32$\times$10$^{-2}$ 
				   \\
				\hline	
			\end{tabular}
		\end{center}\caption{\label{tab:9}
			The photon-Pomeron and-Reggeon induced cross sections in $\mu$b for the $\eta_{c}$ proton-nucleus productions at LHC with forward detector acceptance, $0.0015<\xi_{1}<0.5$.}
	\end{table*}

\begin{figure}[htp]
		\centering
		\includegraphics[height=4.8cm,width=5.9cm]{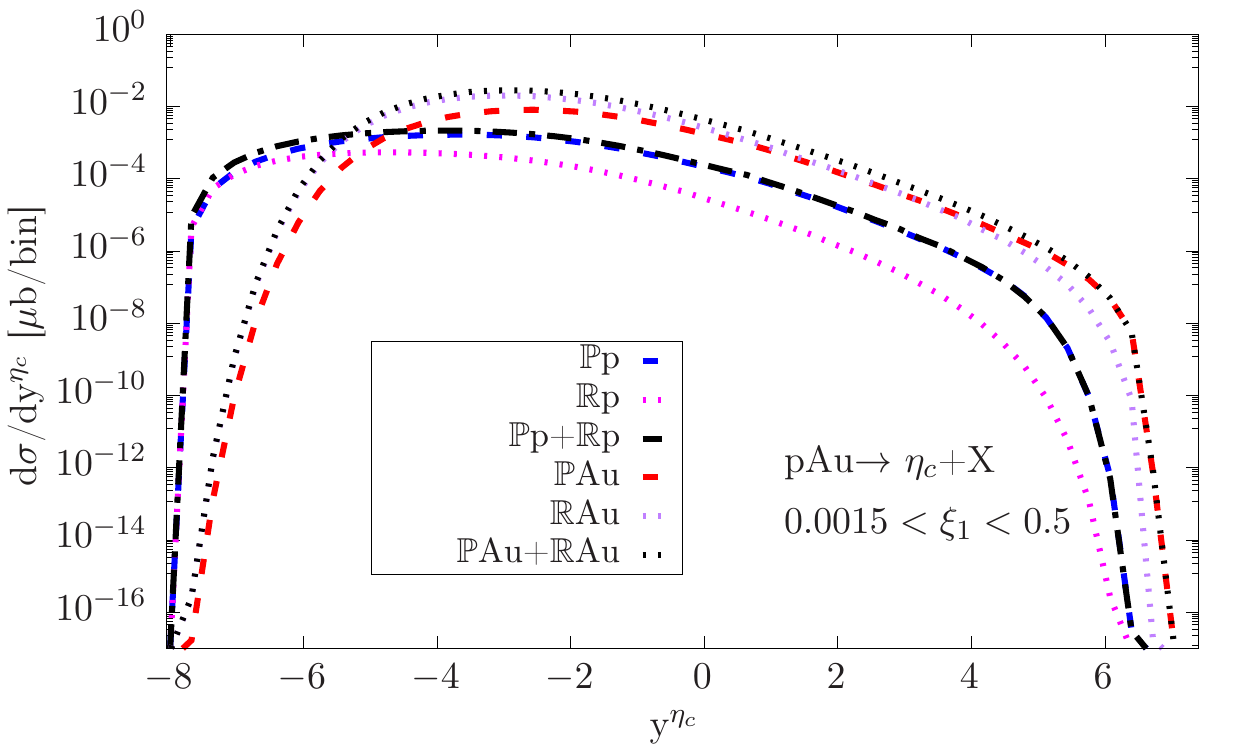}
		\includegraphics[height=4.8cm,width=5.9cm]{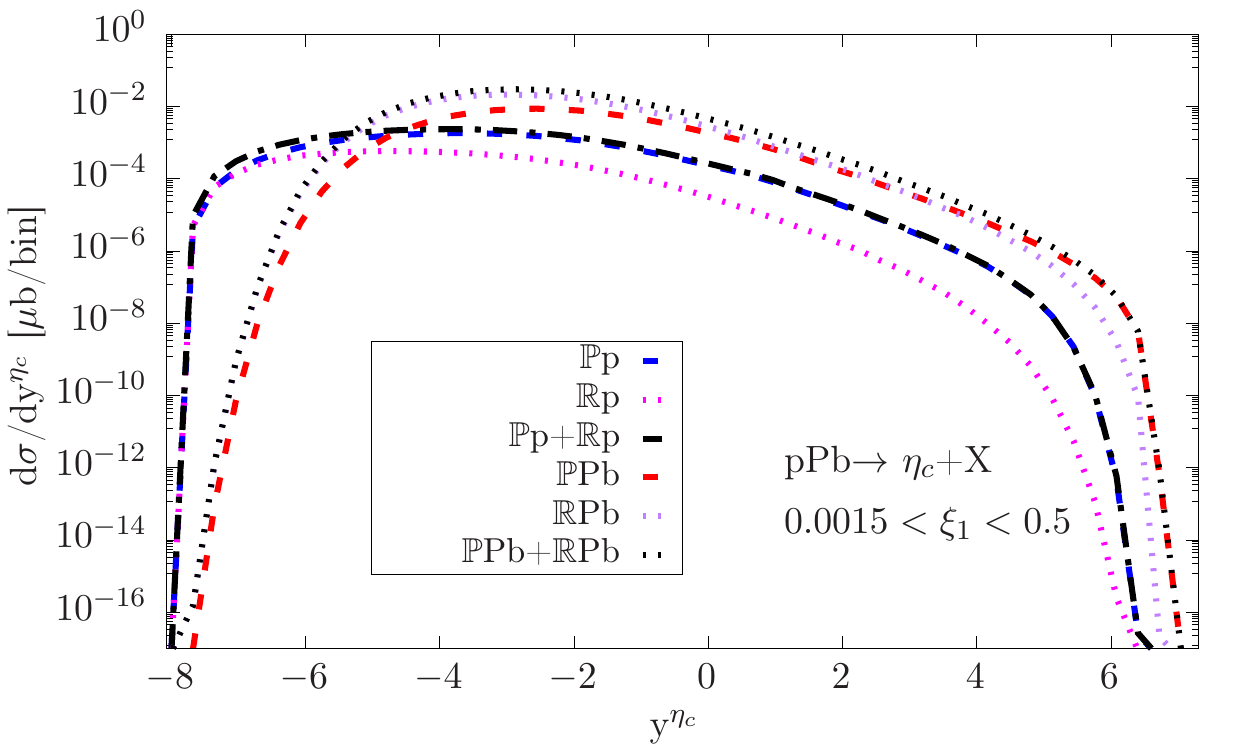}
		\includegraphics[height=4.8cm,width=5.9cm]{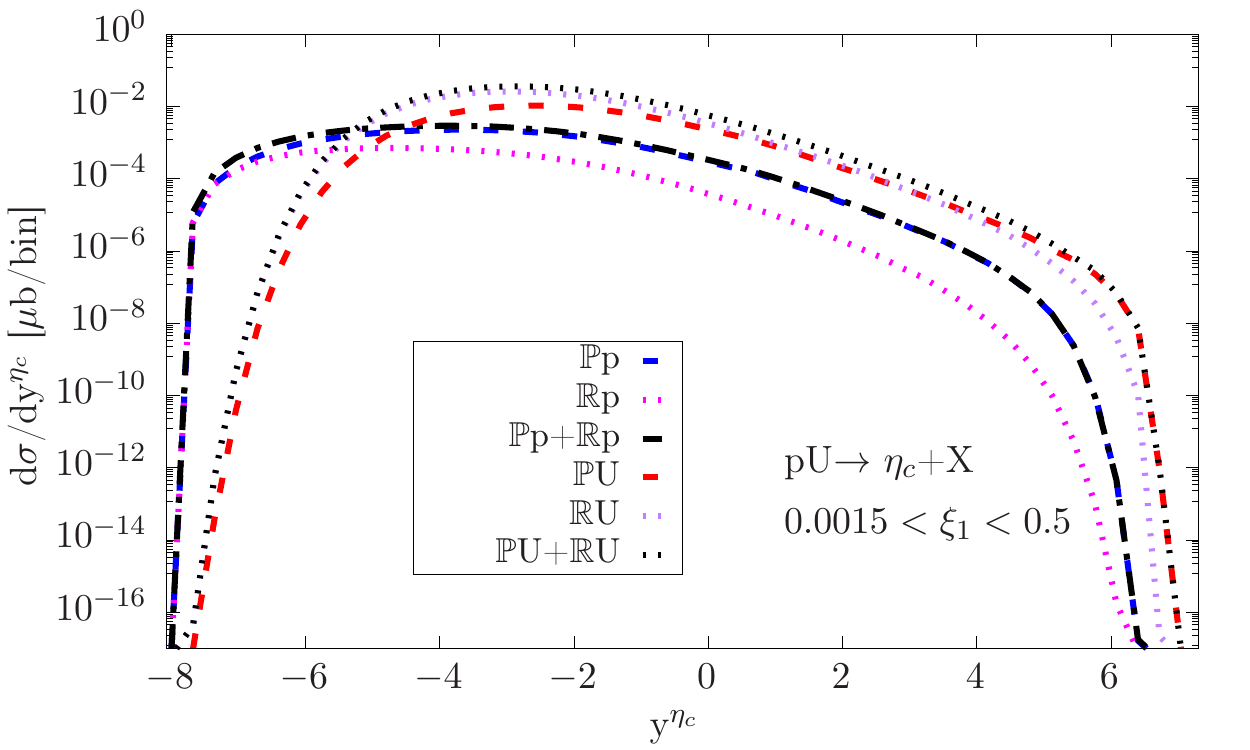}\\
		\caption{\normalsize (color online) The $\rm y^{\eta_{c}}$ distributions for the $\mathbb{P}$p (blue dashed line), $\mathbb{R}$p (magenta dotted line), $\mathbb{P}$p+$\mathbb{R}$p (black  dash dotted line), $\mathbb{P}$A (red  dashed  line),
			$\mathbb{R}$A (purple dotted line), $\mathbb{P}$A+$\mathbb{R}$A (black dotted line)
			) and $\rm gg$ (cyan dash dotted line)  in  photon and Pomeron induced processes  for pA collisions.}
		\label{fig20:limits}
	\end{figure}

The $\eta_{c}$ rapidity distributions in photon-Pomeron and-Reggeon induced  processes for two circumstances for $\rm pA$ mode are plotted in Fig.\ref{fig20:limits} for heavy nuclei. The  Pomeron distribution is substantial than that of the Reggeon distribution in $\rm pA$ mode ($\mathbb{P}$p,$\mathbb{R}$p). However, the Reggeon distribution dominates over  Pomeron distribution $\rm pA$ mode ($\mathbb{P}$A,$\mathbb{R}$A). The asymmetric rapidity distributions in two situations have maximums shifted to  backward rapidities with respect to the mid-rapidity because of  the photon and pomeron fluxes are different for a proton and a nucleus. The first circumstance distribution dominates over the second circumstance distribution and they keep the same order of magnitude for the middle heavy nucleus.

\begin{figure}[htp]
	\centering
	\includegraphics[height=4.8cm,width=5.9cm]{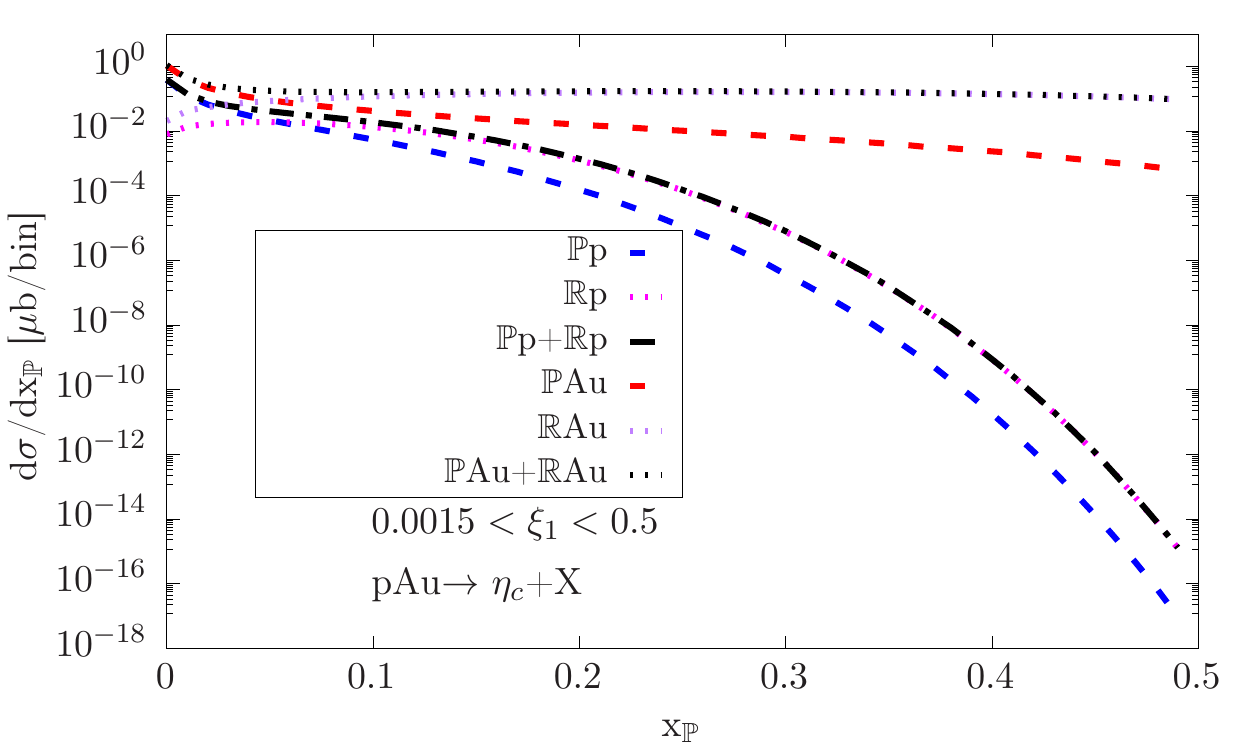}
	\includegraphics[height=4.8cm,width=5.9cm]{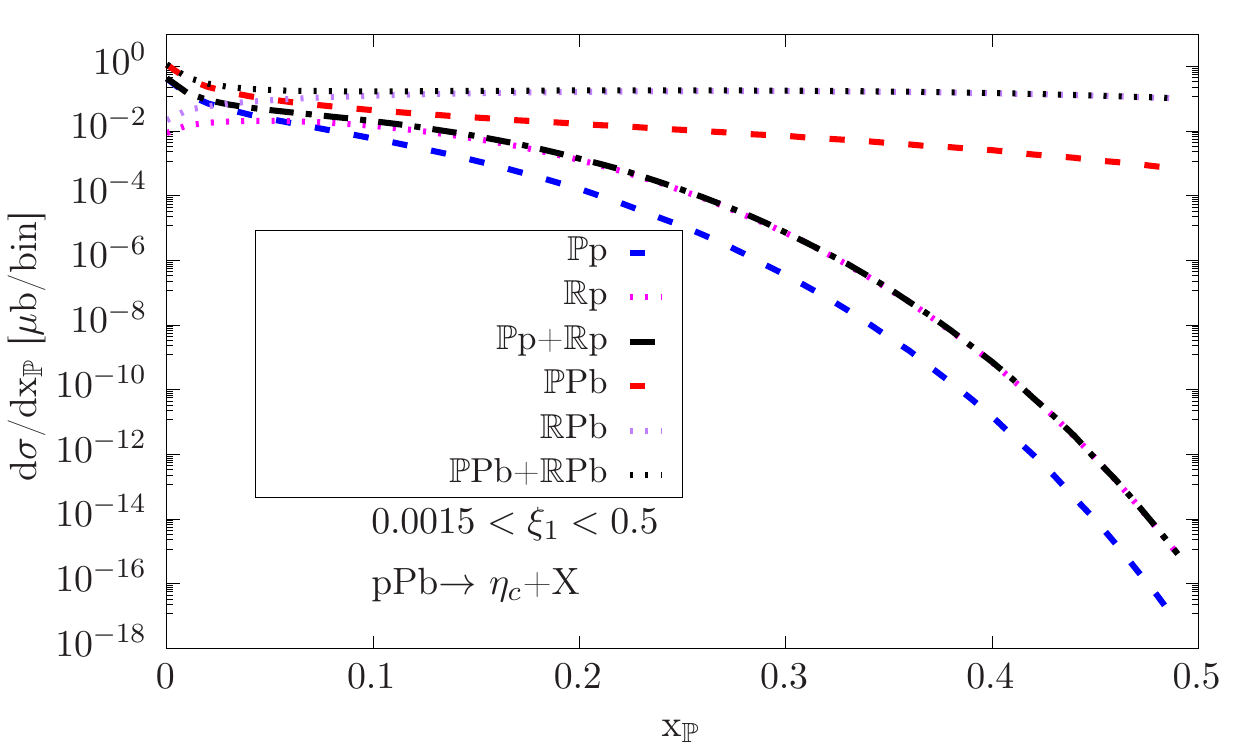}
	\includegraphics[height=4.8cm,width=5.9cm]{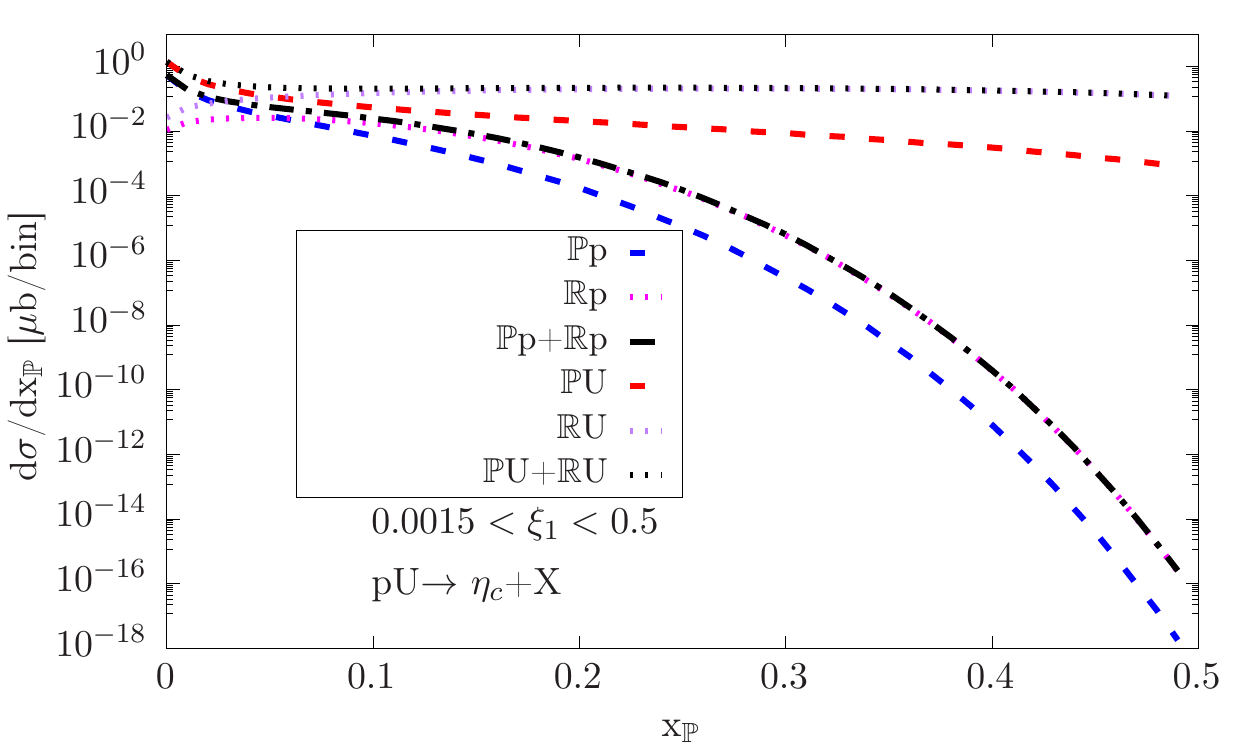}\\
	\caption{ \normalsize (color online)
		The $x_{\mathbb{P}}$ distributions for the $\mathbb{P}$p (blue dashed line), $\mathbb{R}$p (magenta dotted line), $\mathbb{P}$p+$\mathbb{R}$p (black  dash dotted line), $\mathbb{P}$A (red dashed line), $\mathbb{R}$A (purple dotted line) and $\mathbb{P}$A+$\mathbb{R}$A (black dotted line)  in  photon and Pomeron induced processes  for pA collisions.}
	\label{fig21:limits}
\end{figure}

The Fig.\ref{fig21:limits} shows $x_{\mathbb{P}}$ distributions of the $\rm pA$ collision for heavy nuclei for two scenarios as explained above. For the first and second scenario, the Pomeron distribution is larger than that of Reggeon one for small $x_{\mathbb{P}}$ and their distribution swaps for large $x_{\mathbb{P}}$ where the Reggeon distribution dominates. For the first scenario, the Pomeron distribution decreases for small and large $x_{\mathbb{P}}$ while Reggeon distribution increases for smallv$x_{\mathbb{P}}$ and become flat for large $x_{\mathbb{P}}$. For the second scenario,
the Pomeron and Reggeon distributions decrease for small and large $x_{\mathbb{P}}$. The total distribution of the first case dominates clearly over the total distribution of the second case for large $x_{\mathbb{P}}$.

\begin{figure}[htp]
	\centering
	\includegraphics[height=4.8cm,width=5.9cm]{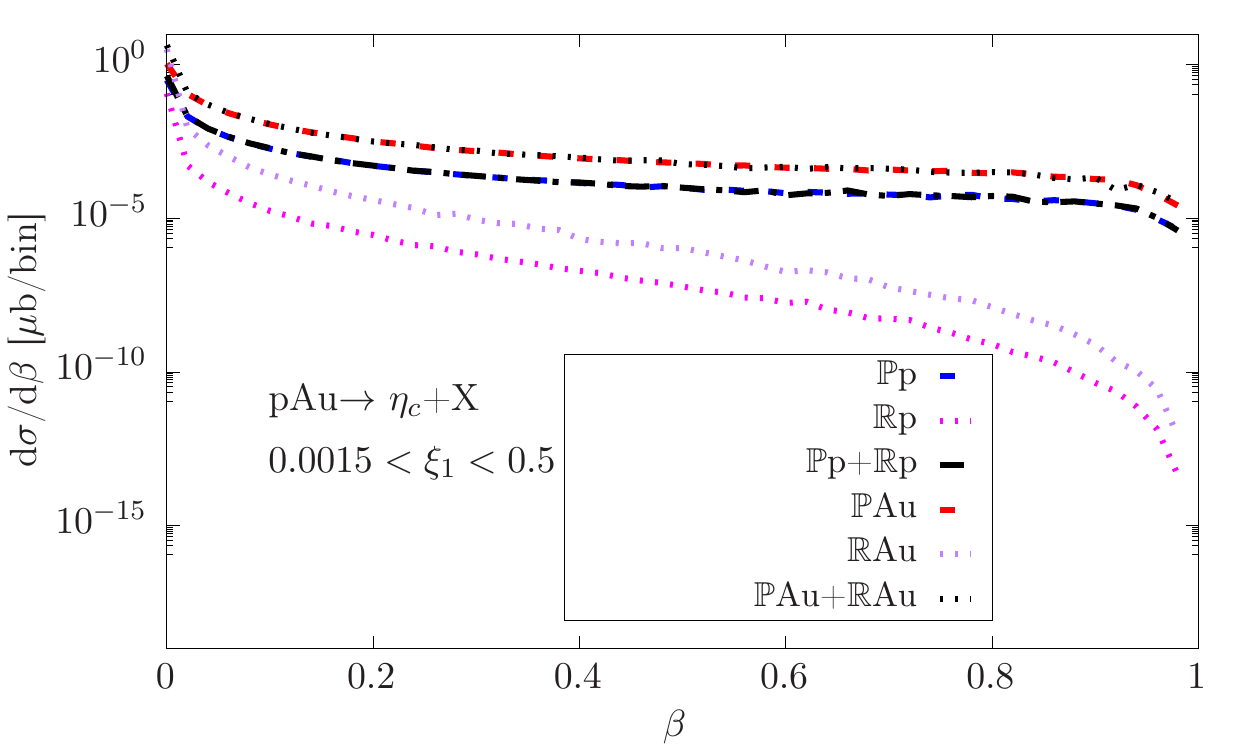}
	\includegraphics[height=4.8cm,width=5.9cm]{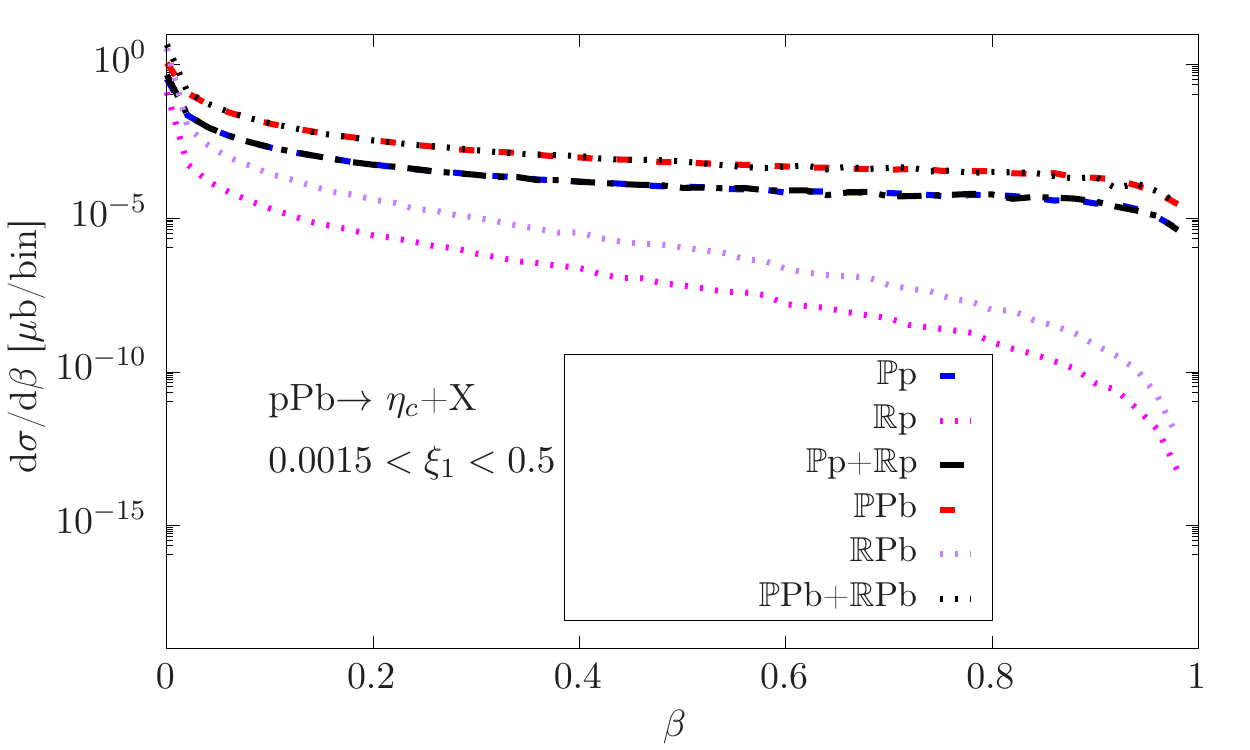}
	\includegraphics[height=4.8cm,width=5.9cm]{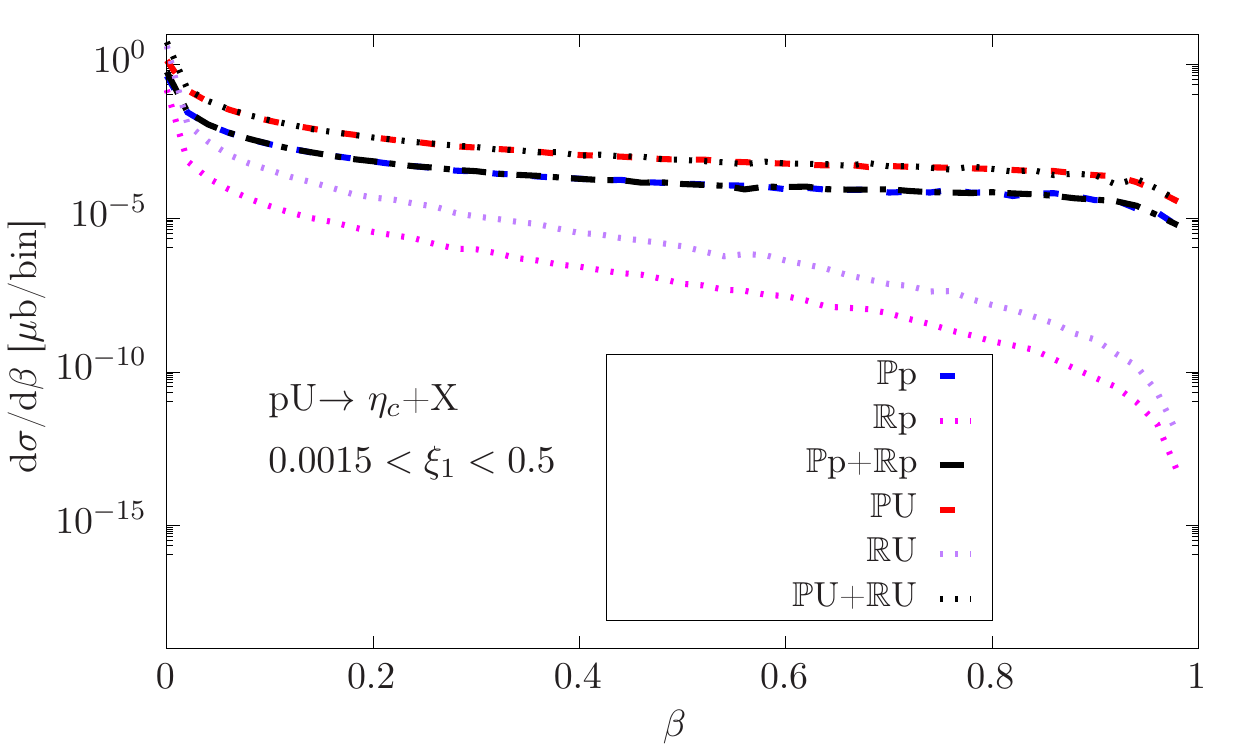}\\
	\caption{ \normalsize (color online)
		The $\beta$ distributions for the $\mathbb{P}$p (blue dashed line), $\mathbb{R}$p (magenta dotted line), $\mathbb{P}$p+$\mathbb{R}$p (black dash dotted line), $\mathbb{P}$A (red dashed line), $\mathbb{R}$A (purple dotted line) and $\mathbb{P}$A+$\mathbb{R}$A (black dotted line)  in  photon and Pomeron induced processes  for pA collisions.}
	\label{fig22:limits}
\end{figure}

The gluon longitudinal momentum fraction with respect to the exchanged Pomeron or Reggeon distributions of $\eta_{c}$ for two scenarios in $\rm pA$ mode through photon and Pomeron induced processes are plotted in Fig.\ref{fig22:limits} for heavy nuclei. The Pomeron distribution in two scenarios are dominant over  Reggeon distribution. The both distributions present the concavity and convexity at their end points. As we can notice, the Reggeon distributions decrease for small and large $\beta$ whereas the Pomeron distributions also decrease, shows the flatness in-between value of small and large $\beta$ and, falls again for large $\beta$.

\subsection{Uncertainties in diffractive process}

The total $\eta_{c}$ cross section prediction suffer from uncertainties within the used calculation framework. The uncertainty can originate from the gluon and photon distributions inside the proton and nucleus. The nuclear shadowing effect parametrization can be considered as an uncertainty source. Due to their different values in the literature, the heavy quark mass, the long distance matrix elements, the factorization scale or renormalization scale may be give the uncertainties in the modeling. \cite{Bi:2016vbt}. The gap survival probability may present the highest error of around $\pm 50$\% \cite{Kaidalov:2003fw} or  30 \% \cite{Luszczak:2014mta} in the total $\eta_{c}$  cross section  prediction and should be multiplied by a factor of $\rm 4$  at the CMS collider \cite{Chatrchyan:2012vc}. The findings from the use of two distinct diffractive Fits, H1 2006 dPDF Fit A and H1 2006 dPDF Fit B, give a slight change in the  total $\eta_{c}$ cross section prediction. The poor knowledge of the gluon density at high $\beta$ gives error of  25 \% which takes into account the uncertainty of QCD fits. It is related only to the gluon density from Pomeron $\rm g^{\mathbb{P}}(\frac{x_{g}}{x_{\mathbb{P}}},Q^{2})$, which is multiplied by an uncertainty factor $\rm (1-\beta)^{\nu}$ with $\rm \nu = -0.5$ or 0.5 \cite{Kepka:2007nr,Royon:2006by}. This high  $\beta$ region is of a specific attention for the large hadron collider(LHC) since it corresponds to for instance a direct background to the investigation of exclusive events\cite{Royon:2006jf}. The Reggeon contribution  error range is unknown in literature. The poor understanding of the gluon distribution and uncertainty in the gluon distribution itself due to infrared region is also another origin of error. The radiative corrections in the higher-order QCD give also extra errors \cite{HarlandLang:2009qe}.

\section{Summary and Conclusion}
\label{Conclusion}

In this work, we compute the production of $\eta_{c}$ via gluon-Pomeron and -Reggeon, photon-Pomeron and -Reggeon, Pomeron-Pomeron, Reggeon-Reggeon, Pomeron-Reggeon and Reggeon-Pomeron processes at the LHC with $\rm \sqrt{s} = 5.02$ TeV energies in $\rm pp$, $\rm pA$ and $\rm AA$ modes. Considering the NRQCD formalism along with the Regge theory formalism, we predict the total cross sections and the differential cross sections. Our results show that the contribution of Reggeon-Reggeon and Reggeon-Pomeron are non negligible for a chosen forward detector acceptance, and therefore this study can be useful to better constrain the Reggeon parton content and correct the experimental model with the specific choice of the mode at LHC.

\begin{acknowledgments}
Hao Sun is supported by the National Natural Science Foundation of China (Grant No.11675033) and by the Fundamental Research Funds for the Central Universities (Grant No. DUT18LK27).
\end{acknowledgments}

\bibliography{v3}

\end{document}